\documentclass[preprint,12pt,authoryear]{elsarticle}

\usepackage[margin=1in]{geometry}
\usepackage{graphicx}
\usepackage{newtxtext}
\usepackage{newtxmath}
\usepackage{natbib}
\usepackage{hyperref}
\hypersetup{
    colorlinks = true,
    urlcolor   = blue,
    citecolor  = black,
}

\newcommand{\RomanNumeralCaps}[1]
\linenumbers

\usepackage{bm}
\usepackage{multirow}
\usepackage{color}
\usepackage{psfrag}
\usepackage{scalerel}
\usepackage{siunitx}
\usepackage{stackengine,wasysym}
\usepackage{subcaption}
\usepackage{threeparttable}
\usepackage{tikz}
\usepackage{titlesec}
\usepackage{upgreek}

\usepackage[toc,page]{appendix}
\definecolor{myOrange}{HTML}{E69F00}
\definecolor{blueGreen}{HTML}{009e73}

\title{On the settling and clustering behavior of polydisperse gas-solid flows with application to pyroclastic density currents}

\author[label1]{E. Foster}
\author[label2,label3]{E.C.P. Breard}
\author[label1]{S. Beetham\footnote{Cooresponding author: sbeetham@oakland.edu}}
\address[label1]{Oakland University, Department of Mechanical Engineering}
\address[label2]{University of Edinburgh, School of Geosciences}
\address[label3]{University of Oregon, Department of Earth Sciences}

\begin{document}
\maketitle

Sedimenting flows occur in a wide range of society-critical systems, such as circulating fluidized bed reactors and particle-laden gravity currents such as pyroclastic density currents (PDCs), the most hazardous volcanic process. In these systems, where mass loading is sufficiently high ($\gg\mathcal{O}(1)$), momentum coupling between the phases gives rise to mesoscale behavior, such as formation of coherent structures. This heterogeneity has been demonstrated to generate and sustain turbulence in the carrier phase and directly impact large-scale quantities of interest, such as settling time, degree of heat transfer and chemical conversion efficiency. While contemporary work has explored the physical processes underpinning these multiphase phenomenon for \emph{mono}disperse assemblies of particles, the way in which \emph{poly}dispersity alters flow behavior has been largely understudied. Since all real-world flows consist of a polydisperse particulate phase, in which parameters, such as particle diameter, vary widely across the ensemble, understanding the role of polydispersity in gas-solid systems is critical for informing closures that are increasingly accurate and robust. For example, the sedimentation of pyroclastic ejecta is a critical process of PDCs and the mechanisms in which this process occurs have important implications for the extent and speed at which PDCs propagate. To this end, this work characterizes the sedimentation behavior of two polydispersed gas-solid flows, with statistical and physical properties of the particulate phase sampled from historical PDC ejecta. Highly resolved data for both polydisperse distributions of particles at two volume fractions (1\% and 10\%) is collected using an Euler-Lagrange framework and is compared with monodisperse configurations of particles with diameters equivalent to the arithmetic mean of the polydisperse configurations. From this data, we find that polydispersity has an important impact on cluster formation and structure and that this is most pronounced for more dilute flows. At higher volume fraction, the effect of polydispersity is less pronounced. In addition, we propose a new metric for predicting the degree of clustering, termed `surface loading', and a new model for the coefficient of drag as a function of particle diameter that produces accurate predictions for settling velocity given the high-fidelity data presented.


\section{Introduction}
\label{sec:Introduction}

% Hook: Sedimenting flows occur in a wide range of applications. All flows of interest include particles of variable properties (density, size, shape, etc.). Understanding them and accurately predicting their behavior is paramount to a similarly wide range of applications 
Sedimenting multiphase flows occur in a wide range of society critical applications, such as the upgrading of biomass into biofuels in circulating fluidized bed reactors~\citep{Grace2007, Mettler2012}, transport of contaminants within ground water~\citep{Miller1998} or the atmosphere~\citep{Ravishankara1997, Agranat2020} and several other environmentally-relevant systems, such as volcanic processes~\citep{Lube2020} and avalanches~\citep{Gardezi2022, Cicoira2022, Sovilla2018, Zhuang2023}. All of these inherently multiphase systems represent problems of great societal interest that are also historically challenging to study, both computationally and experimentally. It is these challenges that, to date, have hindered our ability to fully understand and accurately predict their behavior, thus impeding more optimal engineering designs and systems, as well as more effective procedures and standards for safety and risk mitigation.

% Problem: Multiphase flows are inherently challenging, especially given the wide range of length and time scales present and the numerous and varying physical processes present at different scales. These complications only proliferate, when the particle phase is considered to be variable. With increasing computational capabilities, recent work has begun to understand underlying physical processes at a range of scales and propose reduced order models, however the vast majority of this work has been on overly simplified systems of monodisperse (particles with uniform size, shape and density) particles. 

As is often the case in multiphase systems of interest, when the mass loading (i.e., the ratio of the specific mass of the disperse phase to the specific mass of the carrier phase, $\varphi = (\rho_p \alpha_p)/(\rho_f \alpha_f)$ is sufficiently high, e.g., $\varphi \gg 1$, the particle phase and fluid phase are considered to be strongly coupled~\citep{Elghobashi1992, Bosse2006, Breard2017, Breard2018}. This implies that the fluid's behavior is significantly altered by the presence of particles and that the fluid likewise has a non-trivial effect on the particles. As has been observed both experimentally~\citep{ Breard2017, Breard2018,Sovilla2018} and computationally~\citep{Capecelatro2014, Lube2020}, coupling between the phases gives rise to heterogeneity in the flow in the form of coherent structures, such as clusters. This heterogeneity arises due to the reduced drag felt by particles~\citep{Capecelatro2015} and represents a substantial departure from that of an equivalent uncorrelated (homogeneous) flow, e.g., a flow with a uniform or random dispersion of particles. As one may expect, this behavior has an important effect on nearly all quantities of interest for a multiphase system, such as settling velocity and degree of mixing in the flow. Several recent works have highlighted this effect and shown that heterogeneity can impede chemical conversion~\citep{Beetham2019}, alter thermal exchange efficiency~\citep{Guo2019} and entrance lengths~\citep{Beetham2023} and enhance the mean particle settling velocity~\citep{Bosse2006, Wang1993, Aliseda2002, Ferrante2003, Yang2003, Bosse2006}. 

In addition to the complexity generated by heterogeneity in sedimenting flows and the coupling between the phases, another key challenge is the breadth of length- and time-scales at which physical processes occur. At the particle scale, which is often on the order of micrometers, small wakes and boundary layer-induced instabilities are formed at the surface of particles and in systems that exhibit heat and mass transfer or are chemically reactive, these physical processes take place at very small time scales. These multi-physics processes drive the need for high computational resolution both in space and time in order to fully resolve relevant physics. Computational studies that aim to capture this level of detail often require grids fine enough that there are tens of grid points across the particle diameter~\citep{Subramaniam2020}. Juxtaposed with this, nearly all systems of societal interest span length scales on orders relevant to a reactor or volcanic topography (meters or kilometers) and time scales on the order of minutes or hours. In light of this, it is obvious that even with modern computing capabilities, fully-resolved computations of full-scale systems of interest is intractable. Because of this, several computational strategies exist for simulating multiphase flows at a range of scales. Particle resolved Direct Numerical Simulation (PR-DNS), while model free, is useful for studying behavior at the \emph{microscale} (i.e., systems on the order of hundreds of particles) and for developing the models required by coarser-grained methodologies, such as highly-resolved Euler--Lagrange simulations, where the fluid is treated in the Eulerian sense, particles are individually tracked as Lagrangian bodies and the phases are coupled through interphase exhange terms~\citep{Capecelatro2015}. In these coarser-grained simulations, the grid is on the order of 2-3 times larger than the particles, and models (informed by PR-DNS) are required to capture sub-grid behavior. Importantly, these methodologies allow simulations at the \emph{mesoscale} with hundreds of millions of particles~\citep{Capecelatro2014,Beetham2021}. Systems at this size are sufficiently large to observe the mesoscale heterogeneity that microscale PR-DNS simulations cannot. While this is a dramatic improvement in the scale of systems that can be considered, mesoscale systems are also computationally limited, making them incapable of tractably considering systems at scales of interest (i.e., full reactor scale or the several kilometers required to capture  volcano or avalanche runout). Thus, in the same way that PR-DNS data is useful for informing model closures for mesoscale (Euler-Euler or Euler-Lagrange) simulations, this resulting data can then in turn be used to formulate closures for even coarser-grained strategies, such as the multiphase Reynolds Averaged Navier--Stokes (Multiphase RANS) equations~\citep{Beetham2021}. 

It is worth emphasizing that this incremental and careful approach--moving strategically from PR-DNS to highly resolved Euler-Lagrange to even coarser grained Multiphase RANS simulations--is a necessity for retaining relevant mesoscale physics. While early work in multiphase flow modeling attempted to augment existing single-phase turbulence models~\citep{Sinclair1989, Dasgupta19986, Rao2012, Sundaram1994, Zeng2006, Cao1995}, this approach fails beyond very dilute suspensions of particles. This primarily owes to the fact that in multiphase systems, turbulent energy is generated at the scale of the particles and \emph{cascades up} to generate large-scale turbulence. This phenomenon is in direct opposition to classical turbulence theory~\citep{Kolmogorov1941} and underscores the shortcomings of extending single-phase turbulence models to multiphase flows. To this end, work by \citet{Fox2014} pointed out that deriving the multiphase RANS equations directly from the microscale equations (Navier--Stokes) is incapable of capturing the momentum exchange between the phases. Instead, it is necessary to formulate the multiphase RANS equations by averaging the mesoscale equations. This strategy, while exact, leads to several unclosed terms that require modeling. While recent work~\citep{Beetham2021} has proposed closures for these terms, these closures are limited to unbounded flows of spherical, monodisperse particles of a single density ratio and the development accurate, widely-applicable is still very much in its infancy. 

While a large body of work in the last decades~\citep{Balachandar2010, Subramaniam2020} has shed light on the physics underlying multiphase processes as well as formulations for guiding reduced order modeling of their behavior, most contemporary studies consider particles that are monodisperse. In contrast to this, all real-world flows of interest are comprised of \emph{polydispersed} particles, where particle parameters often span a wide range of diameters, shapes, densities, etc. We note that some recent studies have considered particulate phases that were not monodisperse(see, e.g.,~\citep{Fox2024}), however, several of these studies often were limited to bi-disperse assemblies of particles~\citep{Municchi2017}, quantities of particles that were too few to capture mesoscale behavior or were limited to particles too small to be considered strongly coupled~\citep{Islam2019}.  Thus, this area of multiphase research still remains understudied and macroscale models for mesoscale behavior are remain elusive. 

One society-critical example of a polydisperse gravity-driven, sedimenting flow, and the motivation for this work, is pyroclastic density currents (PDCs)--the fast-moving, gravity-driven flow of particulate matter resulting from the collapse of an ejected volcanic column, collapse of a volcanic lava (dome) or proximal material perched on steep slopes~\citep{Lube2020}. PDCs are the most destructive volcanic process and can cause extensive damage to human settlements, infrastructure, and ecosystems~\citep{Lube2020, Breard2017, Breard2023}. Understanding the physics behind pyroclastic density currents (PDCs) and accurately predicting their behavior is essential due to their impact on society and the environment. However, as discussed previously, their inherent complexity, owing to their multiphase nature, makes accurate prediction and the formulation of robust models challenging.

It is instructive to note here, that a PDC is typically considered to be comprised of three regions ranging from very dilute (and loosely coupled) to dense (granular flow dominant) (see Fig.~\ref{fig:AnatomyofAPDC})--the co-PDC ash-cloud, the dilute ash-cloud and the concentrated basal underflow. Given the definition of mass loading, the dilute ash-cloud layer exhibits a mass loading $\gg 1$ throughout, implying that cluster induced turbulence (CIT) will dominate the flow physics there. Further, the ash-cloud layer includes regions of the flow with a mass loading much smaller than unity. In this layer, buoyancy dominates flow behavior. Finally, in the concentrated basal underflow layer, the mass loading and volume fraction of particles is so high that granular methods must be used to assess the physical processes at play. 

Historically the study of PDCs has spanned field measurements, analog experiments, and computational methods. Direct observation of PDCs is typically done post-eruption, providing an average of intrinsically transient behavior. However, recent large-scale experiments\citep{Breard2017, Breard2018, Lube2015, Breard2019, Dellino2007, Brosch2022} have been conducted to understand the behavior of PDCs and propose models for how PDC properties relate to quantities of interest. These experiments are capable of providing real-time measurements but the challenges associated with experimentally probing gas-solid flows remain and results are still more coarsely-grained than computational methods.  

To this end, recent 1D \citep{Pouget2016, Degruyter2012, Folch2016, Dufek2015}, 2D \citep{Williams2008, Barsotti2008, Pfeiffer2005}, and 3D multiphase flow models \citep{Lube2020, Neri2007, Dufek20072, Costa2016, Cao2021} have been used to quantify the important physical processes of PDCs that are often inaccessible by large scale experimentation and develop improved reduced order forecasting models. To be successful, these models require accurate predictions of polydisperse settling behavior. While current state-of-the-art models can reveal some aspects of the internal structure of PDCs, they have not captured the heterogeneous effects of clustering to date. This is largely because highly resolved studies of the settling behavior and underlying physical processes present in polydisperse assemblies of particles has been, to date, largely understudied. 

In this work, we present an analysis of the clustering and settling behavior observed in high-fidelity Euler-Lagrange simulations for two polydisperse assemblies of particles, each at two volume fractions. Here, simulation parameters are chosen for consistency with PDCs, however, as previously described, these results have a much broader range of applications and are not limited to only PDC applications. In addition, comparisons are drawn against the behavior observed in analogous monodisperse configurations in an effort to isolate the effect of polydispersity on heterogeneous multiphase behavior. This work represents, to the authors' knowledge, the most highly-resolved study of polydisperse clustering and settling behavior to date and an initial step toward improved reduced-order models. 

\section{Methodology} 
\begin{figure}
    \centering
    \includegraphics[width=.99\textwidth]{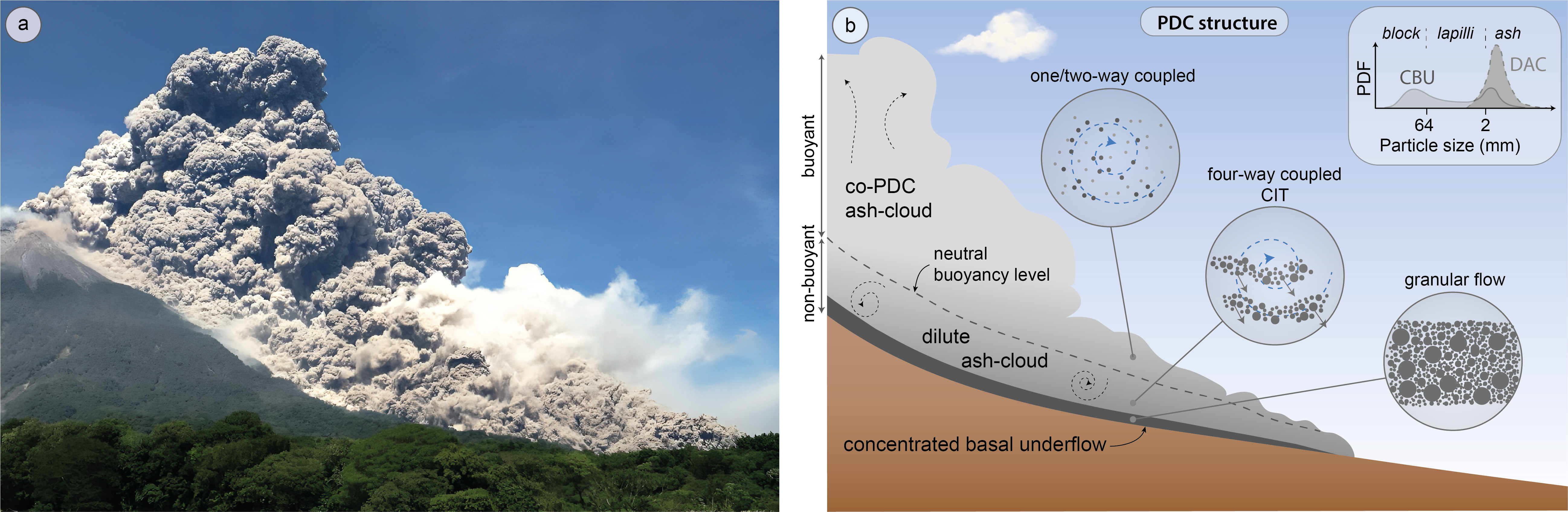}
    %\caption{Anatomy of PDCs}
\caption{a: Pyroclastic density current on September 13 2012 at Fuego volcano (Guatemala, photo courtesy of Vinicio Bejarano). b: Sketch of the anatomy of a pyroclastic density current, with concentrated basal underflow (CBU) fed by settling of particulates from the dilute ash-cloud (DAC) where abundant mesoscale clusters with cluster-induced turbulence (CIT) occurs. The upper part of the PDC is made of the co-PDC ash-cloud, which forms thermals that rise buoyantly and feeds co-PDC plumes that can reach heights up to tens of kilometers in the atmosphere.}
    \label{fig:AnatomyofAPDC}
\end{figure}

In general, PDCs can be partitioned into three flow regions: An upper dilute buoyant region, a dilute to intermediate density current region, and a concentrated basal underflow region (see Fig.~\ref{fig:AnatomyofAPDC}). These regions are coupled and, as such, evolve together. In the uppermost dilute region, the flow is buoyant, forming a turbulent thermal cloud where particles are dispersed at very low mass loading, resulting in one-way coupled behavior with the fluid phase (i.e., the particles tend to act primarily as tracers \citep{Elghobashi1991} and are unlikely to become correlated with one another, outside the mechanism of preferential concentrations due to underlying turbulent structures). In contrast, the bottom-most concentrated region is comprised of very high concentrations of particles ($\alpha_p \gtrapprox 0.3$) and is dominated by particle-particle interactions. Recent work~\citep{ Lube2020, Breard2016} showed that within pyroclastic density currents, the non-turbulent underflow and fully turbulent ash-cloud areas were connected by an intermediate zone characterized by cluster-induced turbulence (CIT). In this region, the dispersed and continuous phases are two-way coupled due to sufficiently high concentrations of particles and mass loadings higher than unity throughout the layer. This strong coupling between the fluid and particles leads to the formation of mesoscale structures in the form of clusters~\citep{Ferrante2003}, thereby resulting in altered particle settling when compared with a homogeneous particulate phase \citep{Lube2020, Breard2016}. The mechanism for this development of heterogeneity is primarily due to the reduced drag particles are subjected to due to the presence of other nearby particles. When drag is reduced, these particles can attain higher settling velocities and become correlated with their neighbors, thereby forming coherent structures, such as clusters. Because this intermediate region is critical in determining the sedimentation rate into the concentrated basal underflow (CBU) region, it serves as the motivation and context for this work. 

\subsection{System description}
\label{sec:systemDescription}
In particular, we aim to quantify the effect of polydispersity on clustering and settling behavior. To this end, we consider four configurations containing a polydisperse assembly of particles with parameters sampled from historical PDC ejecta data. In each of these configurations, the particles are assumed to be rigid spheres with diameter, $d_p$, and density, $\rho_{p}$. The gas has a constant density, $\rho_{f}$, and dynamic viscosity, $\nu_{f}$, which are specified based on the properties of air at 400$^o$C, the average temperature in the intermediate region \citep{Breard2017, Lube2020, Breard2016}. 

Particle density is constant across all configurations and is consistent with the density typically found in the intermediate layer of PDCs \citep{Lube2015, Breard2016}. Particle diameter distributions are chosen following the data presented in \citep{Lube2014}. In this study, we consider two lognormal distributions corresponding to $(\phi, \phi_{\text{sorting}}) = (0, 1)$ and $(1.5, 1)$ (for a description of $\phi$ and $\phi_{\text{sorting}}$, refer to Sec.~\ref{Appendix:sorting}), denoted as $A$ and $B$ throughout the manuscript. While the details of both distributions are summarized in Tab.~\ref{tab:Parameters}, it is important to note that the particles for each configuration are selected from the described distribution between the values of $d_{\text{min}}$ and $d_{\text{max}}$. This results in the mean and standard deviation of the particle assemblies in the simulations that deviate from the mean and standard deviation for a lognormal distribution described by $\mu$ and $\sigma$ with infinite support. 

\begin{table}
    \centering
    \vspace{-1em}
\begin{tabular}{l l l c c}
\multicolumn{3}{c}{\includegraphics[width=0.45\textwidth]{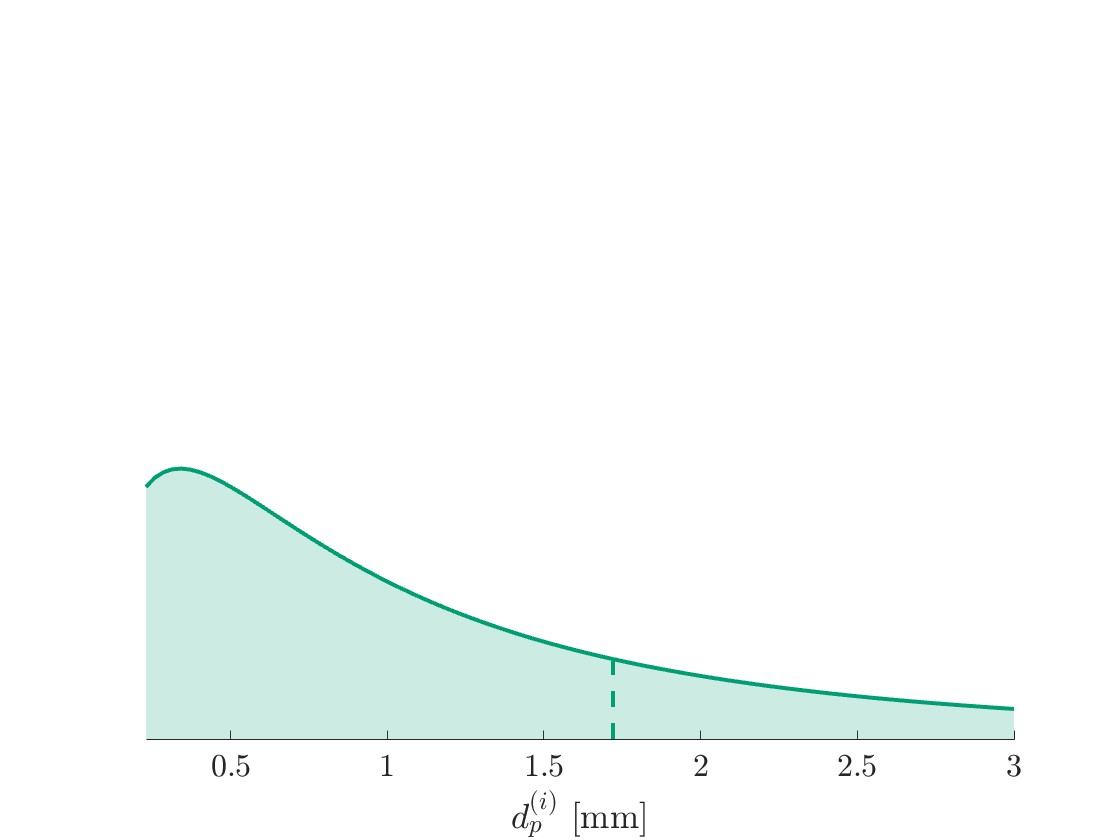} }& \multicolumn{2}{c}{\includegraphics[width=0.45\textwidth]{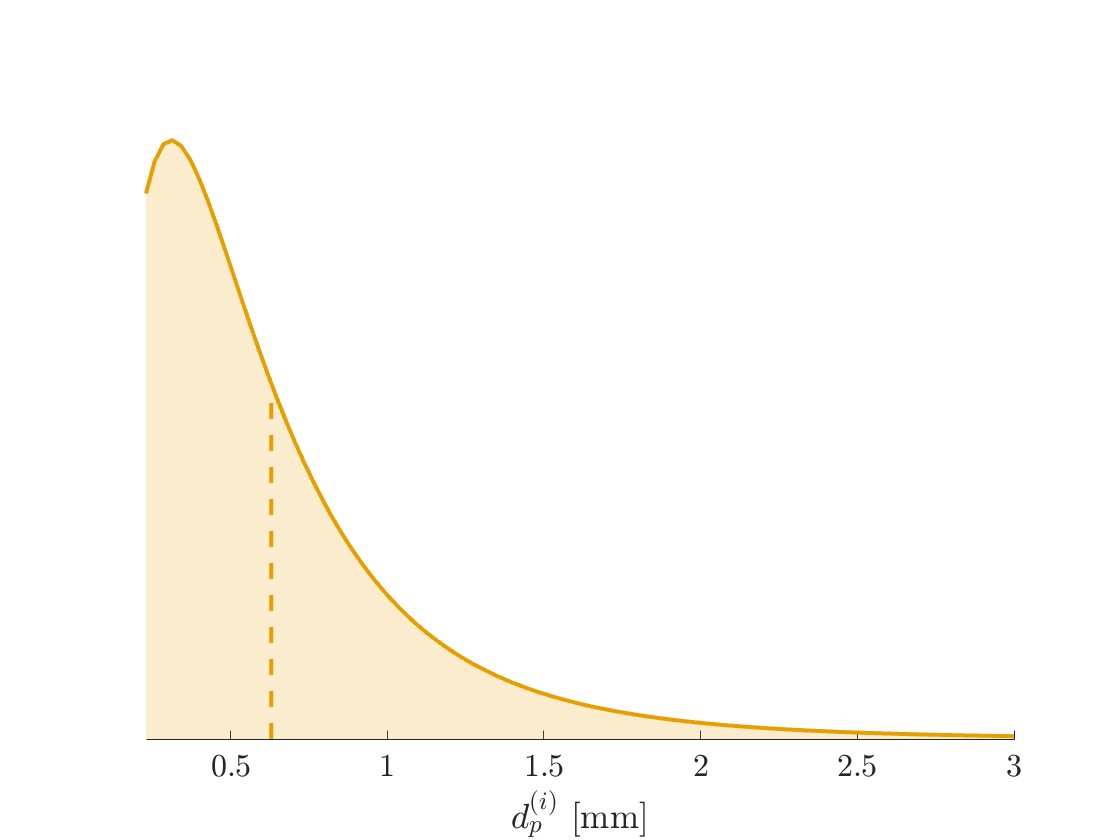}}\\ 
\multicolumn{3}{c}{\textcolor{blueGreen}{\textbf{Distribution $A$}}} &\multicolumn{2}{c}{\textcolor{myOrange}{\textbf{Distribution $B$}}}
\end{tabular}
\vspace{1em}
  \resizebox{0.8\textwidth}{!}{%
  \begin{tabular}{lllcc}
&\multicolumn{4}{c}{\textbf{Polydisperse particle-phase properties}}\\
\hline
 & & &{\textcolor{blueGreen}{\textbf{Distribution $A$}}} &{\textcolor{myOrange}{\textbf{Distribution $B$}}}\\
 lognormal & $\mu$ & & 0.00 & -0.69 \\   [-0.25ex]
    parameters & $\sigma$& & 1.04 &  \;0.69   \\ 
    skewness & skew$(d_p)$ & &0.8 & 1.8\\ 
    kurtosis & kurt$(d_p)$ & &2.8 & 7.1\\[1ex]
    cut off & $d_{\text{min}}$ &[mm]& {0.23} & 0.23  \\
    diameters & $d_{\text{max}}$ & [mm] & 3.00 & 3.00\\ [1ex]
    arithmetic mean diameter & $d_{p,10} = \langle d_p\rangle$ & [mm] & 1.11 & 0.69 \\ [0.5ex]
    standard deviation & $\sqrt{\langle d_p^{\prime 2}\rangle}$ &[mm] & 0.70 & 0.43 \\[2 ex]
 particle relaxation time & $\tau_{p,10}$& [s] & {9.47}& 3.61 \\
 Stokes settling velocity & $\mathcal{V}_{0,10}$ &[m/s] & {-0.15} & -0.06 \\
 Particle Reynolds number & Re$_{p,10}$ && {4.83} & 1.14 \\
 Froude Number & Fr$_{10}$ & &{$1.3\times10^3$} & 3.0$\times10^{2}$ \\
    volume fraction & $\langle \alpha_p \rangle$ & & (0.01, 0.1) & (0.01, 0.1) \\
    Number of particles & $N_p$ & & (19\;627, 198\;369) & (75\;695, 747\;431) \\ [1ex]
    &\multicolumn{4}{c}{\textbf{Monodisperse particle-phase properties}}\\
\hline
 & & &{\textcolor{blueGreen}{\textbf{Distribution $A_0$}}} &{\textcolor{myOrange}{\textbf{Distribution $B_0$}}}\\
particle diameter & $d_p$ & [mm] & 1.72 & 0.64 \\
particle relaxation time & $\tau_p$ & [s] & 22.74 & 3.12\\
Stokes settling velocity & $\mathcal{V}_{0}$ &[m/s] & -0.38 & -0.05 \\
 Particle Reynolds number & Re$_{p}$ & & 18.00 & 0.91 \\
 Froude Number & Fr & & 4.9$\times10^{3}$& 2.5$\times10^2$   \\
    volume fraction & $\langle \alpha_p \rangle$ & & (0.01, 0.1) & (0.01, 0.1) \\
    Number of particles & $N_p$ & & (12\;790, 127\;898) &  (251\;876, 2\;518\;757)\\ [1ex]
    &\multicolumn{4}{c}{\textbf{Distribution independent properties}}\\
\hline
fluid density & $\rho_f$ & [kg/m$^{3}$]& \multicolumn{2}{c}{0.5 }   \\
    fluid dynamic viscosity & $\mu_f$ & [kg /(m s)]& \multicolumn{2}{c}{1.85 $\times 10^{-5}$ } \\ 
    particle density & $\rho_p$ & [kg/m$^{3}$]& \multicolumn{2}{c}{2500} \\
    gravity & $\bm{g}$ & [m/s$^{2}$]&\multicolumn{2}{c}{(-0.016, 0, 0) } \\
    \emph{a priori} mass loading & $\varphi_0$ &  & \multicolumn{2}{c}{(50.27, 552.7)} \\ [2ex]
    Domain size & $L_x\times L_y \times L_z$ & [mm] & \multicolumn{2}{c}{($603.7\times75.10\times75.10$)}\\
    \end{tabular}}
\caption{\small Summary of simulation parameters under consideration. The pdfs shown above the table represent the distribution from which particles were sampled and are specified according to lognormal parameters $\mu$ and $\sigma$. The dashed vertical lines represent the diameter for the corresponding monodisperse simulations. } 
\label{tab:Parameters}
\end{table}

For each configuration $A$ and $B$, we draw comparisons with a monodisperse analog, denoted as $A_0$ and $B_0$, where the diameter of the particles in each of these is equivalent to the mean diameter of distributions $A$ and $B$, respectively. Here, these diameters are specified by the expected diameter of the distributions $A$ and $B$ with full support.  

In addition to studying the effect of two different degrees of polydispersity and two analogous monodisperse configurations, we also consider each configuration at two volume fractions. For consistency with the conditions typical of the intermediate region of PDCs \citep{Breard2016, Breard2017, Breard2018, Dufek20071, Lube2015, Lube2020}, we consider two global particle-phase volume fractions: $\langle \alpha_p \rangle = 0.01$ and $0.1$. Here, angled brackets denote a domain average and the particle volume fraction, $\alpha_p$, is defined as the ratio of the volume occupied by particles to the volume of the total domain. 

To isolate the effect of polydispersity on clustering and settling behavior, we consider the eight assemblies of particles described above in an unbounded domain and subjected to gravity. Due to the relative simplicity of such a configuration, these flows can be characterized by a small set relevant non-dimensional quantities: the particle Reynolds number, Re$_p$, the Froude number, Fr, and the Stokes number, St. The particle Reynolds number, Re$_p$, is defined as $\mathcal{V}_0 d_p/\nu_f$, where $\mathcal{V}_0$ is the Stokes settling velocity for a given particle diameter, defined as $\mathcal{V}_0 = \tau_p g$, where $\tau_p = \rho_p  d_p ^2/(18 \mu_f)$ is the Stokes settling time. The Froude number, which quantifies the balance between gravitational and inertial forces, is defined as Fr$=\tau_p^2 g/d_p$. A list of these non-dimensional numbers and other simulation parameters are summarized in Tab. \ref{tab:Parameters}. Note that quantities appended with a subscript of `10' corresponds to that quantity evaluated with a particle diameter equal to $D_{10} = \langle d_p \rangle$, the arithmetic mean particle diameter. A complete description of the definitions of statistical diameters can be found in ~\ref{appendix:Diameters}.

In this work, we utilize an Euler--Lagrange framework to capture high fidelity behavior of each of the particle configurations under study. The mathematical formulation for this framework is summarized in the subsequent sections. 

\subsection{Eulerian fluid phase description}
\label{sec:EulerEquations}
To account for the presence of particles in the fluid phase without resolving the boundary layers around individual particles, we consider the volume-filtered, incompressible Navier–Stokes equations~\citep{Anderson1967}. This procedure replaces the point microscale variables (fluid velocity, pressure, etc.) with smooth, locally filtered fields and requires models to be used for sub-grid-scale behavior, such as particle drag and particle-particle collision forces. Vectors and tensors in equations are represented in bold text. The volume-filtered conservation of mass and momentum equations are given by
\begin{equation} \label{eq:CMF}
\frac{\partial \left( \alpha_f \rho_f \right)}{\partial t} + \nabla \cdot (\alpha_f \rho_f \bm{u}_f) = 0
\end{equation}
and
\begin{equation} \label{eq:CMF2}
%\frac{\partial \left( \alpha_f \bm{u}_f \right)}{\partial t}+ \nabla \cdot (\alpha_f \bm{u}_f \otimes \bm{u}_f)= -\frac{1}{\rho_f} \boldsymbol{\nabla} p_f + \nabla \cdot \boldsymbol{\sigma}_f - \frac{\rho_p}{\rho_f} \alpha_p \bm{\mathcal{A}} + \alpha_f \bm{g},
\frac{\partial \left( \alpha_f \rho_f \bm{u}_f \right)}{\partial t}+ \nabla \cdot (\alpha_f \rho_f \bm{u}_f \otimes \bm{u}_f)=  \nabla\cdot \boldsymbol{\tau}_f + \alpha_f \rho_f \bm{g} + \bm{\mathcal{F}} + \bm{F}_{\text{mfr}},
\end{equation}
where $\bm{\mathcal{F}}$ accounts for the two-way coupling between the fluid and particle phases and is defined in Sec.~\ref{sec:interphaseExchange}. $\bm{F}_{\text{mfr}}$ is a source term imposed to ensure the system maintains a net constant mass flow rate in order to achieve a statistically stationary state. The fluid-phase viscous stress tensor is defined as 
\begin{equation} \label{eq:FFVS}
\boldsymbol{\tau}_f= -p_f \mathbb{I} + \mu_f [\nabla \bm{u}_f + (\nabla \bm{u}_f)^T - \frac{2}{3} \nabla \cdot \bm{u}_f \mathbb{I}],
\end{equation}
where $\mathbb{I}$ is the identity matrix. 

\subsection{Lagrangian particle phase equations}
\label{sec:ParticleEqs} 
The dispersed phase is solved using Lagrangian mechanics. The position and velocity of the particles are advanced according to Newton’s second law
\begin{align} 
\label{eq:ParticlePosition}
\frac{\text{d}\bm{x}_p^{(i)}}{\text{d}t}&=\bm{u}_p^{(i)}  \textrm{ and } \\
m_p^{(i)}\frac{\text{d}\bm{u}_p^{(i)}}{\text{d}t} &= \bm{F}^{(i)}_{\text{inter}}+\bm{F}^{(i)}_{\text{col}} + m_p^{(i)} \bm{g} \label{eq:partVel}
%\rho_p \nabla \cdot \boldsymbol{\tau}_f \lbrack \bm{x}_p^{(i)} \rbrack +\bm{f}^{(i)}_{\text{drag}} + \frac{1}{m_p}\bm{F}_c^{(i)} + \bm{g},
\end{align}
where the superscript $(i)$ denotes the $i$-th particle. Throughout the rest of this section, square brackets indicate a fluid quantity evaluated at the center of the $i$-th particle's location. 

As shown in Eq.~\ref{eq:partVel}, particles are subject to three forces: the force due to interphase momentum exchange, $\bm{F}_{\text{inter}}$, the force due to inter-particle collisions, $\bm{F}_{col}$, and the body force.  Here, the interphase exchange term is given by, 
\begin{equation} 
\label{eq:InterExch2}
\bm{F}_{\text{inter}}^{(i)}= V_p^{(i)} \nabla \cdot \boldsymbol{\tau}_f\lbrack \bm{x}_p^{(i)}\rbrack + \frac{m_p^{(i)} \alpha_f\lbrack \bm{x}_p^{(i)}\rbrack F_d\left(\alpha_f, \text{Re}^{(i)}_p\right)}{\tau_p^{(i)}} (\bm{u}_f\lbrack\bm{x}_p^{(i)}\rbrack - \bm{u}_p^{(i)}),  
\end{equation} 
where the rightmost term is the drag force felt by the $i$-th particle. Here, ${F}_d$ is the non-dimensional drag correction of \citet{Tenneti2011} that takes into account local volume fraction and Reynolds number effects and is given by, 
\begin{equation} 
\label{eq:Nonddrag}
F_d(\alpha_f , Re_p) =\frac{1 +0.15Re_p^{0.687}}{\alpha_f^2}  + \alpha_fF_1(\alpha_f) + \alpha_fF_2(\alpha_f , Re_p).
\end{equation}
Here, the local particle Reynolds number is defined as 
\begin{equation} 
\label{eq:ReynP}
\text{Re}_p^{(i)} = \frac{\alpha_f\lbrack\bm{x}_p^{(i)}\rbrack \left\vert \bm{u}_f\lbrack \bm{x}^{(i)} \rbrack - \bm{v}_p^{(i)}\right\vert d_p}{\nu_f}
\end{equation}
and the remaining two terms in \ref{eq:Nonddrag} are given by
\begin{equation}
F_1(\alpha_f)= \frac{5.81 \alpha_p}{\alpha_f^3} +  \frac{0.48 \alpha_p^{1/3}}{\alpha_f^4}
\end{equation}
and
\begin{equation}
F_2(\alpha_f, \text{Re}_p)= \alpha_p^{3} \text{Re}_p \left(0.95 + \frac{0.61 \alpha_p^3}{\alpha_f^2}\right).
\end{equation}

The force of collisions is accounted for using a soft-sphere collision model~\citep{CundallandStrack1979} and particles are treated as inelastic and frictional with a coefficient of restitution of 0.85 and coefficient of friction of 0.1.

\subsection{Two-way coupling}
\label{sec:interphaseExchange}
The fluid-phase equations introduced in Sec.~\ref{sec:EulerEquations} contains two interphase exchange terms: the particle volume fraction, $\alpha_f$, and the momentum exchange term, $\boldsymbol{\mathcal{F}}$. Each of these terms requires projection of the Lagrangian particle information to the Eulerian mesh. We make use of the two-step filtering approach proposed by \citet{Capecelatro2013}. In this approach, the particle-localized data is extrapolated to the nearest grid points and is implicitly smoothed using a Gaussian filter kernel (denoted $\mathcal{G}$ in Eqs.~\ref{eq:volfrac} and \ref{eq:momentumexchange}, below) with a width equal to 7 times the mean particle diameter. 

Given this, the particle phase volume fraction is defined as 
\begin{equation} 
\alpha_f = 1- \sum_{i=1}^{N_p} \mathcal{G}\left(\vert\bm{x} - \bm{x}^{(i)}_p \vert\right) V_p^{(i)} \label{eq:volfrac},
\end{equation}
where $N_p$ is the total number of particles and $V^{(i)}_p$ is the volume of the $i$-th particle. 

Similarly, the interphase momentum exchange is defined as 
\begin{equation}
\label{eq:momentumexchange}
\bm{\mathcal{F}} = - \sum_{i=1}^{N_p}\mathcal{G}\left(\vert\bm{x} - \bm{x}^{(i)}_p \vert\right) \bm{F}_{\text{inter}}.
\end{equation} 

\subsection{Computational methodology} 
\label{Computational methodology}
The equations presented in Secs.~\ref{sec:EulerEquations} through \ref{sec:interphaseExchange} are evolved using an in-house code, NGA ~\citep{Desjardins2008}, a fully conservative, low-Mach-number finite volume solver. A pressure Poisson equation is solved to enforce continuity via fast Fourier transforms in all three periodic directions. The fluid equations are solved on a staggered grid with second-order spatial accuracy and advanced in time with second-order accuracy using the semi-implicit Crank–Nicolson scheme of \citet{Pierce2001}. Lagrangian particles are integrated using a second-order Runge–Kutta method. Fluid quantities appearing in Sec.~\ref{sec:ParticleEqs} are evaluated at the position of each particle via trilinear interpolation and particle data are projected onto the Eulerian mesh using the two-step filtering process described in \citet{Capecelatro2013}. 

Gravity is oriented in the negative $x$ direction and is specified such that flow statistics are properly resolved. The domain size and grid spacing is specified such that clustering statistics are properly resolved. Namely, the grid spacing is set to be no greater than 1.75 $d_{p,\text{min}}$ and the domain length in the gravity-aligned direction is speci several times larger than the expected length of clusters. These domain parameters are summarized in Tab.~\ref{tab:Parameters}. 

It is important to note that while prior work~\citep{Capecelatro2016} has suggested the domain size for gravity-driven gas-solid flows should be at least 32$\tau_p^2 g$ long in the streamwise direction in order to properly resolve flow statistics, this guidance originated from monodisperse particles of a smaller diameter (90 $\mu$m) than the mean particles considered in this work. Thus, we have considered a domain length large enough to contain a sufficient number of particles to observe clustering behavior.

The domain for all the configurations under study are triply periodic. Particles are initially randomly distributed in the domain and the fluid phase is initially quiescent. After a transient period, the flow reaches a statistically stationary state, which is assessed by monitoring the mean particle settling velocity and the mean variance in particle volume fraction. 

\section{Phased averaged quantities of interest}
\label{sec:PhaseAveragedEqs}
As described in Sec.~\ref{sec:Introduction}, coarse-grained models often make use of averaging. Thus, phase-averaged terms of high-fidelity data are useful for constructing improved models. In this section, we present several flow equations and quantities to aid in the quantification of the configurations under study and lay the foundation for improved models. This section serves as a brief summary, however, more details regarding the quantities presented here can be found in \citet{Capecelatro2015} and \citet{Beetham2021}. 

For multiphase flows, it is convenient to introduce a phase average after taking the Reynolds average of the volume- filtered questions. These phase averages are analogous to Favr\'{e} averages for variable-density flows and are denoted by $\langle(\cdot) \rangle_p = \langle \alpha_p (\cdot) \rangle / \langle \alpha_p \rangle$ and $\langle(\cdot) \rangle_f = \langle \alpha_f (\cdot) \rangle / \langle \alpha_f \rangle$. Phase averages are denoted with angled brackets and a subscript $f$ or $p$ to indicate the phase with respect to which the average has been taken. Angled brackets without a subscript indicate a standard Reynolds average, which consists of an average overall spatial dimensions in this work. Fluctuations about particle phase-averaged quantities are denoted with two primes, e.g., $\bm{u}_p^{\prime\prime}=\bm{u}_p- \langle \bm{u}_p \rangle_p$, with $\langle \bm{u}_p^{\prime\prime} \rangle_p=0$. Fluctuations from fluid phase-averaged quantities are denoted with three primes, e.g., $\bm{u}_f^{\prime\prime\prime}=\bm{u}_f- \langle \bm{u}_f \rangle_f$.

For fully-developed gravity-driven flow, phase-averaged variables are statistically stationary. As a consequence of this, the phase-averaged continuity equation implies $\alpha_f$ is constant. Further, the fluid-phase momentum equation reduces to $\langle u_f \rangle_f=0$. Taking the phase average of Eq.~\ref{eq:partVel} projected to the Eulerian mesh results in the particle phase-averaged momentum equation. The $x$-direction component of this expression, the only non-zero component, is given as   

\begin{equation} \label{eq:PV}
\frac {\partial \langle u_p \rangle_p}{\partial t} = \frac{1}{\tau_p^*}(\langle{u_f}\rangle_{p}-\langle{u_p}\rangle_{p})  + \frac{1}{\rho_p} \left(\left \langle \frac{\partial \sigma_{f,xi}}{\partial x_i}\right \rangle_p - \left\langle \frac{\partial p_f}{\partial x} \right\rangle_p \right) +g.
\end{equation}

Here, we note that the transport of the phase-averaged particle velocity results in a balance between the forces exchanged by the fluid (i.e. drag model and resolved surface stresses) and gravity. Prior work~\citep{Capecelatro2015, Beetham2021} has shown that the fluid stress and fluid pressure gradient terms are small enough for gas-solid flows to be reasonably neglected. This implies that at steady state, $\langle u_p \rangle_p \approx \langle u_f \rangle_p + \tau_p^{\star} g$. Here, $\langle u_f\rangle_p$ is often thought of as the fluid velocity \emph{seen by the particles}. It is also notable that we incorporate the nonlinearities associated with the drag in $\tau_p^*=\langle\tau_p\rangle/ \langle F_d \rangle_p$, where $\langle F_d \rangle_p(\langle \alpha_f \rangle , \langle Re_p \rangle)$ and $\langle \tau_p \rangle(\langle d_p \rangle)$ are the nonlinear drag correction and particle response time with averaged flow quantities used as argument to Eq.~\ref{eq:Nonddrag} and the definition of $\tau_p$. In the following sections, we will make use of this relationship to propose a model for settling velocity that follows the formulation that the mean settling velocity is equal to the Stokes settling velocity plus an unclosed term. 

Phase averaging also gives rise to two Reynolds stress tensors, one in each of the particle and fluid phases, $\langle \bm{u}_p^{\prime \prime} \bm{u}_p^{\prime \prime}  \rangle_p$ and  $\langle \bm{u}_f^{\prime \prime \prime} \bm{u}_f^{\prime \prime \prime}  \rangle_f$, respectively. Analogous to single-phase turbulence, the trace of each of these tensors yields the turbulent kinetic energy in the particle and fluid phases, respectively. These are denoted as $k_p$ and $k_f$ and defined as 
\begin{align}
k_p &= \frac{1}{2}\langle \bm{u}_p^{\prime \prime} \cdot \bm{u}_p^{\prime \prime} \rangle_p \quad \text{and} \\ 
k_f &= \frac{1}{2}\langle \bm{u}_f^{\prime \prime \prime} \cdot \bm{u}_f^{\prime \prime \prime} \rangle_f.
\end{align}

It is notable that in the case of the particle phase averages, Lagrangian particle quantities, such as $\bm{u}_p$, must be evaluated on the Eulerian mesh. This is done by trilinear interpolation and the application of a Gaussian filter kernel, as described in Sec.~\ref{sec:interphaseExchange} and \citet{Capecelatro2013}. 

In addition to the phase-wise Reynolds stresses, it is also important to note that the Reynolds average of the inner product of the particle velocity fluctuations gives way to the total energy fluctuations in the particle phase, or the total granular energy, $\kappa_p$, given by
\begin{equation}
\kappa_p = \frac{1}{2}\langle \bm{u}_p^{\prime} \cdot \bm{u}_p^{\prime} \rangle. 
\end{equation}
Here, angled brackets denote a Reynolds decomposition according to $\bm{u}_p = \langle \bm{u}_p \rangle + \bm{u}_p^{\prime}$, with a single prime denoting a fluctuation from the Reynolds averaged quantity.  

When comparing the particle turbulent kinetic energy, $k_p$, and the total granular energy, $\kappa_p$, it is observed that there is an additional term that represents the uncorrelated, random component to the total particle energy that exists at the particle scale. This is the information that is lost when filtering Lagrangian particle data to the Eulerian mesh to evaluate $k_p$ and is termed `granular temperature', $\Theta$. This yields an expression for the total granular energy, 
\begin{equation}
\kappa_p = k_p + \frac{3}{2} \langle \Theta \rangle_p,
\end{equation}
that is equal to the sum of the correlated and uncorrelated contributions, $k_p$ and $3\langle \Theta\rangle_p/2$. 

We compute $\Theta_p$ directly, by computing the volume-filtered particle volume fraction and velocity. Here, the particle volume, the product of velocity and particle volume and the product of velocity squared and particle volume are extrapolated to the Eulerian grid via trilinear interpolation. These fields are then divided by the Eulerian cell volume and filtered using the Gaussian kernel described in Sec.~\ref{sec:interphaseExchange}, thus yielding the smoothed Eulerian quantities for particle volume fraction, $\tilde{\alpha_p}$, particle velocity, $\tilde{\bm{u}}_p$, and particle velocity squared, $\widetilde{\bm{u}^2}_p$. From this, the granular temperature is computed according to 

\begin{equation}
   \Theta_p = \frac{1}{3}\left\lbrack \frac{\text{tr} \left(\widetilde{\bm{u}^2}_p \right)}{\tilde{\alpha_p}} - \frac{\text{tr}\left(\tilde{\bm{u}}_p^2\right)}{\tilde{\alpha}_p} \right\rbrack.
\end{equation}

These quantities will be leveraged in the following discussion of our findings and utilized to propose models that capture polydisperse effects for both clustering and settling behavior. 

\begin{figure}
\centering
\begin{tabular}{c c c c @{\hskip 0.25in}c c c c}
\multicolumn{2}{c}{(a) $\langle \alpha_p \rangle = 0.01$} & 
\multicolumn{2}{c}{(b) $\langle \alpha_p \rangle = 0.1$} &
\multicolumn{2}{c}{(c) $\langle \alpha_p \rangle = 0.01$} & 
\multicolumn{2}{c}{(d) $\langle \alpha_p \rangle = 0.1$} \\ [0.5ex]
Dist. $A$ & Dist. $A_0$ & Dist. $A$ & Dist. $A_0$ & Dist. $B$ & Dist. $B_0$ & Dist. $B$ & Dist. $B_0$ \\
\includegraphics[height = 0.7\textwidth]{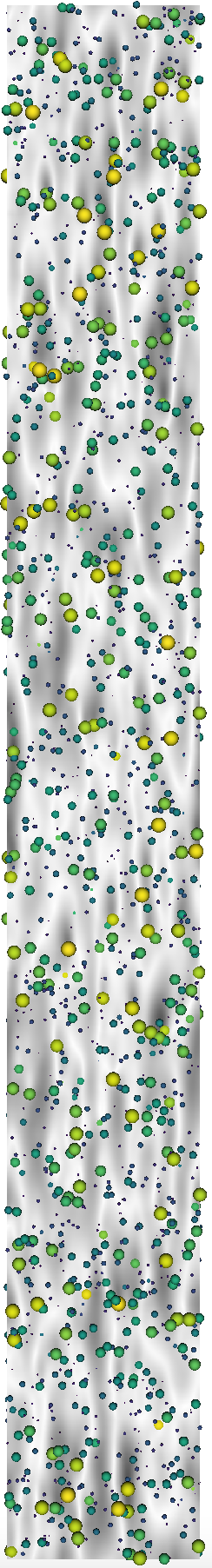} &
\includegraphics[height = 0.7\textwidth]{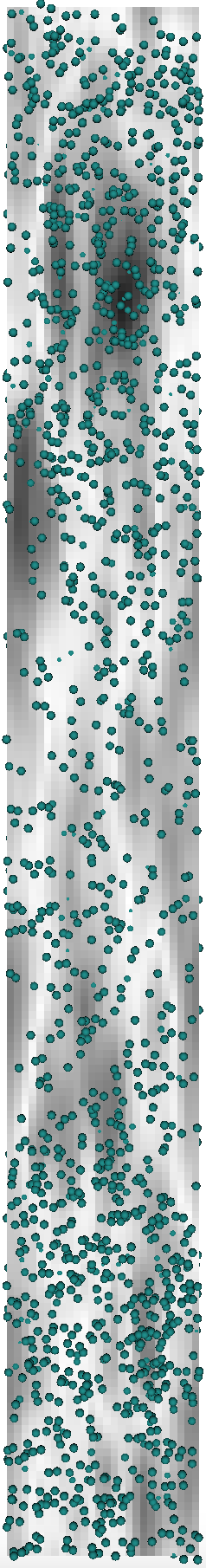}&
\includegraphics[height = 0.7\textwidth]{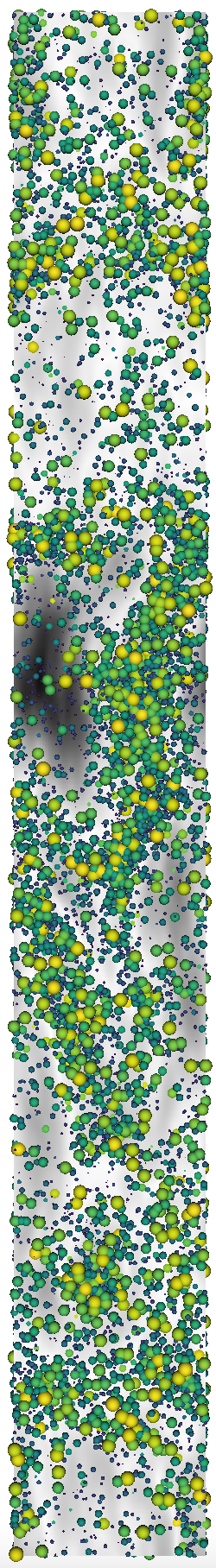} & 
\includegraphics[height = 0.7\textwidth]{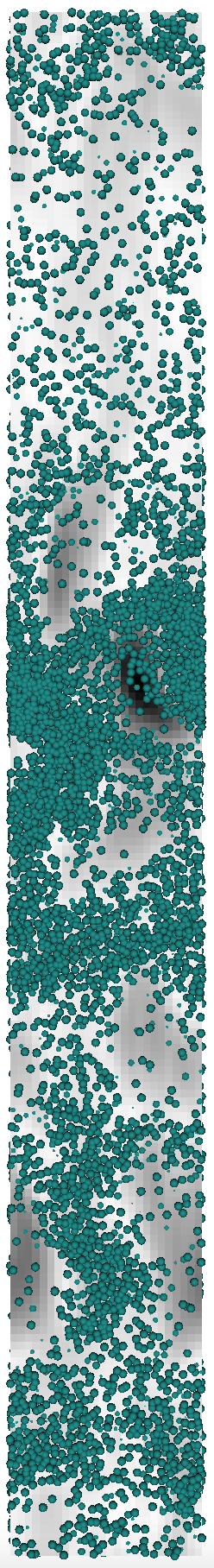}&
\includegraphics[height = 0.7\textwidth]{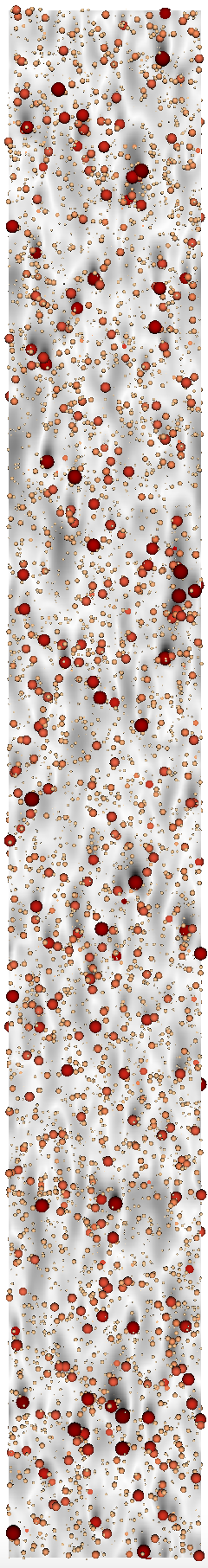} 
&\includegraphics[height = 0.7\textwidth]{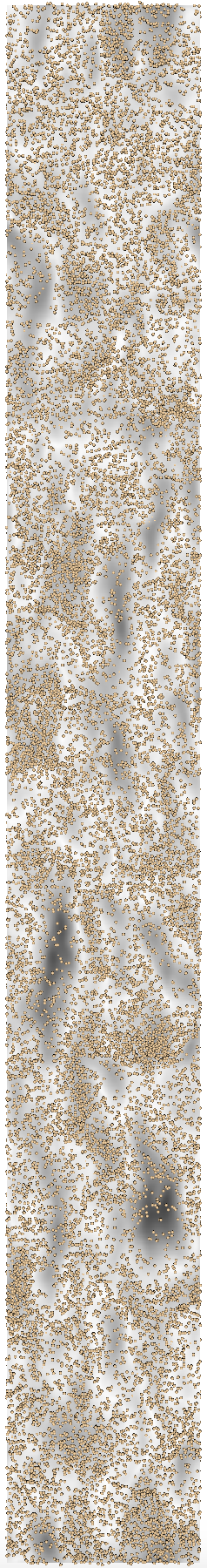}&
\includegraphics[height = 0.7\textwidth]{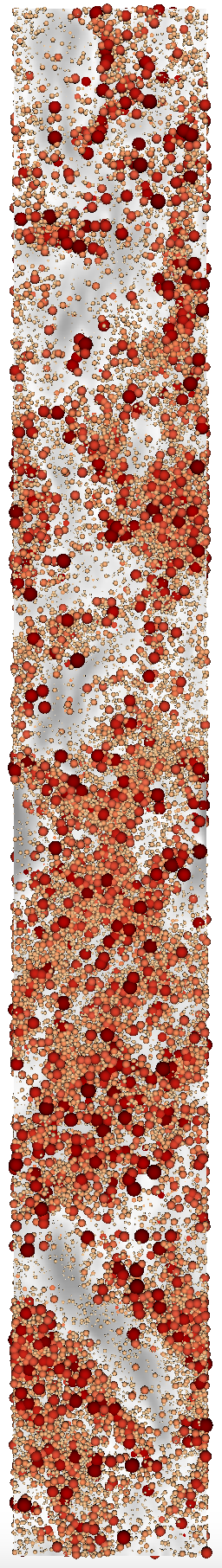} & 
\includegraphics[height = 0.7\textwidth]{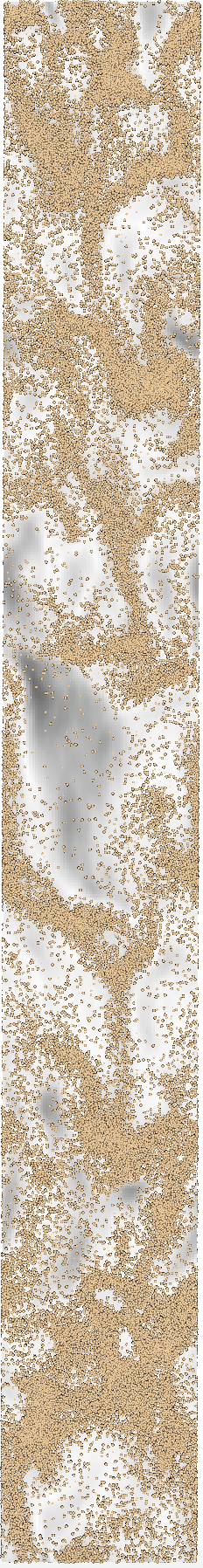} \\
\multicolumn{2}{c}{$u_f / \mathcal{V}_{0,10}$} & \multicolumn{2}{c}{$u_f /\mathcal{V}_{0,10}$} &\multicolumn{2}{c}{$u_f /\mathcal{V}_{0,10}$} & \multicolumn{2}{c}{$u_f /\mathcal{V}_{0,10}$} \\

\multicolumn{2}{c}{\includegraphics[angle=-90,origin=c,height=0.1\textwidth]{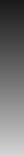}}
& \multicolumn{2}{c} {\includegraphics[angle=-90,origin=c,height=0.1\textwidth]{velcolorbar.png}}
&\multicolumn{2}{c}{\includegraphics[angle=-90,origin=c,height=0.1\textwidth]{velcolorbar.png}}
&\multicolumn{2}{c}{\includegraphics[angle=-90,origin=c,height=0.1\textwidth]{velcolorbar.png}}\\ [-7.5ex]
\multicolumn{2}{c}{$0.0 {\hskip 0.9in} 0.75$}
& \multicolumn{2}{c} {$0.0 {\hskip 0.9in} 3.50$}
&\multicolumn{2}{c}{$0.0 {\hskip 0.9in} 0.75$}
&\multicolumn{2}{c}{$0.0 {\hskip 0.9in} 3.50$}\\
\multicolumn{4}{c}{\includegraphics[angle=-90,origin=c,height=0.1\textwidth]{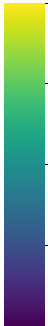}}
&\multicolumn{4}{c}{\includegraphics[angle=-90,origin=c,height=0.1\textwidth]{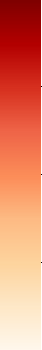}}\\ [-7.5ex]
\multicolumn{4}{c}{$0.23$ mm {\hskip 0.35in} $3.0$ mm} & \multicolumn{4}{c}{$0.23$ mm {\hskip 0.35in} $3.0$ mm} 
\end{tabular}
    \caption{Representative snapshots from the statistically stationary period of each configuration under study. Each slice is an $x$–$y$ plane at $z = 0$. Fluid phase velocity (gray) is normalized with the polydisperse Stokes velocity, $\mathcal{V}_{0,10}$ and particles are colored by diameter (from blue (small) to yellow (large) for Dist. A and from pink (small) to red (large) for Dist. B). }
    \label{fig:SSD}
\end{figure}

\section{Results} 
\label{sec:results} 
In this work, simulations are evolved from their initial, uncorrelated condition to a statistically stationary state, which is determined based on the temporal variation of the volume mean settling velocity and volume mean variance of volume fraction. After a stationary state is maintained for several characteristic time scales, $\tau_p$, representative snapshots are used for analysis. Representative $x-y$ planes of these snapshots are shown in Fig.~\ref{fig:SSD}.  

In our analysis, we examine the effect of both polydispersity and volume fraction on clustering behavior and how this in turn implicates overall settling behavior. In addressing both clustering and setting phenomenon, we draw comparisons between observations in the polydisperse configurations and analogous monodisperse configurations as well as between the two volume fractions.  

First, we begin this discussion with averaged behavior on the scale of the full domain. In doing so, we leverage several mean quantities that are often useful in characterizing heterogeneity. We present these quantities first, to paint a broad picture of the clustering and settling behavior of the configurations studied, but note that examining mean flow quantities in isolation paints an incomplete picture. Then, to present a more complete analysis of flow phenomenon, we present a statistical analysis on a particle-wise basis to illuminate the physics underlying domain-scale behavior. 

\subsection{Characterization of mean flow quantities}
\label{sec:MeanFlow}
Mean degree of clustering is often characterized using the parameter, $D$, introduced by \citet{Eaton1994} and defined as  
\begin{equation} 
\mathcal{D} = \frac{\left(\sqrt{\langle \alpha_p^{\prime 2}\rangle} -\sqrt{\langle \alpha_p^{\prime 2}\rangle}\Big\vert_0 \right)}{\langle \alpha_p \rangle} \approx \frac{\sqrt{\langle \alpha_p^{\prime 2}\rangle} }{\langle \alpha_p \rangle},
\end{equation}
where $\langle \alpha_p^{'2}\rangle$ is the variance of the particle volume fraction of the fully developed configuration, also denoted as $\text{var}(\alpha_p)$ in this work, and $\langle \alpha_p^{\prime 2}\rangle\Big\vert_{0}$ is the variance of an uncorrelated assembly of particles, assumed here to be null. Given these definitions, the numerator of $D$ represents the deviation of the standard deviation of particle volume fraction from a random distribution of particles. As a collection of particles becomes increasingly correlated and heterogeneity in the flow develops, this quantity similarly increases.

While $\mathcal{D}$ has been traditionally used to characterize clustering behavior in gas-solid flows, this description is statistically incomplete when the solid phase is polydisperse. The third and fourth moment means of the filtered particle phase volume fraction are also relevant for characterizing the extent of clustering. These quantities are tabulated in Tab.~\ref{tab:SvClustering} for each of the configurations and are defined as 
\begin{align}
    \text{skew}(\alpha_p) &= \frac{\sum\limits_{i=1}^{N_{\text{cells}}} \left({\alpha}_p^{(i)} - \langle {\alpha}_p\rangle \right)^3}{(N_{\text{cells}}-1)\left(\text{var}(\alpha_p)\right)^{3/2}}\quad \text{and} \\
     \text{kurt}(\alpha_p) &= \frac{\sum\limits_{i=1}^{N_{\text{cells}}} \left({\alpha}_p^{(i)} - \langle {\alpha}_p\rangle \right)^3}{(N_{\text{cells}}-1)\left(\text{var}(\alpha_p)\right)^{2}}.
\end{align}

We note that skewness represents the asymmetry of the distribution of volume fraction, with $\text{skew}(\alpha_p)=0$ denoting perfect symmetry in the data. Values for skewness less than and greater than null signifying asymmetry skewed toward volume fractions more dilute and more concentrated than the global mean, respectively. In other words, the value of skewness quantifies the propensity for a given distribution of particles to produce correlated regions that are either very dense (positive skewness) or very dilute (negative skewness). Kurtosis is a measure of the `tailedness' of the distribution and indicates how the volume fraction is distributed between the mean and tails. Null represents a normal distribution, while positive (leptokurtic) and negative (platykurtic) values are representative of distributions that are more tightly and loosely spread about the mean, respectively. 

In computing the variance, skewness and kurtosis of the particle volume fraction for all the configurations under study, we make several observations. First, we find that the variance in volume fraction is substantially higher for dilute polydisperse configurations relative to their monodisperse counterparts. In particular, the ratios of the standard deviation for Dist. $A/A_0$ and $B/B_0$ are 2.07 and 2.73, respectively. Interestingly, the standard deviation in particle volume fraction is nearly equivalent between the polydisperse configurations and their monodisperse counterparts at high volume fraction. The difference between $A$ and $A_0$ is 1.4\% and the difference between $B$ and $B_0$ is 2.5\%. A similar trend is observed when considering the skewness and the kurtosis in the particle volume fraction. Again, the dilute configurations demonstrate much higher deviations in the polydisperse configurations relative to their monodisperse counterparts, and this difference is diminished at higher volume fraction. As previously described, reduced drag is a primary mechanism of clustering. The marked difference in mean clustering behavior between polydispersed and monodispersed assemblies of particles at dilute configurations and relatively little difference at higher concentrations owes to this. Specifically, at higher volume fractions, there are a sufficient number of particles to consistently disturb the flow and produce regions of reduced drag, thereby initiating clustering events. This, then, reduces the importance of larger particles in the flow. In contrast, for more dilute suspensions, when particles are polydispersed, the presence of larger particles induces larger regions of reduced drag compared with a monodispersed analog. This then translates to more regions of heterogeneity and clustering. 

\begin{table}
\centering
    \begin{tabular}{c c | c c c c}
$\langle \alpha_p \rangle$ & Dist. &  $\mathcal{S}$ &$\mathcal{D}$ & skew$(\alpha_p)$ & kurt$(\alpha_p)$ \\
\hline 
 & $A_{0}$ & 2.07 & 0.42 & 0.66 & \;\;3.37  \\
 \multirow{2}{*}{0.01} & $A_{\;}$ & 1.21 & 0.87 & 2.81 & 15.69  \\
 & $B_{0}$ &  0.28 &0.49 & 0.82 & \;\;3.90  \\
 & $B_{\;}$ & 0.46 &1.34 & 6.50 & 80.68  \\[2ex]
 & $A_{0}$ & 2.28 &0.68 & 1.62 & \;5.79  \\ 
 \multirow{2}{*}{0.10} &$A_{\;}$& 1.32 &0.69 & 1.74 & \;7.62  \\
 & $B_{0}$ & 0.31 &0.80 & 1.55 & \;5.95  \\
 & $B_{\;}$ & 0.51 &0.78 & 1.50 & \;5.62  \\
\end{tabular} 
\caption{Summary of surface loading and statistics on volume fraction for all configurations under study.}
\label{tab:SvClustering}
\end{table}

As previously described, $\mathcal{D}$ is routinely used as an \emph{a posteriori} estimator of clustering. We also note that mass loading, $\varphi = \rho_p \langle \alpha_p \rangle/(\rho_f \langle \alpha_f \rangle)$, has historically been used as an \emph{a priori} estimate for predicting clustering. This metric has been shown to be related to $\mathcal{D}$ in the case of monodispersed assemblies of particles (e.g., ~\citet{Capecelatro2015, Beetham2021}), however, it is agnostic to polydispersity in the particle phase. To underscore this, the mass loading of all the configurations $A$, $A_0$, $B$ and $B_0$ is 50.5 at $\langle \alpha_p \rangle = 0.01$ and 555.5 for all configurations at $\langle \alpha_p \rangle = 0.10$.  This, combined with the wide variation in $\mathcal{D}$ across each of these configurations, highlights the notion that mass loading is incapable of providing differentiation in the clustering behavior between assemblies of particles that have the \emph{same} mass loading but have the particle phase mass partitioned \emph{differently}. Thus, while mass loading still may serve as an initial indicator of whether or not clustering will occur, our data motivates the need for an alternative \emph{a priori} quantity that is capable of giving a more complete prediction of clustering behavior.

To this end, we note that a key difference between configurations that contain differing particle distributions at equivalent mass loading is the number of particles present, or alternatively, the amount of particulate surface area present in the flow. We also mention that total particle surface area is likely to be an important predictor of clustering, due to this quantity's role in particle drag. Based on the data collected as a part of this study, we propose an alternate \emph{a priori} predictor for the degree of clustering that can be expected for a given gas-solid flow, based on the degree of surface area contact between the phases. We term this `surface loading' and define it as
\begin{align}
    \mathcal{S} &= \left(\frac{1}{\langle \alpha_f \rangle A_{\text{cross}}}\right)\left(\frac{\rho_p}{\rho_f}\right)\frac{\pi}{4}\underbrace{\frac{1}{N_p }{\sum\limits_{i=1}^{N_p}  \left(d_p^{(i)}\right)^2}}_{\left(D_{20}\right)^2} 
\end{align}
where $A_{\text{cross}}$ is the area of the cross-stream plane and the term in brackets is the square of the surface mean diameter (see \ref{appendix:Diameters} for more details). It is important to note here, that when $\mathcal{S}$ tends to zero, this is indicative of extremely fine, dilute particles and when $\mathcal{S}$ approaches infinity, this represents very concentrated, very large particles. Due to this, we expect that $\mathcal{D}$ should tend to zero in both limits. In addition, there should exist some critical surface loading for which the degree of heterogeneity is a maximum. 

\begin{figure}
\centering
    \includegraphics[width = 0.47\textwidth]{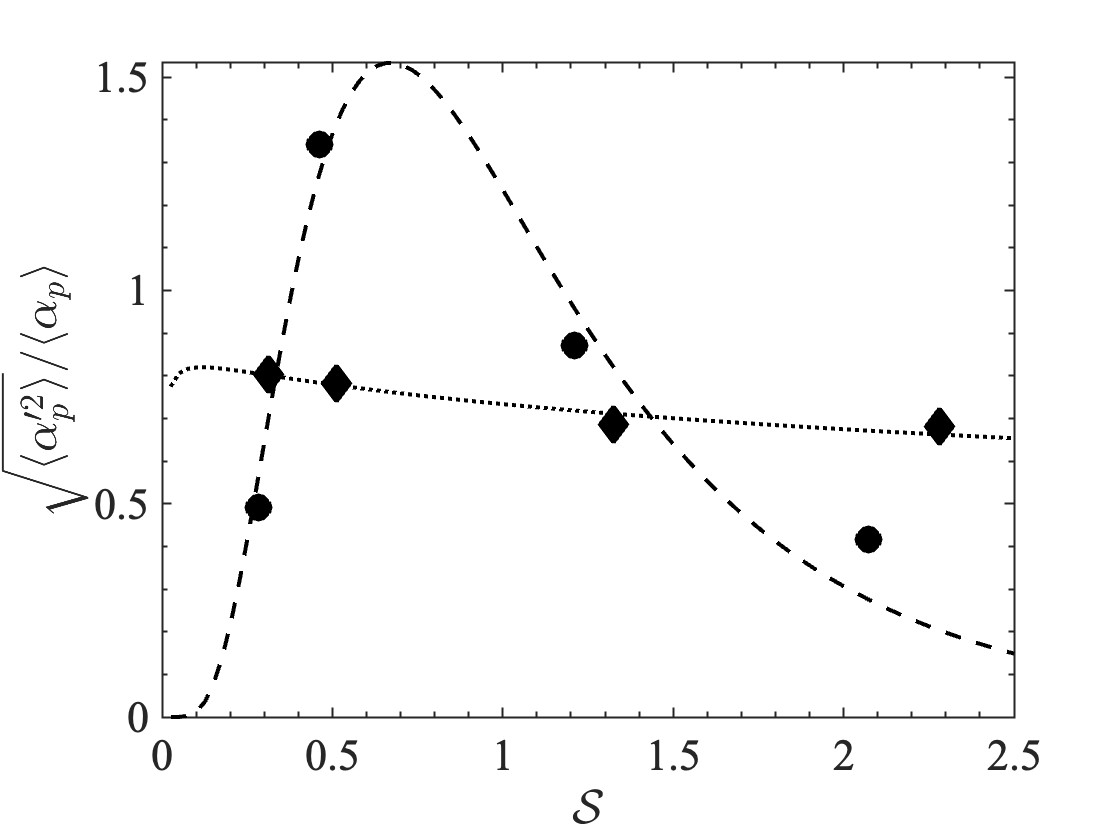}
    \caption{Deviation of normalized particle-phase volume fraction as a function of surface loading. Circles represent data for $\langle \alpha_p \rangle = 0.01$ and squares represent data for $\langle \alpha_p \rangle= 0.10$. The loosely and densely dashed lines represent the model described in Eq.~\ref{eq:Smodel} for $\langle \alpha_p \rangle = 0.01$ and $0.10$, respectively.}
    \label{fig:DvsS}
\end{figure}

Shown in Fig.~\ref{fig:DvsS}, we observe that for each void fraction there is a critical surface loading for which the degree of clustering, $\mathcal{D}$, attains a maximum with respect to the surface loading, $\mathcal{S}$. For the configurations studied here, this maximum occurs at lower surface loading for higher mean volume fraction, but the degree of clustering overall is higher for the dilute suspensions. Additionally, at higher volume fraction we find that $\mathcal{D}$ remains nearly constant, further underscoring that clustering behavior is less sensitive to changes in $\mathcal{S}$ (e.g., polydispersity) at higher volume fraction. 

In light of our data, we propose a model for $\mathcal{D}$ as a function of the surface loading, shown as dashed lines in Fig.~\ref{fig:DvsS}, and defined as  
\begin{align} 
\frac{\sqrt{\langle \alpha_p^{\prime^2} \rangle}}{\langle \alpha_p \rangle} &= \frac{1}{A\; \mathcal{S}} \exp{\left( \frac{-\left(\ln{\left(\mathcal{S}\right)}-B \right)^2}{C}\right)} \label{eq:Smodel}
\end{align} 
where the coefficients $A$, $B$ and $C$, depend upon the mean volume fraction as, 
\begin{align}
A\left(\langle \alpha_p \rangle \right) &= -8.2\langle \alpha_p \rangle + 0.9 \\
B\left(\langle \alpha_p \rangle \right) &=76.0\langle \alpha_p \rangle - 0.8\\
C\left(\langle \alpha_p \rangle \right) &=164.0\langle \alpha_p 
\rangle - 0.9.
\end{align} 
It is important to note, however, that since this study considered only two volume fractions, the dependence of the coefficients $A$, $B$, and $C$ can at most only be described as linear. A more comprehensive model that considers a wider volume fraction is reserved for future work. Further, it is important to note here that prior work~\citep{Beetham2021} has noted that there is also a (likely nonlinear) relationship between $\mathcal{D}$ and the Archimedes (or Froude) number, which has not been captured through $\mathcal{S}$. This dependence is also reserved for future work.   

In addition to $\mathcal{D}$, several other mean quantities are important for describing the bulk hydrodynamic behavior of gas-solid flows. These include the mean settling velocity, $\langle u_p \rangle$, the contributions to granular energy from particle phase turbulent kinetic energy, $k_p$, and granular temperature, $\langle \Theta \rangle_p$. In addition to these, we also consider the magnitude of fluid-phase turbulent kinetic energy, $k_f$ relative to the granular energy. These quantities for each of the distributions considered are summarized in Tab.~\ref{tab:V_star}. 

\begin{table}
    \centering

\begin{tabular}{c c | c c c @{\hskip 0.25in} c c c}
$\langle \alpha_p \rangle$ & Dist. & $\langle u_p \rangle$ & $\mathcal{V}^{\star}_{0}$ & $\langle u_f \rangle_p$ & $k_f/\kappa_p$ & $k_p/\kappa_p$ & $3\langle \Theta \rangle_p /(2\kappa_p)$\\
& & [m/s] & [m/s] & [m/s] & & & \\
%& & [-] & [m/s] & [m$^2$/s$^2$] & [m$^2$/s$^2$] &[m$^2$/s$^2$]  \\
\hline 
 & $A_{0}$ & -0.20 & -0.16 & -0.04 & 0.02 & 0.997 & 0.003\\
 \multirow{2}{*}{0.01} & $A_{\;}$ & -0.17 & -0.10 & -0.07 & 0.02 & 0.98 & 0.02\\
 & $B_{0}$ & -0.05 & -0.04 & -0.01 & 0.23 & 0.97 & 0.03  \\
 & $B_{\;}$ & -0.08 & -0.04 & -0.04 & 0.04 & 0.95 & 0.05 \\ [2ex]
 & $A_{0}$  & -0.16 & -0.10 & -0.06 & 0.31 & 0.99 & 0.01 \\
 \multirow{2}{*}{0.10} & $A_{\;}$ &  -0.14 & -0.05 & -0.09 & 0.32 & 0.99 & 0.01 \\
 & $B_{0}$ & -0.06 & -0.02 & -0.04 & 1.34 & 0.94 & 0.06 \\
 & $B_{\;}$ & -0.08 & -0.02 & -0.06 & 0.56 & 0.97 & 0.03\\
\end{tabular}
\caption{Summary of the domain mean settling velocities and contributions from $\langle u_f \rangle_p$ and $\mathcal{V}_0^{\star}$ along with the relative contributions to total granular energy from particle phase TKE and granular temperature as well as the relative magnitude of fluid phase TKE relative to granular energy.}
\label{tab:V_star} 
\end{table}

\begin{figure}
\centering 
\begin{tabular}{l l l}
(a) & (b) & (c)\\ [-0.25ex]
\includegraphics[width = 0.32\textwidth]{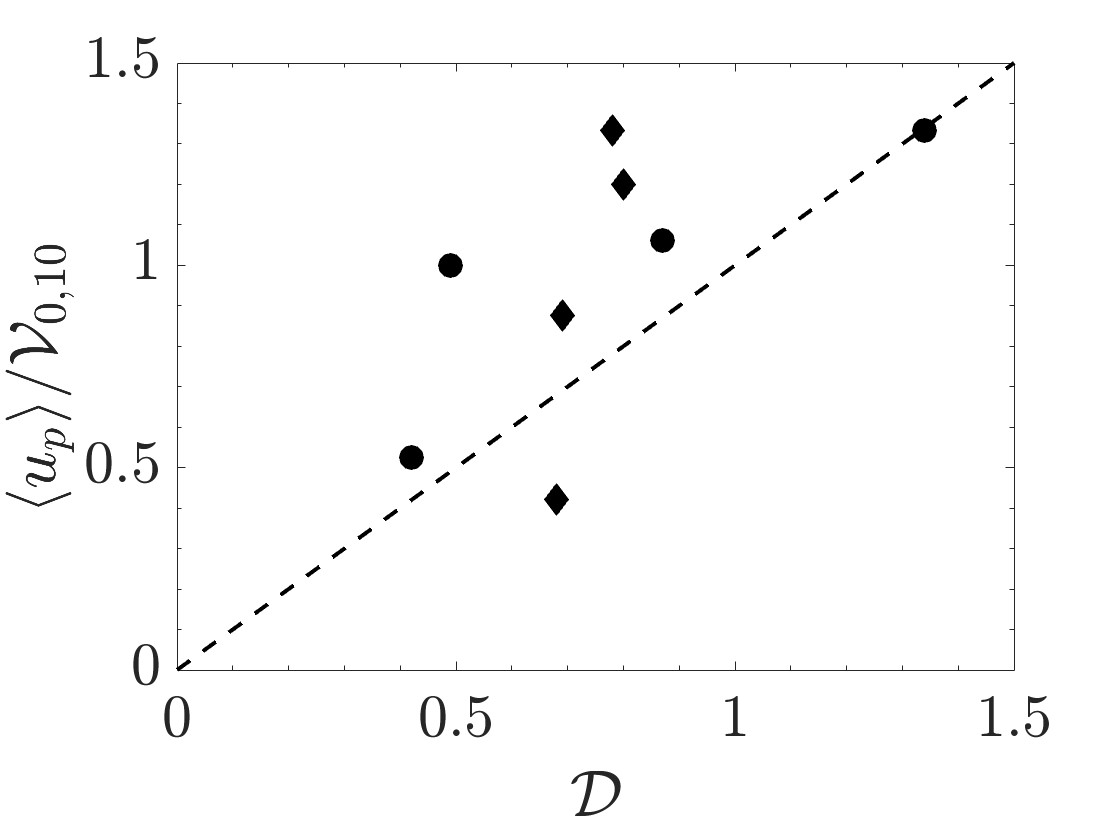} &
\includegraphics[width = 0.32\textwidth]{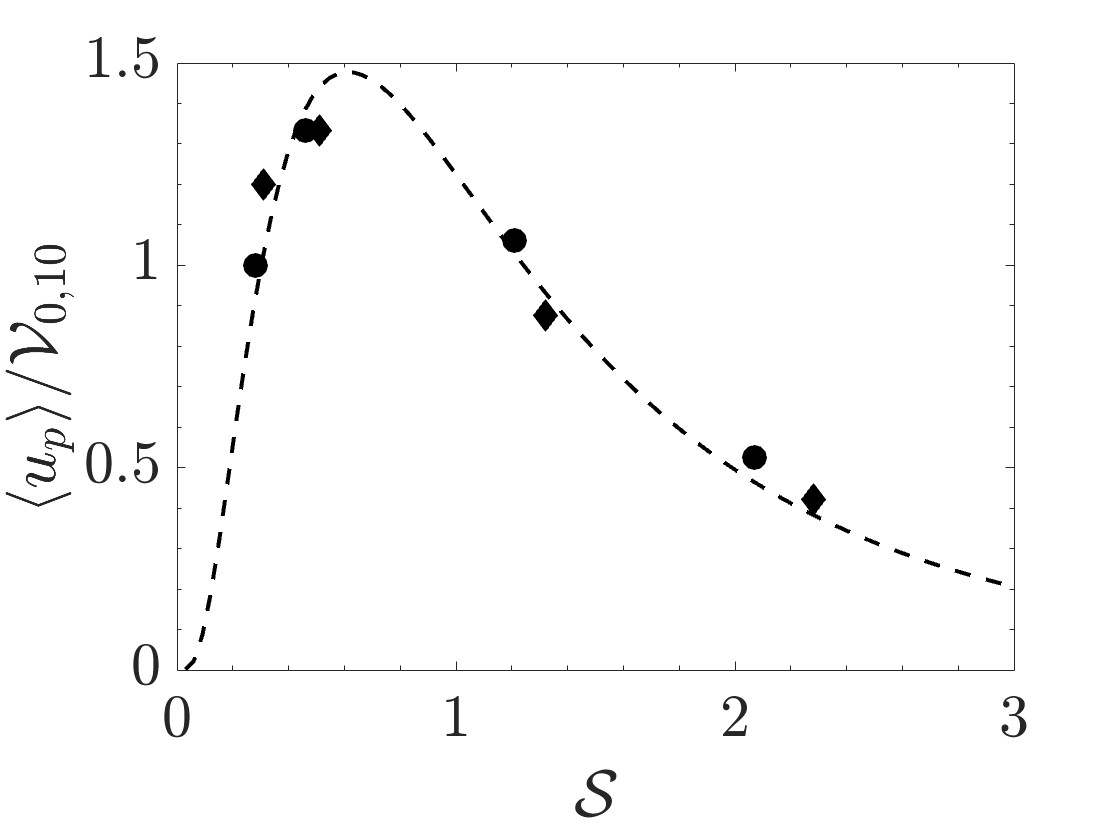} & \includegraphics[width = 0.32\textwidth]{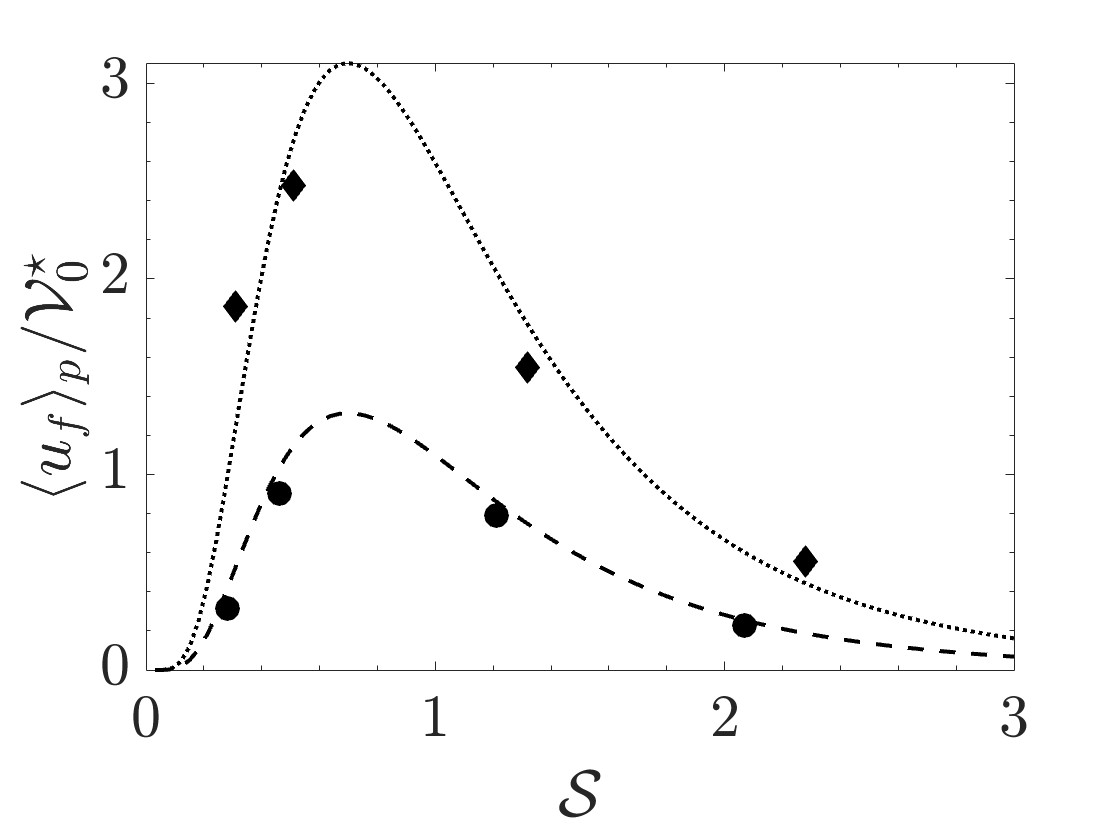}
\end{tabular}
\caption{Summary of the normalized mean settling velocity with respect to (a) degree of clustering, $\mathcal{D}$ and (b) surface loading $\mathcal{S}$. The normalized contribution of $\langle u_f\rangle_p$ relative to $\mathcal{S}$ is shown in (c). Configurations with $\langle \alpha_p \rangle = 0.01$ are denoted with solid circles and configurations with $\langle \alpha_p \rangle = 0.10$ are denoted with solid diamonds. The dashed lines represents a one-to-one correlation in (a) and the model prescribed by Eq.~\ref{eq:settleS} in (b) and Eq.~\ref{eq:ufpS} in (c).}
\label{fig:DvsUp}
\end{figure}

Here, we observe that the normalized mean settling velocity is greater for each of the polydisperse configurations, relative to their analogous monodisperse configuration (i.e., Dist. $A$ has a greater settling velocity as compared to $A_0$), though this effect is more pronounced in the configurations where $\langle \alpha_p \rangle = 0.01$. Additionally, we observe that the dilute configurations demonstrate a greater correlation between the degree of clustering, $\mathcal{D}$, and the normalized mean settling velocity, as shown in Fig.~\ref{fig:DvsUp}. In contrast, the configurations with $\langle \alpha_p \rangle = 0.10$ span a wider range of settling velocities for a much narrow range of $\mathcal{D}$. While the mean settling velocity appears to be nearly one-to-one correlated to the degree of heterogeneity for low volume fractions, the relationship is likely non-linear.    

\begin{equation}
\frac{\langle u_p \rangle}{\mathcal{V}_{0,10}} = \frac{2.5}{\left(B\mathcal{S}\sqrt{2\pi}\right)}\exp\left(-\frac{(\ln(\mathcal{S})-A)^2}{2B^2}\right)\label{eq:settleS}
\end{equation}
where $A = 0.15$ and $B=0.8$. Notably, we observe that our data collapses for both volume fractions studied in this work, resulting in parameters $A$ and $B$ that are constant coefficients. 

As was previously discussed in Sec.~\ref{sec:PhaseAveragedEqs}, the mean settling velocity can also be described as 
\begin{equation}
    \langle u_p \rangle = \langle u_f \rangle_p + \mathcal{V}_0^{\star},
\end{equation}
where $\mathcal{V}_0^{\star}$ is defined as $\tau_p^{\star}g$ and $\tau_p^{\star}$ is the $\tau_p/F$ with mean flow quantities, $\langle d_p \rangle$ and $\langle \alpha_p \rangle$, used as argument to the previously defined expressions for the settling time and drag correction factor. The way settling velocity is partitioned into each of these terms is summarized in Tab.~\ref{tab:V_star} and plotted in Fig.~\ref{fig:DvsUp}. 

When considering the settling velocity in this way, it is apparent that $\mathcal{V}_0^{\star}$ is closed but a model is required for $\langle u_f \rangle_p$. Here, we note that the relative contribution to the unclosed portion of mean settling velocity is 2.4 times larger for the higher volume fraction configurations. In light of this, we introduce an alternate formulation for the mean settling velocity, 

\begin{equation}
\frac{\langle u_p \rangle}{\mathcal{V}^{\star}_{0}} = \frac{C}{\left(0.6\mathcal{S}\sqrt{2\pi}\right)}\exp\left(-\frac{(\ln(\mathcal{S}))^2}{2B^2}\right) + 1, \label{eq:ufpS}
\end{equation} 
where the coefficient $C$ is 1.65 and 3.9 for $\langle \alpha_p \rangle = 0.01$ and $0.10$, respectively. As noted previously, additional studies at a wider range of volume fractions is needed to develop an accurate model for the dependence of $C$ on $\langle \alpha_p \rangle$.

In addition to mean settling velocity, the contributions to turbulent energy in both phases is also helpful in characterizing the flow. In particular, we consider the contributions to total granular energy, $\kappa$, from both the particle phase TKE ($k_p$) as well as from the granular temperature. As shown in Tab.~\ref{tab:V_star}, we observe that for dilute configurations, the polydisperse configurations have greater contributions to the total granular energy from the granular temperature as compared with their monodisperse analogs. Further, we note that Dist. $B$ and $B_0$ have substantially higher granular temperature than Dist. $A$ and $A_0$. Since granular temperature is an indication of granular energy that is uncorrelated from the mean `bulk' granular energy, this suggests that the configurations with smaller particles contain clusters in which particles are more free to move randomly within the cluster. For the denser configurations, the difference in granular temperature between distributions $A$ and $A_0$ is minimal, however, this difference is pronounced between distributions $B$ and $B_0$, though here the granular temperature is greater for the monodisperse analog, in contrast to the behavior observed at lower volume fraction. 

Finally, we also consider the relative magnitude of fluid TKE as compared to the total granular energy (see Tab.~\ref{tab:V_star}). Here, we note that the level of fluid TKE is an order of magnitude larger for distribution $B_0$ at both volume fractions considered. This is likely due to the larger number of particles contained in these configurations, due to their small size. In addition we note that at both volume fractions distribution $B$ exhibits about twice the fluid phase turbulent kinetic energy, again likely due to the greater number of particles present.  

\subsection{A more nuanced portrait of clustering}\label{sec:Clustering}
In the following section, we take a finer-grained approach to characterizing the clustering and settling behavior of the configurations under study. Here, we examine flow behavior using a statistical approach as well as propose improved models for settling in light of this analysis. 

In considering the differences in clustering behavior between all the configurations studied, we first turn our focus to the distribution of particle volume fraction over all Eulerian cells in the domain. In this discussion, we employ the use of isosurfaces, or isocontours, to delineate regions containing particle volume fractions larger by some specified threshold compared to the global average. This requires the specification of a critical volume fraction that denotes regions that are considered to be `clustered'. 

In this work, we identify `clusters' as regions in the flow corresponding to a particle volume fraction greater than 1.5 times the domain mean, $\langle \alpha_p \rangle$. Additionally, we divide the flow into the following regions: 
\begin{enumerate}
\item[] \textbf{Region A}: Most dilute regime, $\alpha_p \leq 1.5 \langle \alpha_p \rangle$ 
\item[] \textbf{Region B}: Loosely clustered regime, $1.5 \langle \alpha_p \rangle < \alpha_p \leq 2.25 \langle \alpha_p \rangle$
\item[] \textbf{Region C}: Moderately clustered regime, $2.25 \langle \alpha_p \rangle < \alpha_p \leq 3.0 \langle \alpha_p \rangle$
\item[] \textbf{Region D}: Densely clustered regime, $\alpha_p > 3.0 \langle \alpha_p \rangle$
\end{enumerate}
We employ these partitions in the remainder of the manuscript, and each of these regions are designated with colors increasing from white (most dilute) to dark gray (most dense). In addition, it is important to identify an appropriate reference volume fraction with respect to which we can normalize particle phase concentrations. A justification for this choice follows in Sec.~\ref{sec:RCP}

\subsubsection{A brief discussion on the close packing potential of polydispersed particles}\label{sec:RCP}
An important comparison we draw when describing clustering is that of the local volume fraction compared with the theoretical close-packing volume fraction, denoted $\alpha_{\text{rcp}}$. As one might anticipate, the packing efficiency observed in clusters, along with their shape, is intimately connected with settling behavior. 

It has been previously established that the random close-packed (RCP) volume fraction of monodisperse spheres is $\alpha_{\text{rcp}} = 0.64$ ~\citep{Farr2009, Farr2013, Desmond2014, Kansal2002}. This value represents the maximum achievable volume fraction for a randomly packed arrangement of spheres. While close packing efficiencies for polydispersed assemblies of spheres are less established in comparison, several theoretical and computational studies have quantified the RCP efficiency for lognormal distributions of spheres~\citep{Farr2009, Farr2013, Desmond2014, Kansal2002}. In ~\citet{Farr2009}, an algorithm that maps 3D configurations of spheres onto a 1D system of rods has shown success in accurately and efficiently estimating the RCP packing efficiency of lognormally distributed spheres. This approach, known as the rod packing (RP) algorithm~\citep{Farr2009}, was subsequently validated~\citep{Farr2013} and also resulted in a closed form expression (Eq.~\ref{eq:phimax}). In this work, \citet{Farr2013} observed that the RCP packing efficiency depended only upon a measure of the distribution `width', denoted $\Tilde{\sigma}$. This parameter is computed using the ratio of the volume moment diameter to the surface moment diameter, as

\begin{equation}
\Tilde{\sigma}=\sqrt{\ln\left(\frac{d_{4,3}}{d_{3,2}}\right)}.
\label{eq:sigd4332}
\end{equation}

Here, we make note that $d_{4,3}$ is sensitive to large particles and $d_{3,2}$ is sensitive to small particles (for more details, see \ref{appendix:Diameters}). When the ratio to these quantities is increasingly larger than 1 this implies a higher proportion of very small particles to larger particles. Naturally, the RCP efficiency for distributions containing a greater number of very small particles (i.e., distributions with increasingly large $\Tilde{\sigma}$), will be higher. The model relating $\alpha_{\text{rcp}}$ to $\Tilde{\sigma}$ based on results of the RP algorithm and developed by \citet{Farr2013} is given as

\begin{equation}
    \alpha_{\text{rcp}}(\Tilde{\sigma})= 1-0.57e^{-\Tilde{\sigma}} + 0.2135e^{\frac{-0.57\Tilde{\sigma}}{0.2135}} + 0.0019(\cos(2\pi(1-e^{-0.75\Tilde{\sigma}^{0.7} - 0.025\Tilde{\sigma}^4}))-1).
    \label{eq:phimax}
\end{equation}

We use this expression to infer the theoretical maximum RCP packing efficiency based on $\Tilde{\sigma}$ of the lognormal distributions under study. The statistical values relating to each of the polydisperse configurations studied are shown in Tab.~\ref{tab:rcpValues}. For completeness, we note that while both volume fractions each for Distributions $A$ and $B$ were sampled from the same lognormal distribution, since the number of discrete particles in the systems differ (fewer for $\langle \alpha_p \rangle = 0.01$ compared with $\langle \alpha_p \rangle = 0.10$), there are minimal differences in the statistical descriptions between the two volume fractions. 

\begin{figure}
\centering
\begin{tabular}{c c c c}
(a) $\alpha^A_{\text{rcp}}=0.6796$&   (b) $\alpha^{A_0}_{\text{rcp}}=0.6400$ & (c) $\alpha^{B}_{\text{rcp}}=0.7036$ & (d) $\alpha^{B_0}_{\text{rcp}}=0.6400$ \\
\includegraphics[height = 0.17\textwidth]{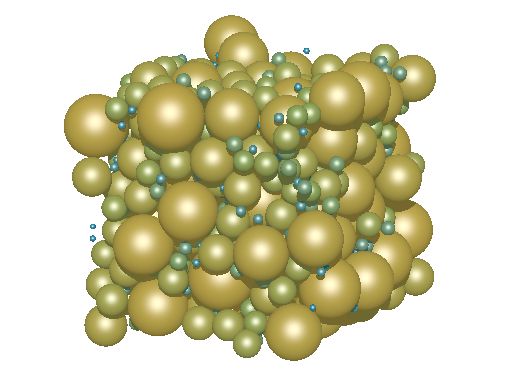}& \includegraphics[height = 0.17\textwidth]{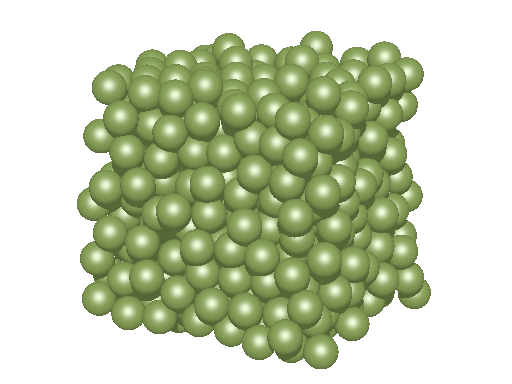}&
\includegraphics[height = 0.17\textwidth]{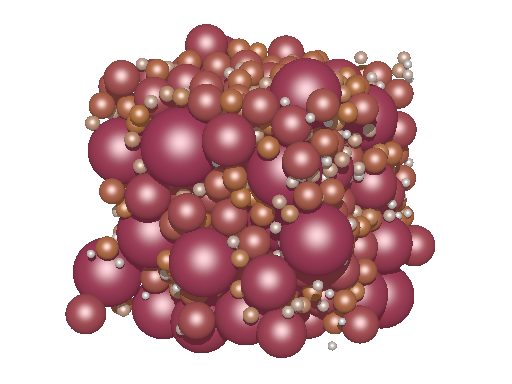}
& \includegraphics[height = 0.17\textwidth]{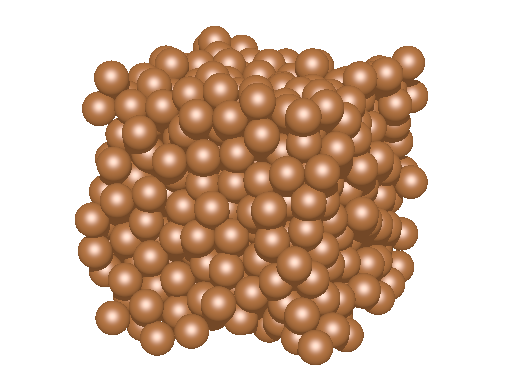}\\
\end{tabular}
    \caption{Random close-packed configurations for the configurations under study. Here $\alpha_{\text{rcp}}$, denotes the RCP volume fraction from the RP algorithm. Visualizations generated by the Kansal Torquatoa Stillinger (KTS) algorithm~\citep{Kansal2002} and RP software~\citep{Farr2013}.}
    \label{fig:RCPviz}
\end{figure}

As one might anticipate, the lognormally distributed particles exhibit a higher maximum RCP volume fraction, as small particles in the distribution occupy voids that would otherwise remain in a monodisperse packing arrangement of particles with a diameter equal to the mean particle diameter. Due to this, we use the RCP packing efficiency of Dist. B, $\alpha_{\text{rcp}}^{B} = 0.7038$ as the normalization factor for volume fraction for all the configurations studied. Visualizations for the RCP packing efficiencies for Distributions $A$, $B$, $A_0$ and $B_0$ are shown, along with their associated $\alpha_{\text{rcp}}$ in Fig.~\ref{fig:RCPviz}.

It is notable that while the values reported in Tab.~\ref{tab:rcpValues} are the random close-packing efficiency limits for the \emph{entire} collection of particles. However when these particles spontaneously cluster, the subsets of particles contained within clusters do not necessarily mirror the overall distribution of particles. In other words, if larger particles are more likely to be found in clusters than smaller particles, then the close-packing efficiency of the subset of particles involved in a cluster will differ from what would be expected for an assembly representative of the full domain. 

\begin{table}
\centering
\begin{tabular}{l c c c c c}
 & & \multicolumn{2}{c}{\textbf{Distribution $\bm{A}$}} & \multicolumn{2}{c}{\textbf{Distribution $\bm{B}$}} \\ [0.5ex]
 & &\multicolumn{4}{c}{\textit{volume fraction,} $\langle \alpha_p \rangle$} \\
 &  &\textit{0.01}  & \textit{0.10} &   \textit{0.01} & \textit{0.10} \\ 
 \hline \\[-1.25ex]
 $d_{1,0}$ & [mm] &1.11 & 1.11 & 0.68 & 0.68 \\
 $d_{2,0}$ & [mm] &1.31 & 1.31 & 0.81 & 0.81 \\
 $d_{3,0}$ & [mm] &1.49 & 1.48 & 0.95 & 0.95 \\
 $d_{3,2}$ & [mm] &1.91 & 1.91 & 1.30 & 1.31 \\
 $d_{4,3}$ & [mm] &2.16 & 2.15 & 1.65 & 1.65 \\ [0.75ex]
$\Tilde{\sigma}$ &  [--] &0.3483 & 0.3499 & 0.4824 & 0.4810 \\
 $\boldsymbol{\alpha}_{\textbf{rcp}}$ &[--] & \textbf{0.6793} & \textbf{0.6796} & \textbf{0.7038} & \textbf{0.7036} \\
\end{tabular}
\caption{Summary of the arithmetic mean, surface area, volume, Sauter and volume moment mean diameters, distribution width, $\Tilde{\sigma}$, and maximum random close packing efficiency resulting from Eq.~\ref{eq:phimax} for the particles studied in each configuration. A complete description of the definitions of the statistical diameters can be found in ~\ref{appendix:Diameters}.}
\label{tab:rcpValues}
\end{table}

\begin{figure}
\centering
\includegraphics[width=0.87\textwidth]{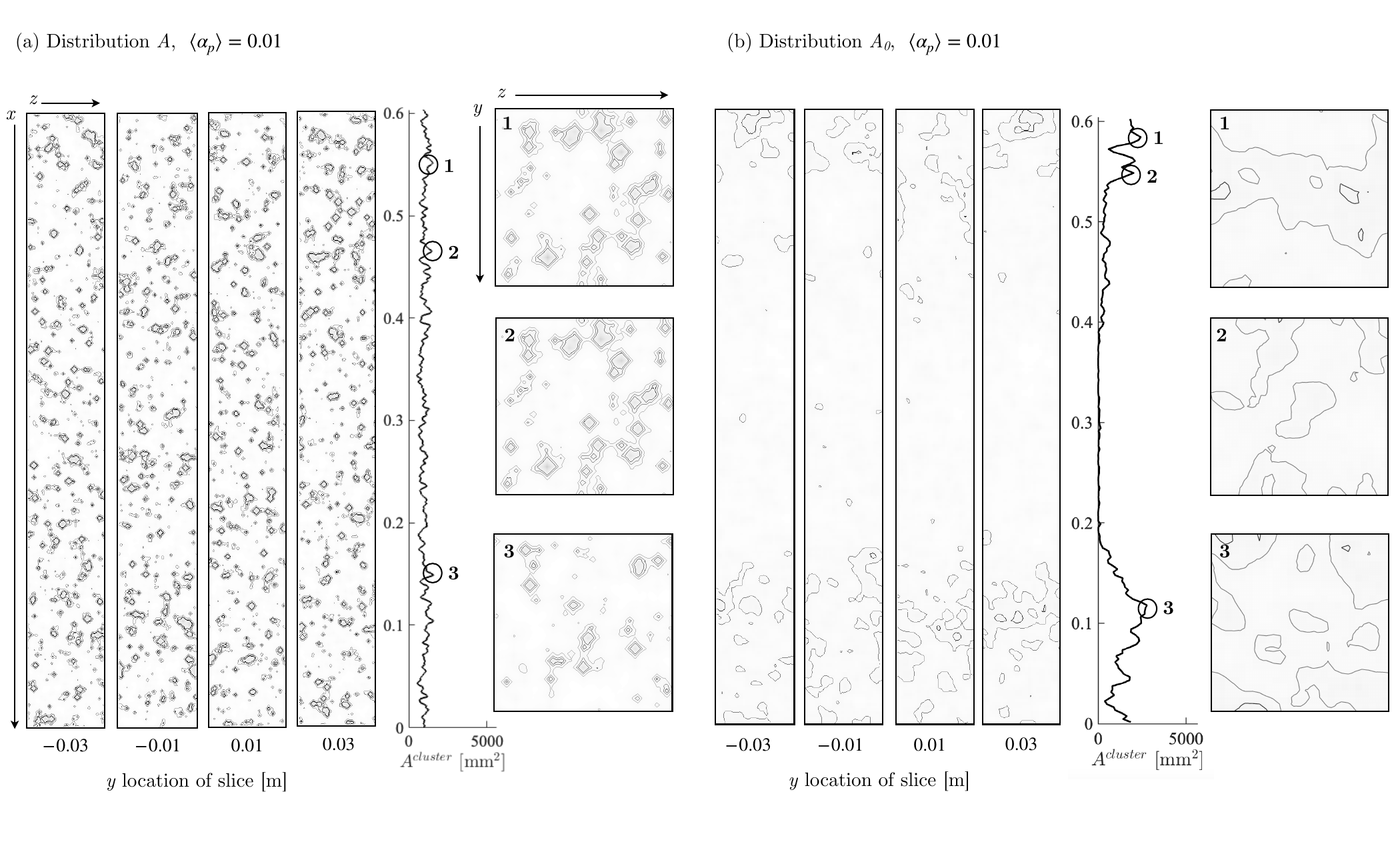} \\ \vspace{-1em}
\includegraphics[width=0.87\textwidth]{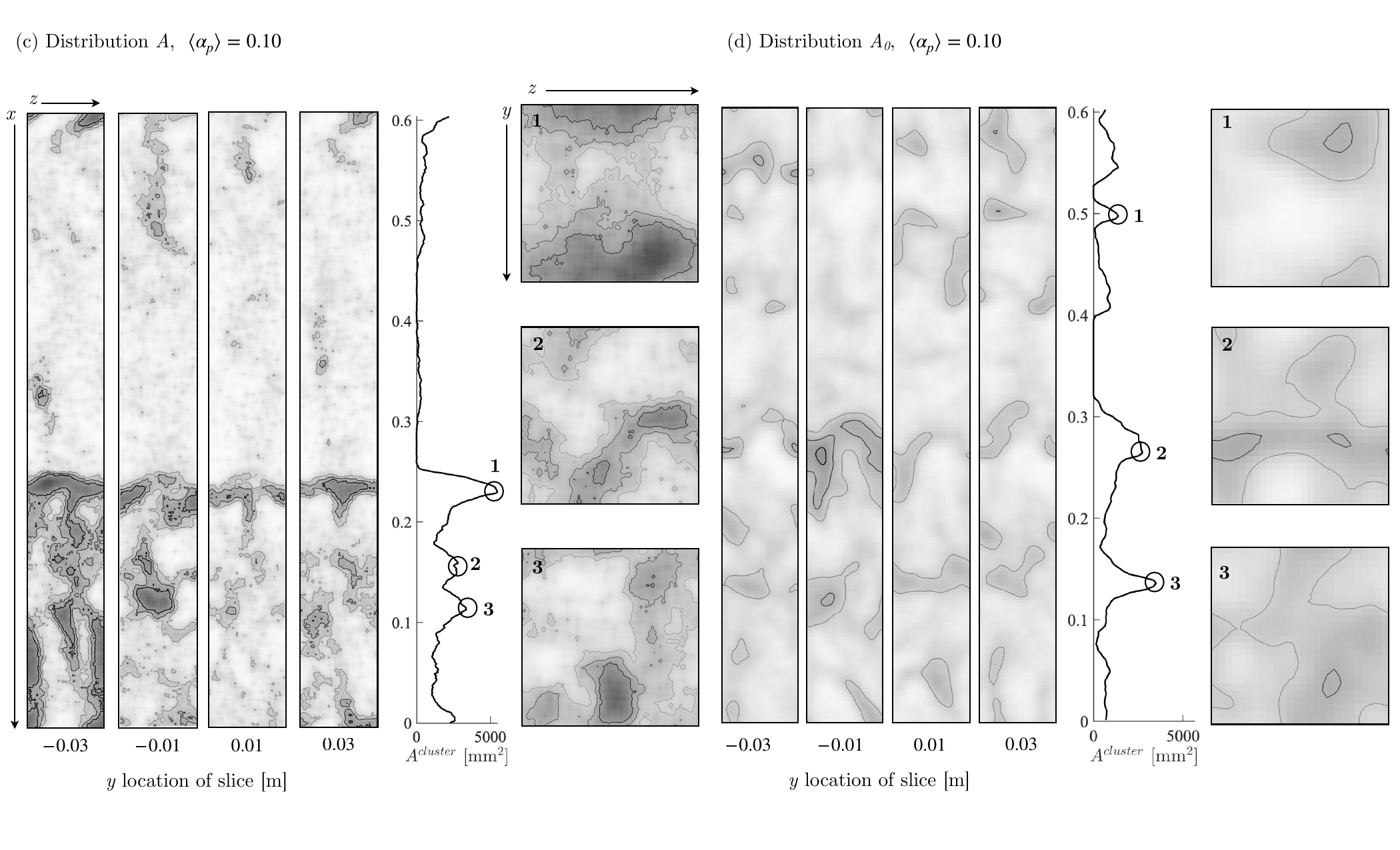}\\ \vspace{-1em}
\includegraphics[width=0.87\textwidth]{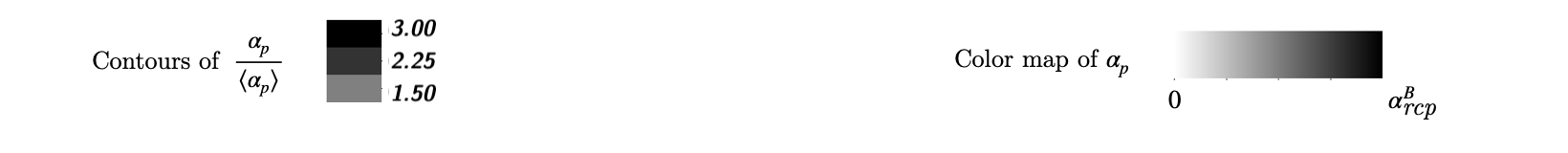}
\caption{Overview of the clustering patterns observed in Distributions $A$ (left figures) and $A_0$ (right figures) at $\langle \alpha_p \rangle = 0.01$ (top row) and $0.10$ (bottom row).}
\label{fig:DistAClustering}
\end{figure}

\begin{figure}
\centering
\includegraphics[width=0.95\textwidth]{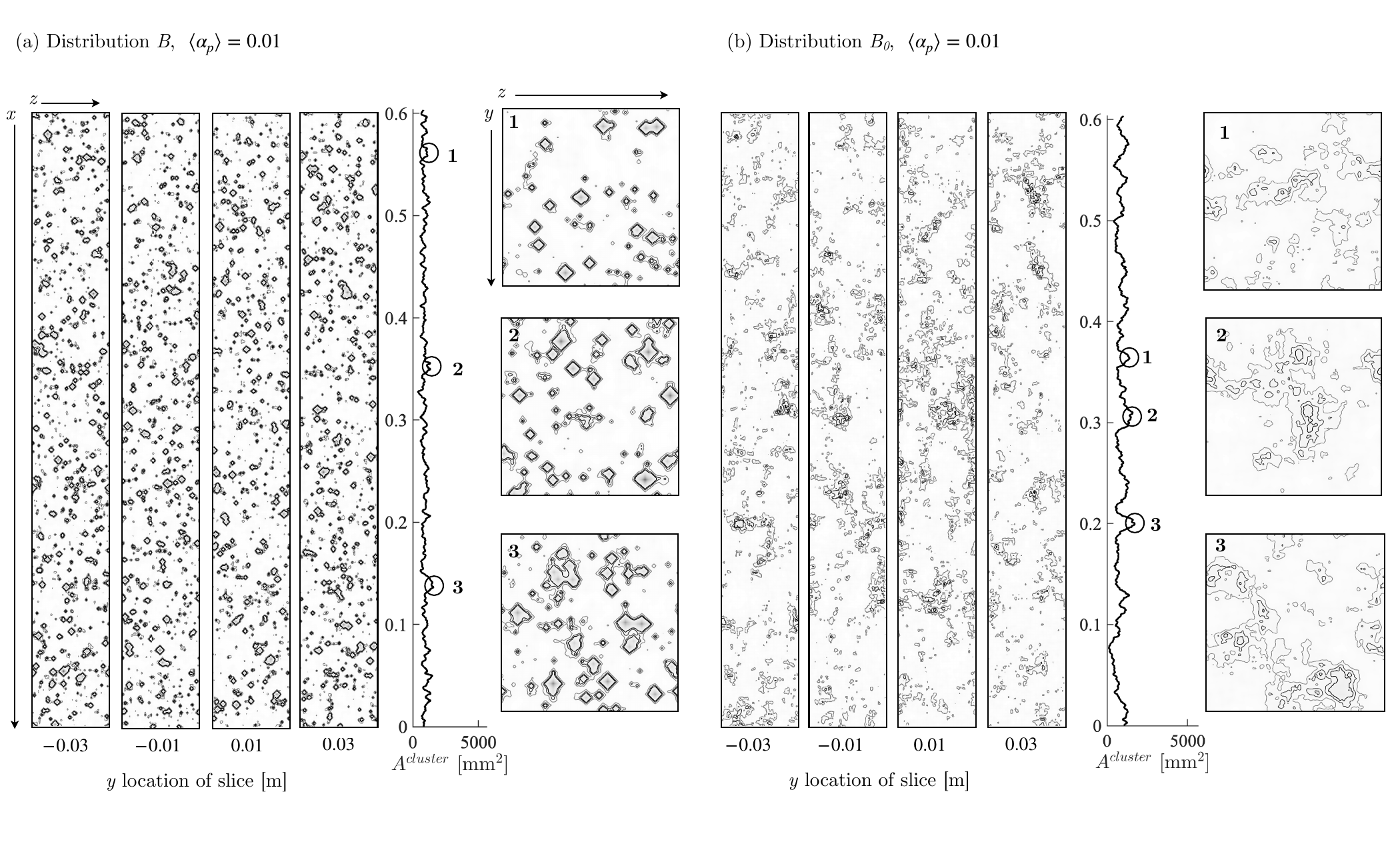} \\ \vspace{-1em}
\includegraphics[width=0.95\textwidth]{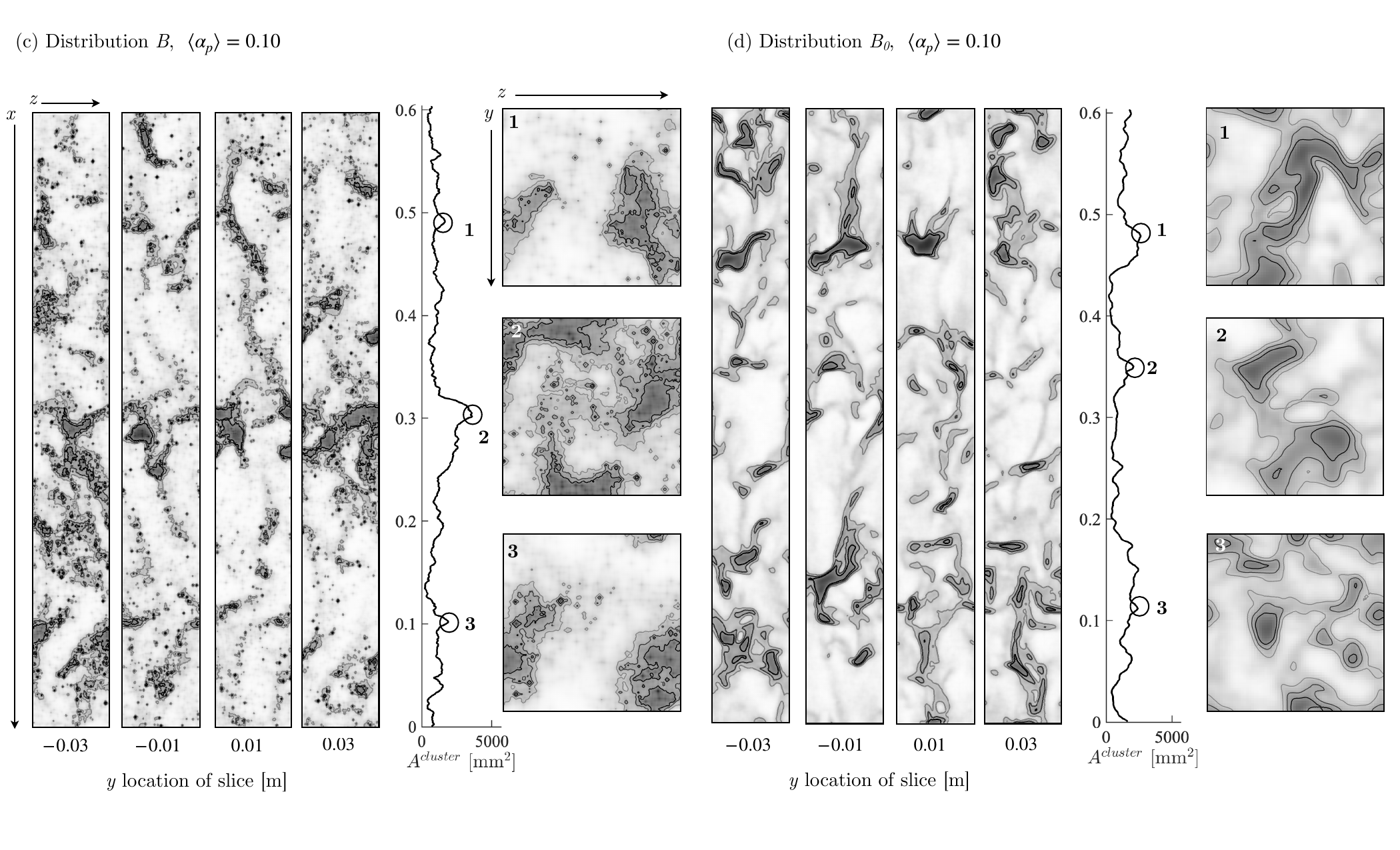}\\ \vspace{-1em}
\includegraphics[width=0.95\textwidth]{ClusterLegend.png}
\caption{Overview of the clustering patterns observed in Distributions $B$ (left figures) and $B_0$ (right figures) at $\langle \alpha_p \rangle = 0.01$ (top row) and $0.10$ (bottom row).}
\label{fig:DistBClustering}
\end{figure}

\begin{figure}
\centering
\begin{tabular}{ c c c c} 
  {Distribution $A_0$} & {Distribution $A$} &{Distribution $B_0$} & {Distribution $B$} \\
\hline \\ [-0.5ex]
\includegraphics[height=0.15\textwidth]{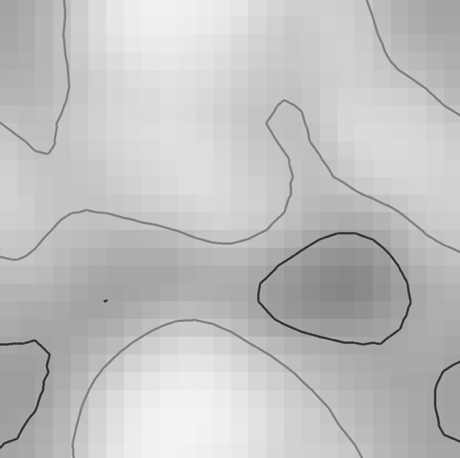} &
\includegraphics[height=0.15\textwidth]{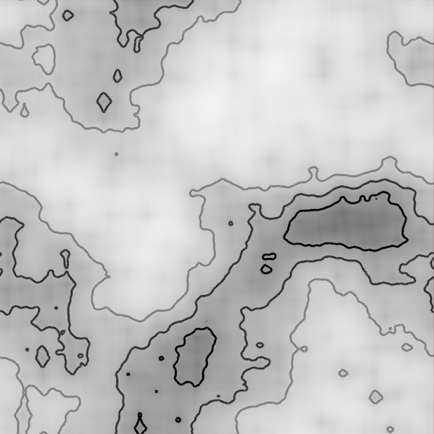} &
\includegraphics[height=0.15\textwidth]{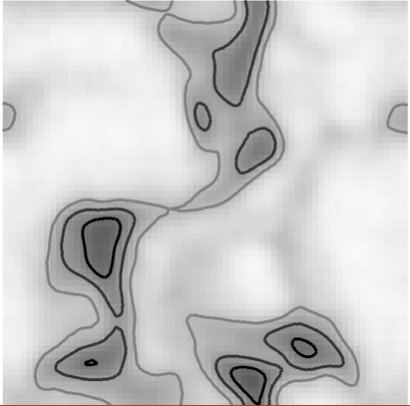} &
\includegraphics[height=0.15\textwidth]{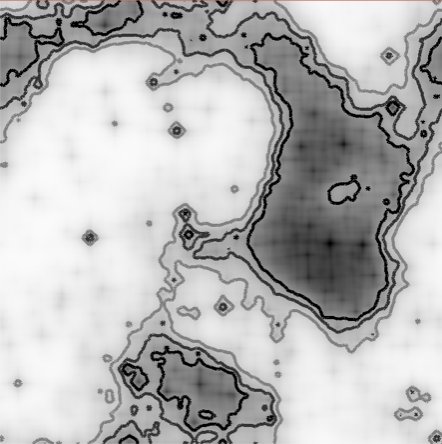}\\
\includegraphics[height=0.15\textwidth]{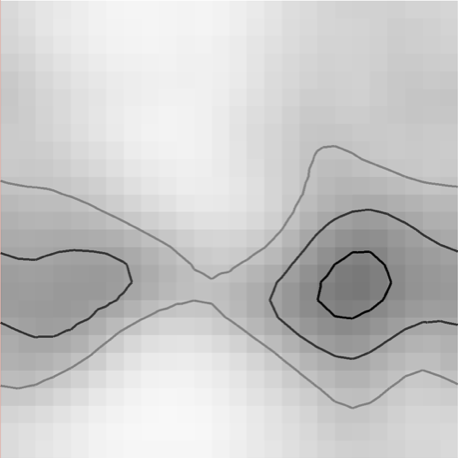} & 
\includegraphics[height=0.15\textwidth]{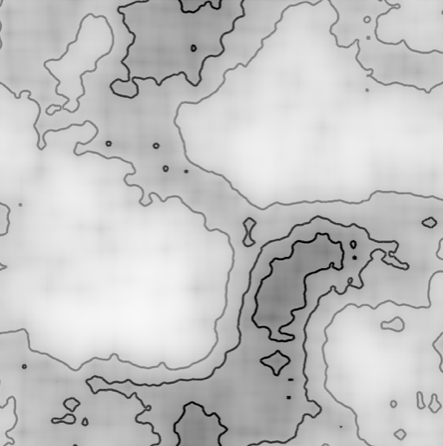} &
\includegraphics[height=0.15\textwidth]{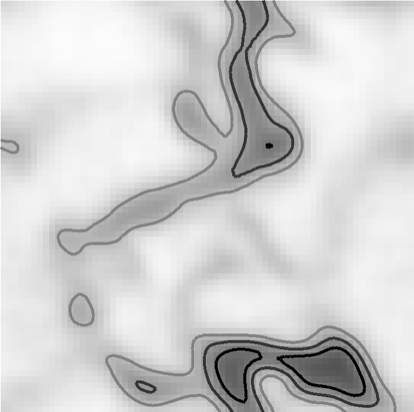} & 
\includegraphics[height=0.15\textwidth]{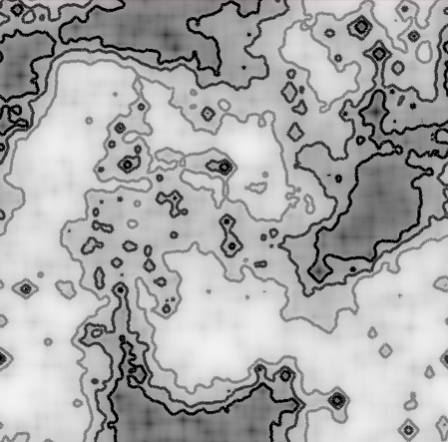} \\
\includegraphics[height=0.15\textwidth]{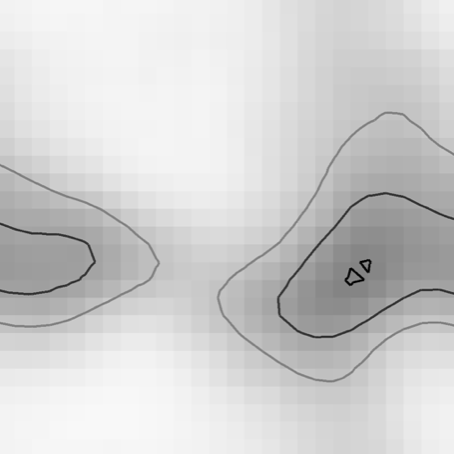}& 
\includegraphics[height=0.15\textwidth]{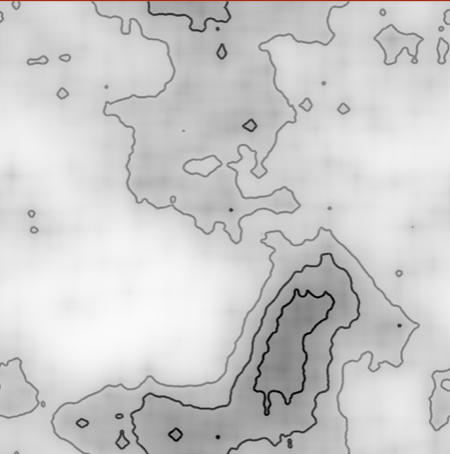} &
\includegraphics[height=0.15\textwidth]{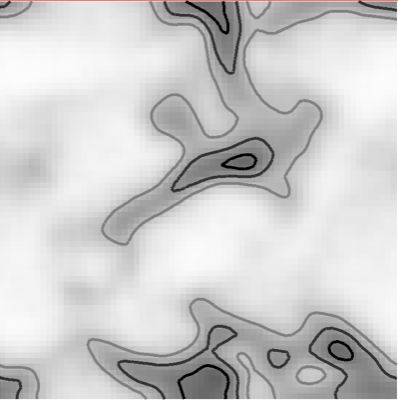}& 
\includegraphics[height=0.15\textwidth]{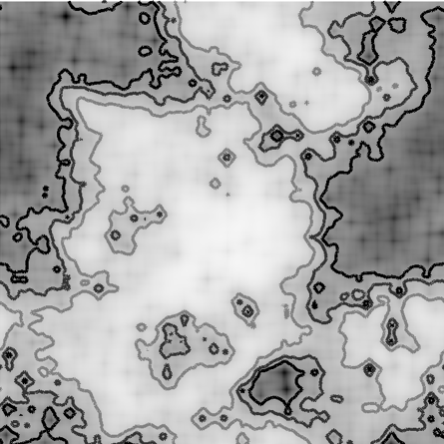}\\
\end{tabular} 
    \caption{Cross sections of particle volume fraction in the $y-z$ plane with contours denoting volume fractions of $(1.5, 2.25, 3.0)\langle\alpha_p \rangle$. The color map represents particle volume fraction and ranges from 0 (white) to $\alpha_{\text{rcp}}$ (black). Cross sections are the same as those detailed in Figs.~\ref{fig:DistAClustering} and \ref{fig:DistBClustering}, but are shown together here to aid in comparisons across the configurations studied at $\langle \alpha_p \rangle = 0.10$.}
    \label{fig:clusteringSummary}
\end{figure}

Following the four region convention for delineating regions of increasing correlation in the flow and normalization with respect to the maximum random close-packing efficiency for Dist. B as described above, we plot the local volume fraction and corresponding contours for four representative $x-z$ planes of each configuration studied (see Figs.~\ref{fig:DistAClustering}--\ref{fig:clusteringSummary}). Qualitatively, we observe that the contours defining clusters in all uniform distributions, $A_0$ and $B_0$, are smoother and achieve lower packing efficiency in their cores. This owes to the increased potential packing efficiency for polydispersed assemblies of particles. 

In comparing these uniform distributions of particles against each other, we observe that the configurations with larger particle diameter (Distribution $A_0$) contain larger regions of correlated particles (i.e., larger clusters), as compared with a smaller particle diameter (Distribution $B_0$) which results in a greater number of smaller clusters. Further, clusters tend to be denser in the case of smaller particles, however, this stems from the difference in the total number of particles present in the domain since monodispersed particles of any size have the same maximum packing efficiency. 

In comparing the polydisperse configurations with each of their analogous uniform distributions (i.e., Distributions $A$ and $A_0$), we observe that at low particle volume fraction, the polydisperse distributions tend to have smaller, more numerous clustered regions with higher packing of particles, compared to their monodisperse counterparts ($A_0$ and $B_0$), which exhibit less densely packed clusters that are more sparsely arranged throughout the domain. In the denser suspensions, cluster shape is qualitatively similar between the polydisperse and monodisperse configurations, however, the polydisperse clusters achieve a higher density of particles than their monodisperse analogs. These observations are illustrated in  Figs.~\ref{fig:DistAClustering} and ~\ref{fig:DistBClustering}, where, $x-z$ planes of the particle volume fraction are shown with the three contours discussed previously. 

In addition to examining stream-wise planes of data, we also consider the cross-stream constitution of clusters. To this end, for each $y-z$ plane along the gravity-aligned direction, $x$, we sum the number of Eulerian cells containing a local particle volume fraction greater than $1.5\langle \alpha_p \rangle$. Multiplying this value by the $y-z$ area of the cells thus yields a measure of total cross-stream cluster area. While this does not delineate whether this cross-sectional area is comprised of one or several clusters, it aids in quantifying the relative structural differences in clustering. In comparing the cluster cross-sectional area curves for $A$ and $A_0$ at $\langle \alpha_p \rangle = 0.01$, we notice that the maximum cluster cross-sectional area is substantially greater for configuration $A_0$, substantiated by the observation that clusters are qualitatively larger in size but fewer in quantity. We also observe that the cross-sectional cluster area for Dist. $A$ is more uniformly distributed throughout the stream-wise direction, indicating that clusters are more evenly spread throughout the domain as compared to Dist. $A_0$. In contrast to these observations at low volume fraction, we observe that the regions of maximum cross-sectional cluster area are comparable between Dist. $A$ and $A_0$ at $\langle \alpha_p \rangle = 0.10$ and that for both configurations, these peaks are observed at similar intervals. 

Some consistency is seen in the Dist. $B$ and $B_0$ configurations, with a few exceptions. First, at $\langle \alpha_p \rangle = 0.01$, we note that Dist. $B_0$ contains relatively larger peak values for cluster cross-sectional area, however, these peaks are similarly distributed as compared with Dist. $A$. At higher volume fraction, $\langle \alpha_p \rangle = 0.10$, the polydisperse assembly, Dist. $B$, achieves slightly higher cross-sectional cluster area values as compared to Dist. $B_0$, which is consistent with the qualitative observation that a few of the polydisperse clusters are aligned more with the cross-stream, than the stream-wise direction.  

Finally, as detailed in Fig.~\ref{fig:clusteringSummary}, we note that the structures of the polydisperse clusters are more fragmented than their monodisperse counterparts. This is partly due to the existence of particles that range from very small to very large particles in the domain but also may point to the formation of larger clusters when two moderately sized clusters merge. 

\begin{figure}
\centering
\begin{tabular}{c c}
(a) Dist. $A_0$, $\langle \alpha_p \rangle = 0.01$ & 
(b) Dist. $A$, $\langle \alpha_p \rangle = 0.01$ \\
\includegraphics[width=0.5\textwidth]{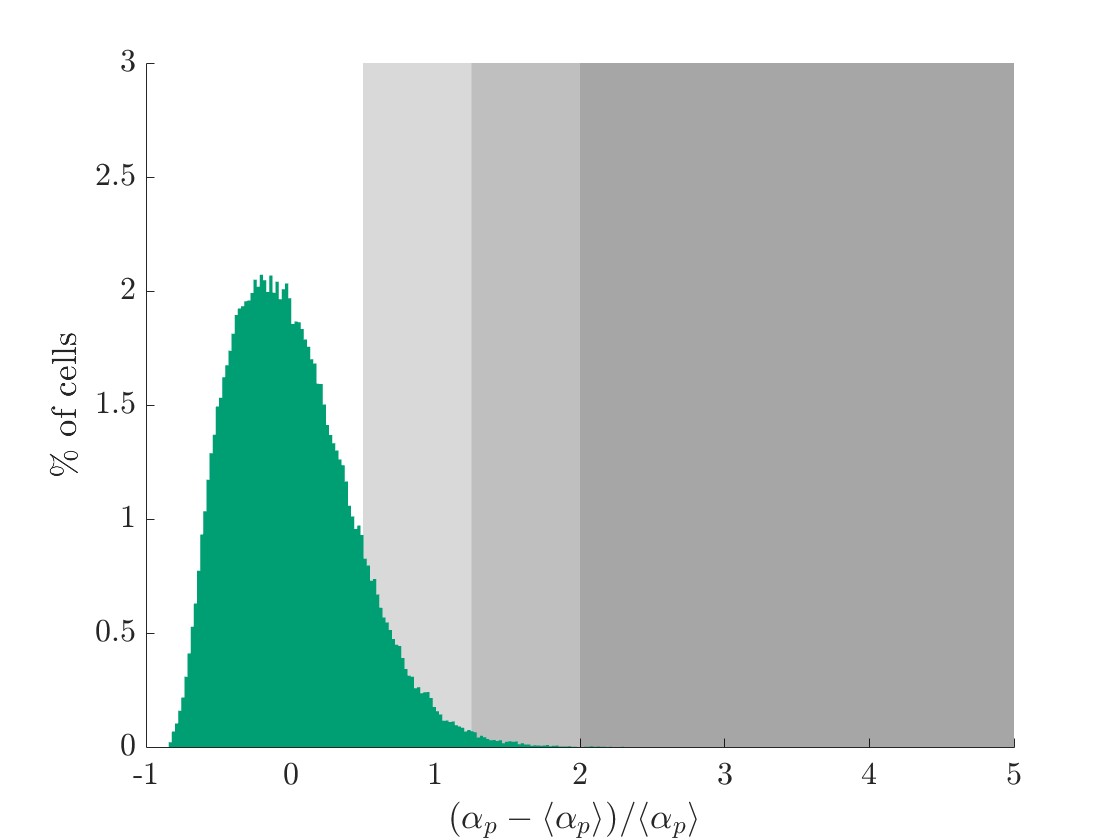} & 
\includegraphics[width=0.5\textwidth]{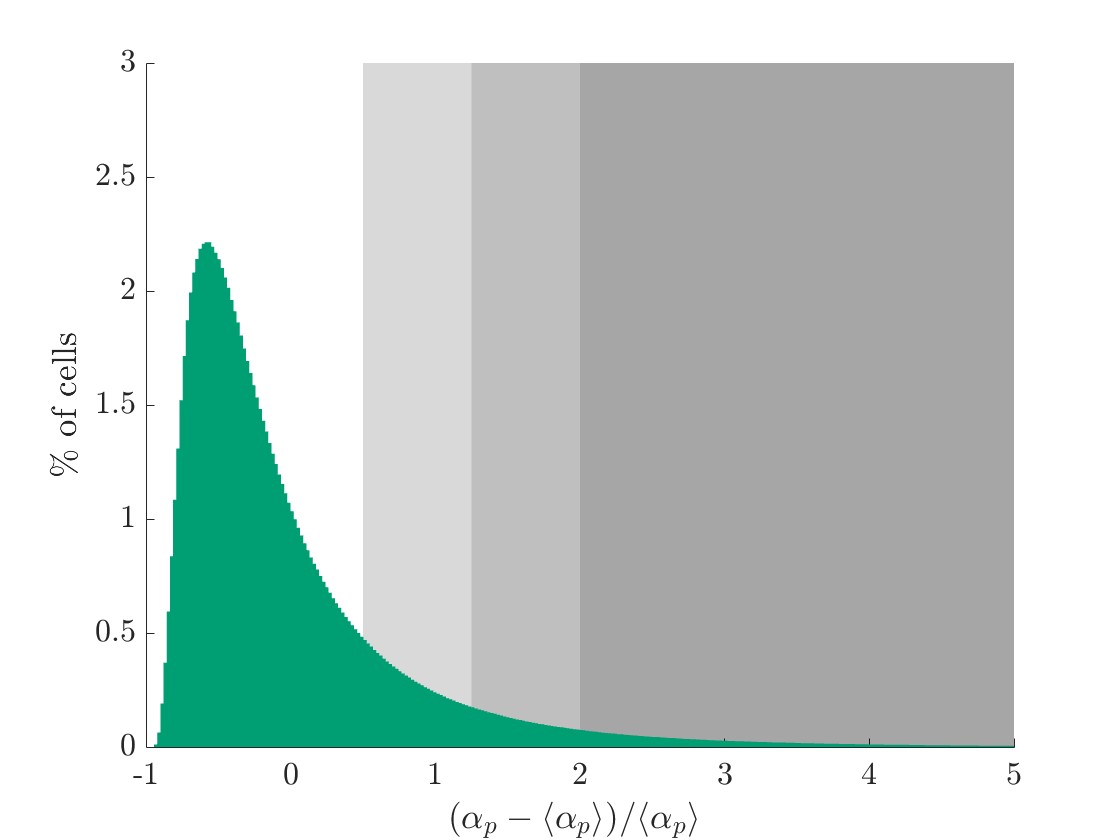} \\
(c) Dist. $B_0$, $\langle \alpha_p \rangle = 0.01$ & 
(d) Dist. $B$, $\langle \alpha_p \rangle = 0.01$ \\
\includegraphics[width=0.5\textwidth]{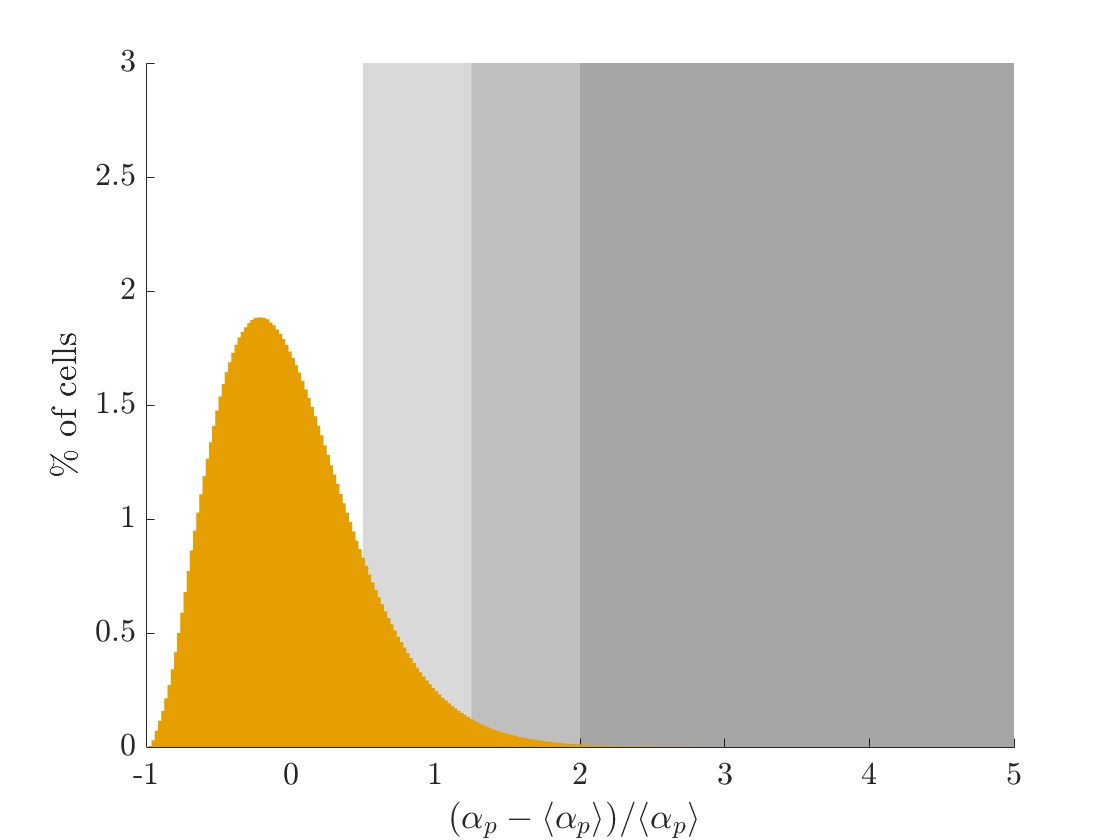} & 
\includegraphics[width=0.5\textwidth]{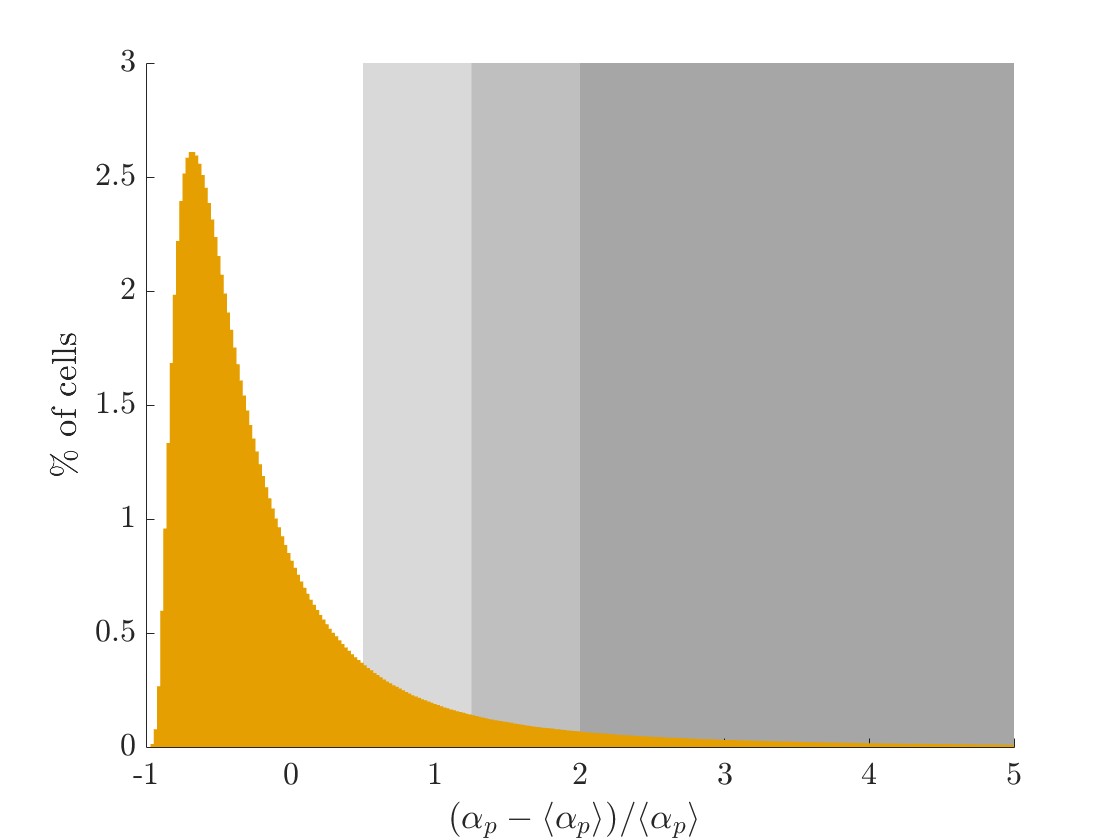}
\end{tabular} 
\caption{Distributions of $\langle \alpha_p \rangle$ for all configurations at the dilute global volume fraction, $\langle \alpha_p \rangle = 0.01.$ The shaded regions indicate regions of the flow for specific ranges of volume fraction: $1.5\geq \alpha_p/\langle \alpha_p \rangle < 2.25 $ (lightest gray), $2.25\geq \alpha_p/\langle \alpha_p \rangle < 3 $ (gray) and $\alpha_p/\langle \alpha_p \rangle \geq 3 $ (darkest gray).}
\label{fig:pdfvf_01}
\end{figure}

\begin{figure}
\centering
\begin{tabular}{c c}
(a) Dist. $A_0$, $\langle \alpha_p \rangle = 0.1$ & 
(b) Dist. $A$, $\langle \alpha_p \rangle = 0.1$ \\
\includegraphics[width=0.5\textwidth]{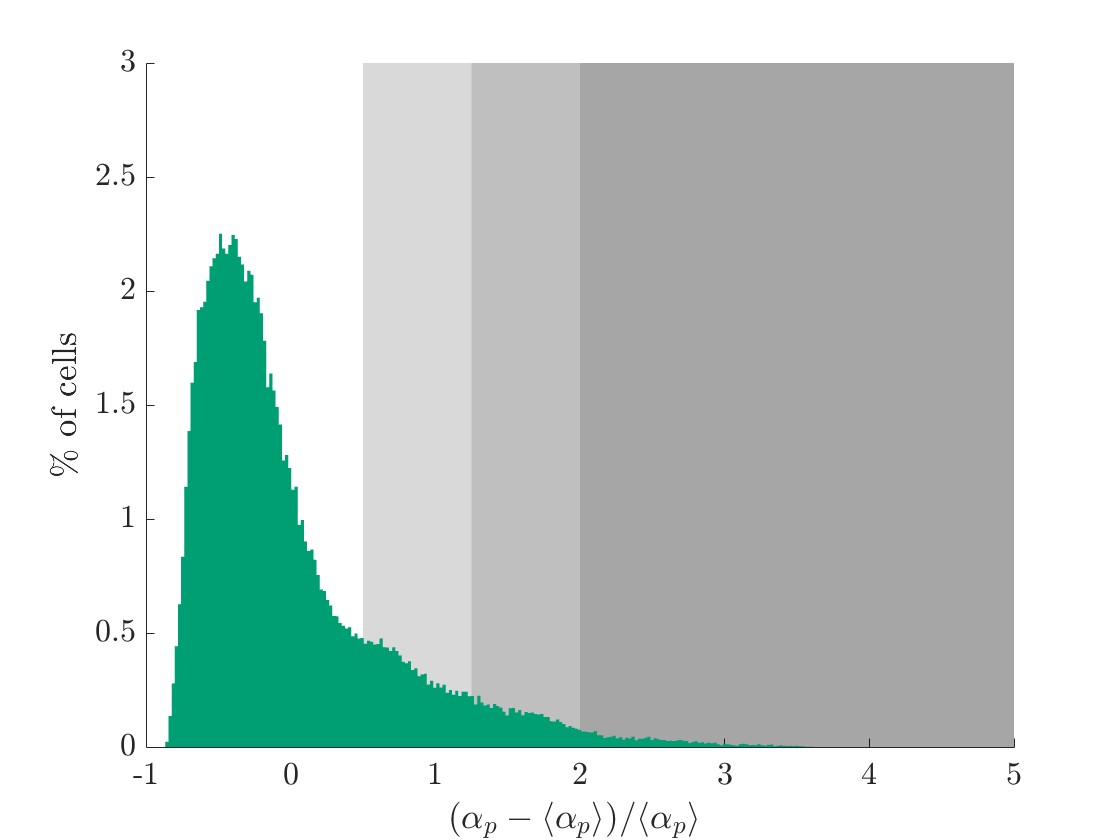} & 
\includegraphics[width=0.5\textwidth]{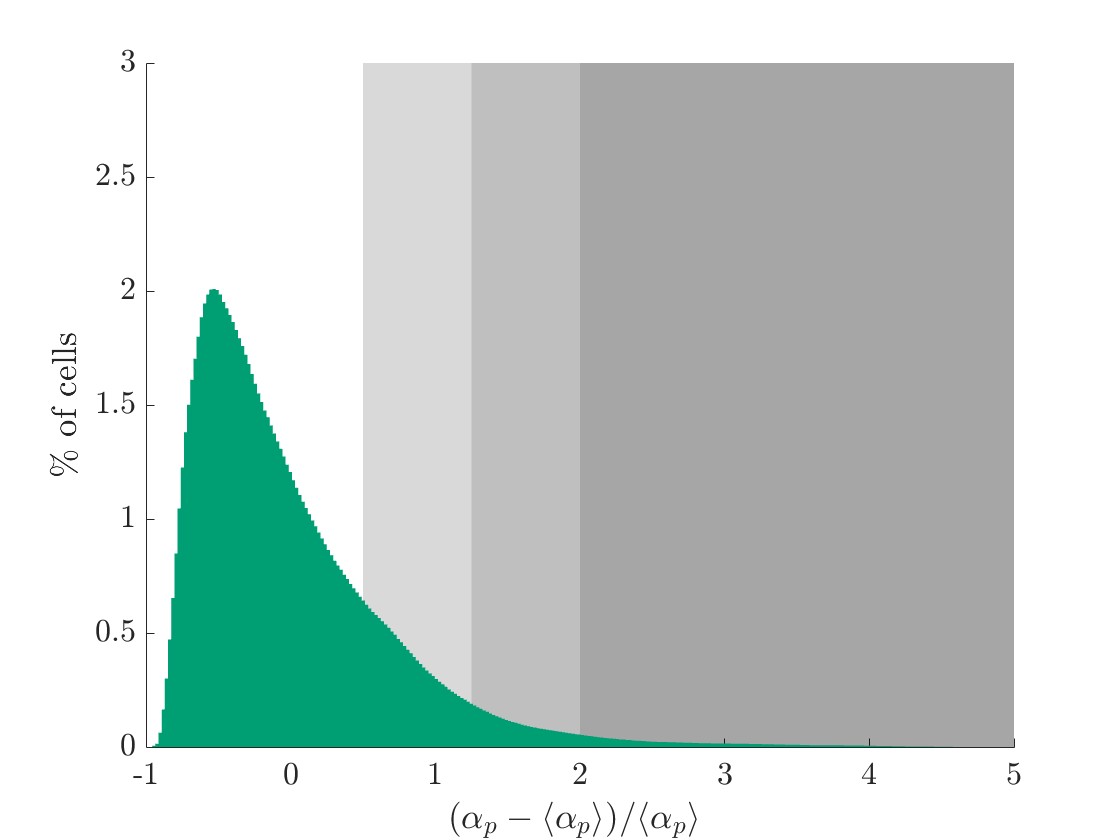} \\
(c) Dist. $B_0$, $\langle \alpha_p \rangle = 0.1$ & 
(d) Dist. $B$, $\langle \alpha_p \rangle = 0.1$ \\
\includegraphics[width=0.5\textwidth]{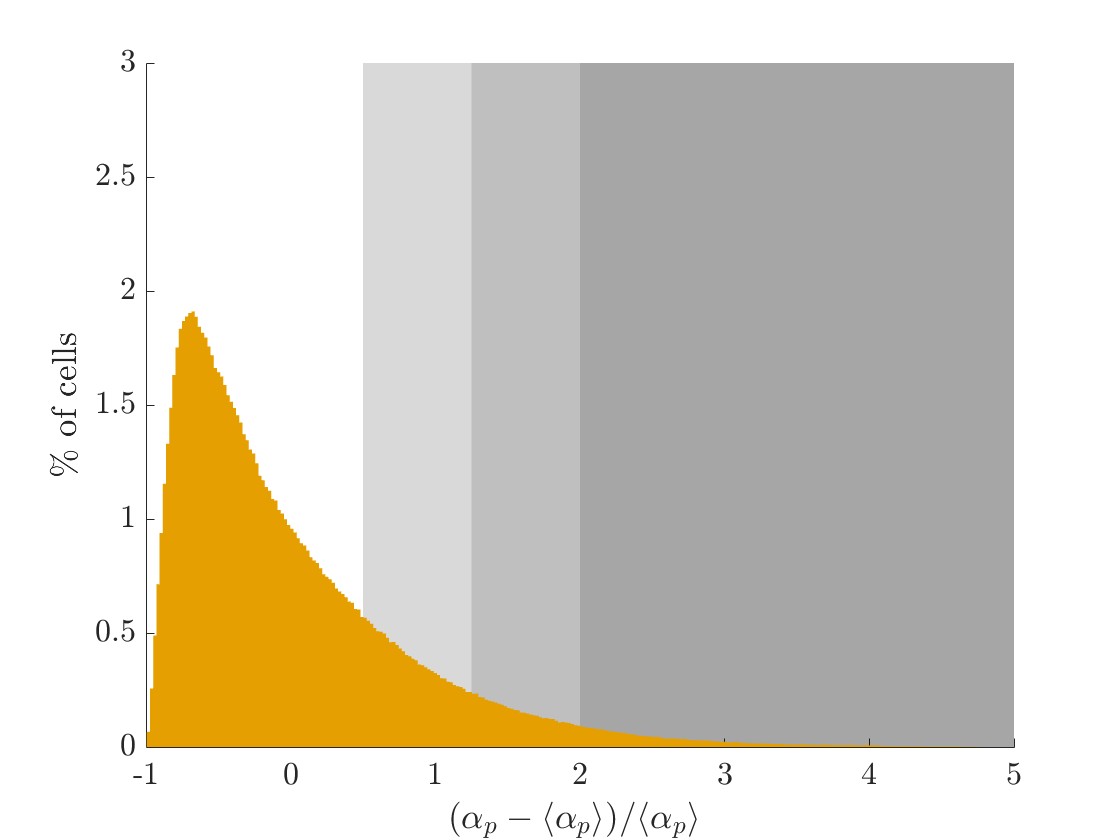} & 
\includegraphics[width=0.5\textwidth]{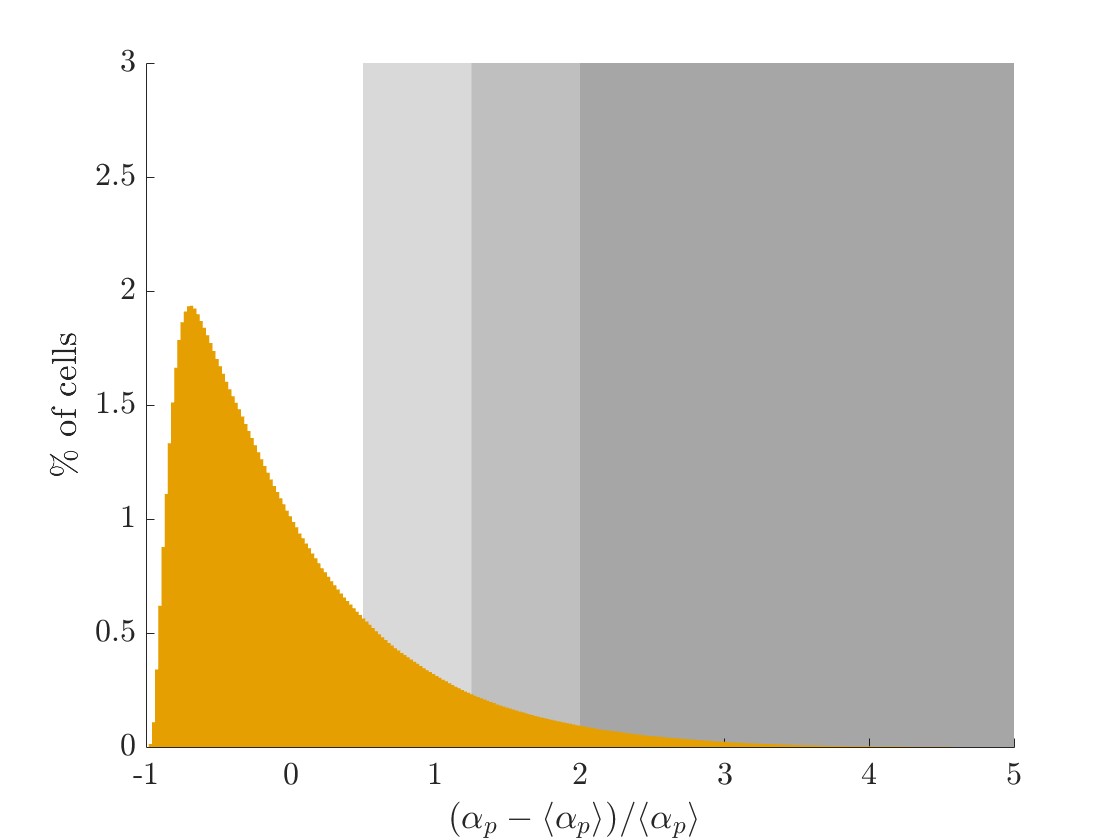}
\end{tabular} 
\caption{Distributions of $\langle \alpha_p \rangle$ for all configurations at the moderately dense global volume fraction, $\langle \alpha_p \rangle = 0.10.$ The shaded regions indicate regions of the flow for specific ranges of volume fraction: $1.5\geq \alpha_p/\langle \alpha_p \rangle < 2.25 $ (lightest gray), $2.25\geq \alpha_p/\langle \alpha_p \rangle < 3 $ (gray) and $\alpha_p/\langle \alpha_p \rangle \geq 3 $ (darkest gray).}
\label{fig:pdfvf_1}
\end{figure}

In addition to the qualitative analysis presented above, we also consider an alternate, quantitative method for parsing particle volume fraction information. Here, we bin Eulerian cells by their volume fraction and generate a domain-wise distribution based on this binning. These distributions are shown in Figs.~\ref{fig:pdfvf_01} and \ref{fig:pdfvf_1} (for reference, the deviation, skewness and kurtosis for the full assemblies of particles have been previously summarized in Tab.~\ref{tab:SvClustering}). In these figures, the shaded regions are consistent with Regions A-D as described previously.

In examining clustering in this way, we note that for $\langle \alpha_p \rangle = 0.01$, the monodisperse configurations have distributions of particle volume fraction that are nearly normally distributed (recall skew$(\alpha_p)$ is 0.66 and 0.82 for Dist. $A_0$ and $B_0$), while the skewness for the polydisperse simulations is much greater in comparison. This implies that the polydisperse configurations attain higher local volume fractions for more regions of the flow compared with their monodisperse analogs. In comparing Dist. $A$ and Dist. $B$ we note that Dist. $B$ contains more nearly void regions compared with $A$ along with a longer tail in the concentrated regime. This implies that Dist $B$ achieves a more stratified version of clustering, where clusters tend to be denser and more regions of the flow are absent of particles. 

Interestingly, at $\langle \alpha_p \rangle = 0.10$, the polydispersed \emph{and} monodispersed assemblies deviate significantly from a normal distribution. Further, the distributions of particle volume fraction are very similar, with the greatest differences occurring between $A$ and $A_0$. This underscores the finding that polydispersity has a much greater effect on clustering behavior at lower volume fraction. We postulate that when the overall number of particles is higher, as is the case for higher volume fraction flows, this allows for a more uniform introduction of flow disturbances, thereby dampening the effect of larger particles compared to smaller particles. 

\begin{figure}
\centering
\begin{tabular}{ l l}
    \includegraphics[height = 0.3\textwidth]{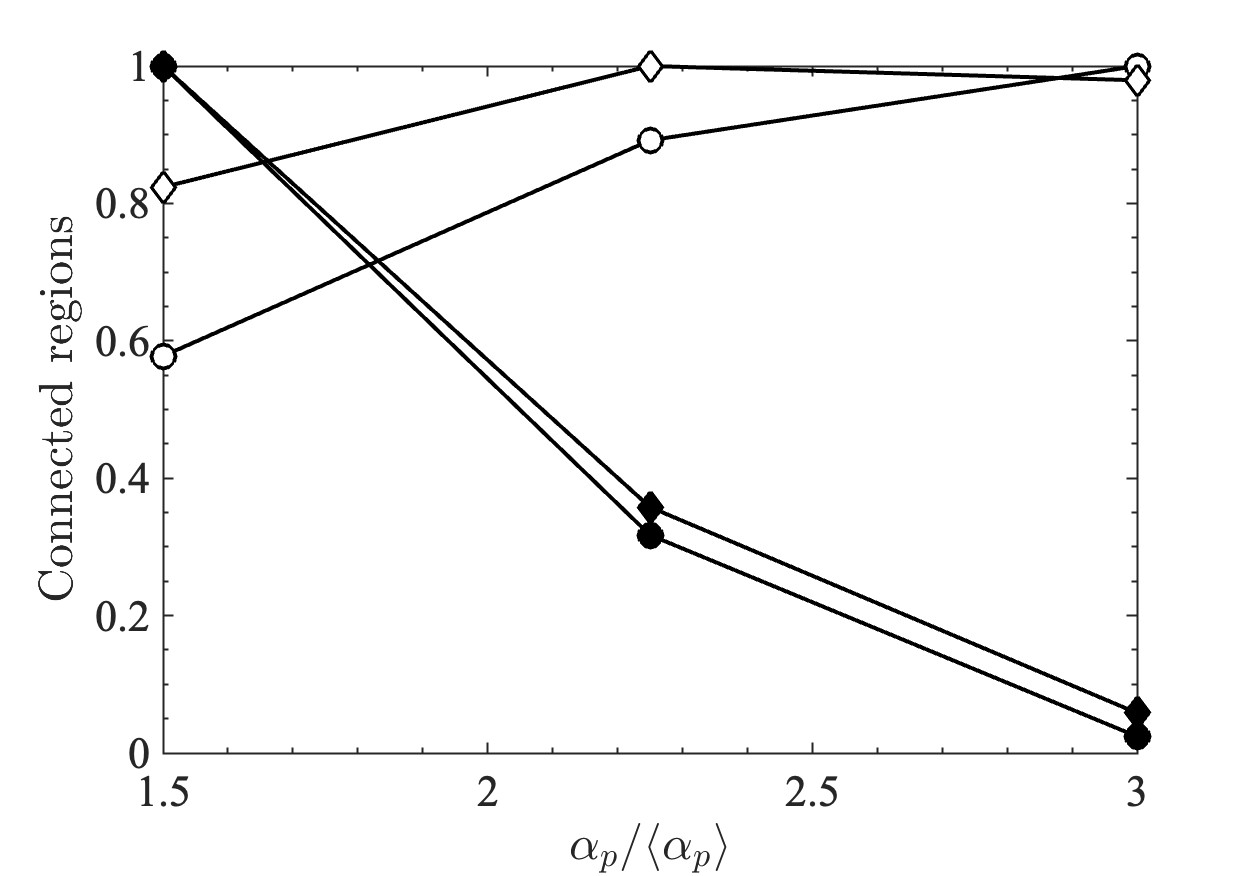} &
   \includegraphics[height = 0.3\textwidth]{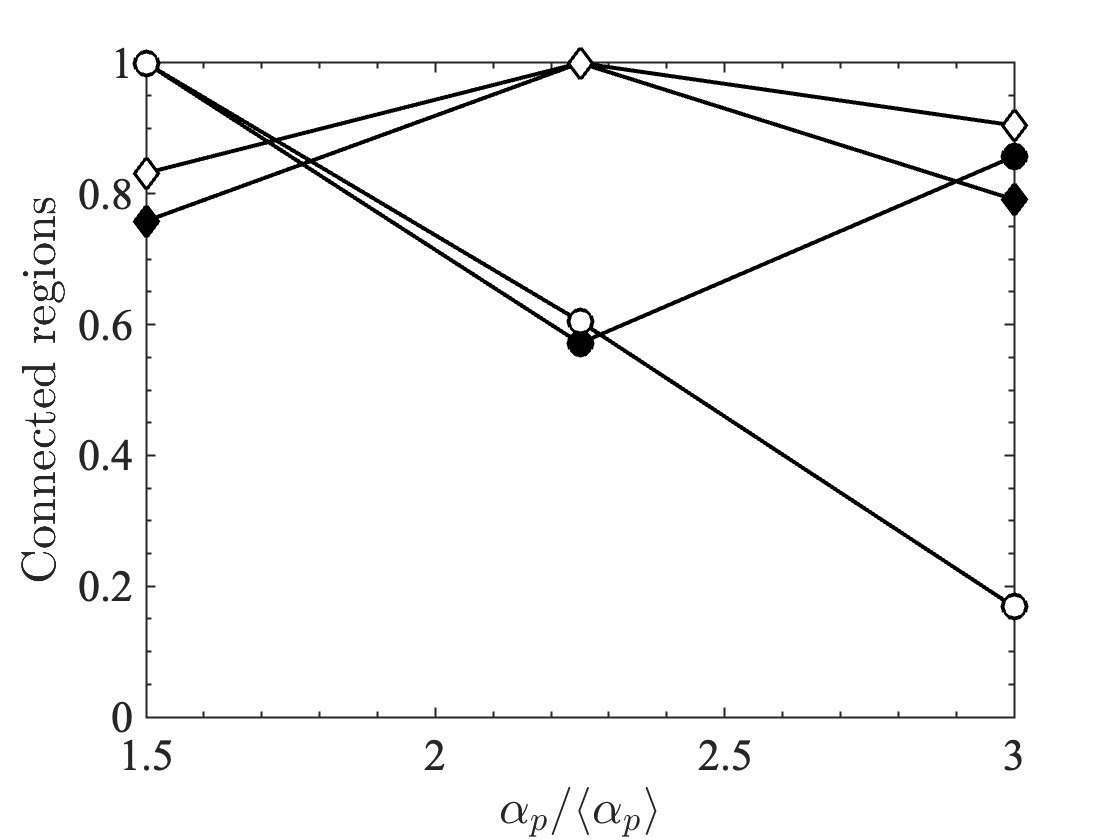} \\
    \end{tabular}
    \caption{Summary of the trend in number of connected regions as a function of the particle volume fraction threshold. To aid in the comparison between configurations, the number of connected regions for each case is normalized by the maximum number of connectivities. Distributions at $\langle \alpha_p \rangle = 0.01$ are shown on the left and $\langle \alpha_p \rangle = 0.10$ are shown on the right. Distributions are denoted as: $A$ (Open circles), $B$ (Open diamonds), $A_0$ (Filled circles) and $B_0$ (Filled diamonds).}
    \label{fig:Connectivity}
\end{figure}

Another important distinction we note in our qualitative observations of clustering is the number of clustered regions in the flow. A quantitative way to illustrate this clustering behavior is to consider the connectivity of the particle volume fraction field. To this end, each Eulerian cell is assigned a binary value corresponding to whether the filtered particle volume fraction is above or below a threshold (corresponding to regions A-D). Cells that have two or more `connected' cells (i.e., either a face, edge or vertex is shared) are considered to be connected and therefore part of a cluster. Connectivity mappings were carried out for all of the configurations at thresholds corresponding to $\alpha_p = (1.5, 2.25, 3.0) \langle \alpha_p\rangle$ (Regions B, C and D). These values are summarized in Tab.~\ref{tab:Connectivity} and illustrated in Fig.~\ref{fig:Connectivity}. 

In keeping with our qualitative assessment of each configuration, we observe that the monodisperse distributions at $\langle \alpha_p \rangle = 0.01$ contain fewer connected regions as the volume fraction threshold increases, whereas the polydisperse cases have an increasing number of connected regions. This implies that polydisperse clusters are comprised of a greater number of fragmented dense regions, whereas monodispersed clusters generally have a minority of regions in clusters that attain a dense packing efficiency. In contrast, at $\langle \alpha_p \rangle =0.10$, no clear trend exists, however, this is likely due to the fact that both $A_0$ and $B_0$ contain relatively very few connected regions in general.

\begin{figure}
\centering
\begin{tabular}{c c c | c c}
& \multicolumn{2}{c}{ $\langle \alpha_p \rangle = 0.01$} & \multicolumn{2}{c}{ $\langle \alpha_p \rangle = 0.10$}\\
& Dist. $A$ & Dist. $B$ & Dist. $A$ & Dist. $B$\\ [1ex]
\hline
\rotatebox{90}{\hspace{1em} Region A} & \includegraphics[width=0.22\textwidth]{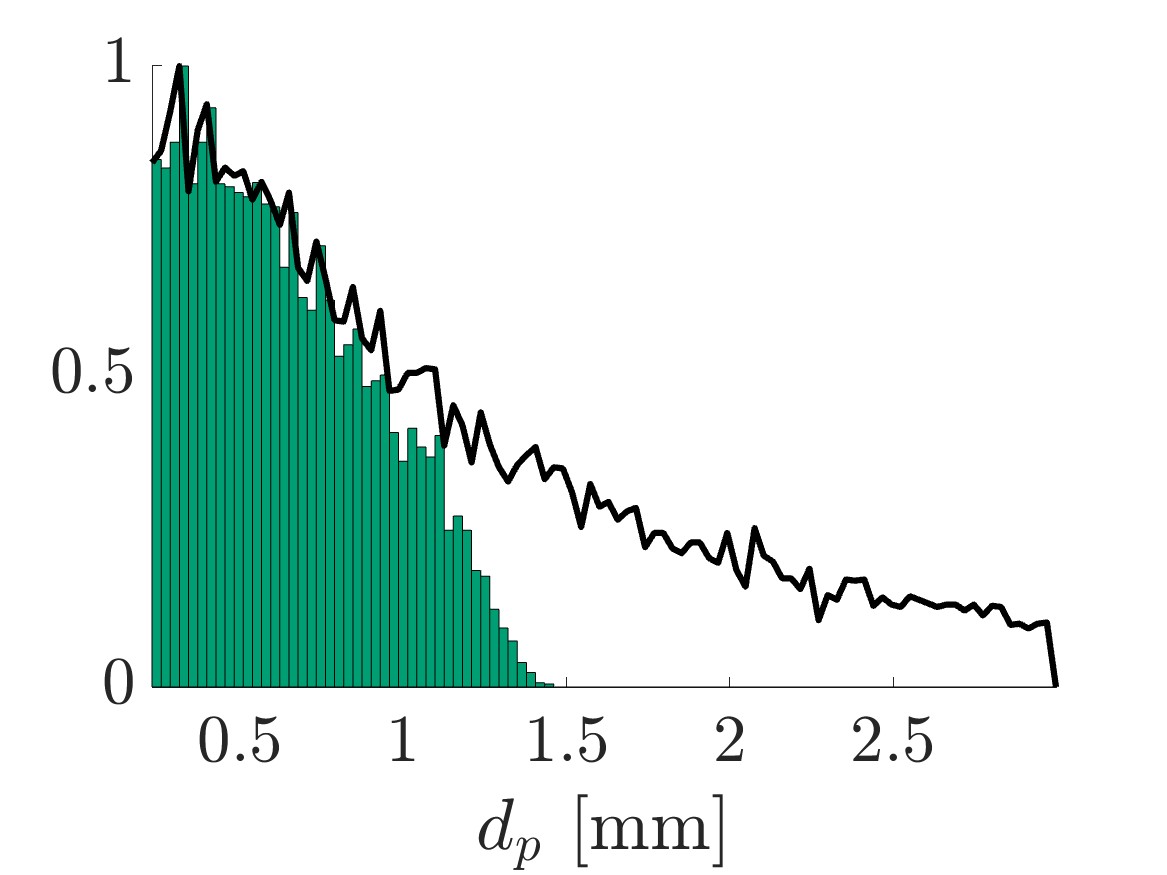} & 
\includegraphics[width=0.22\textwidth]{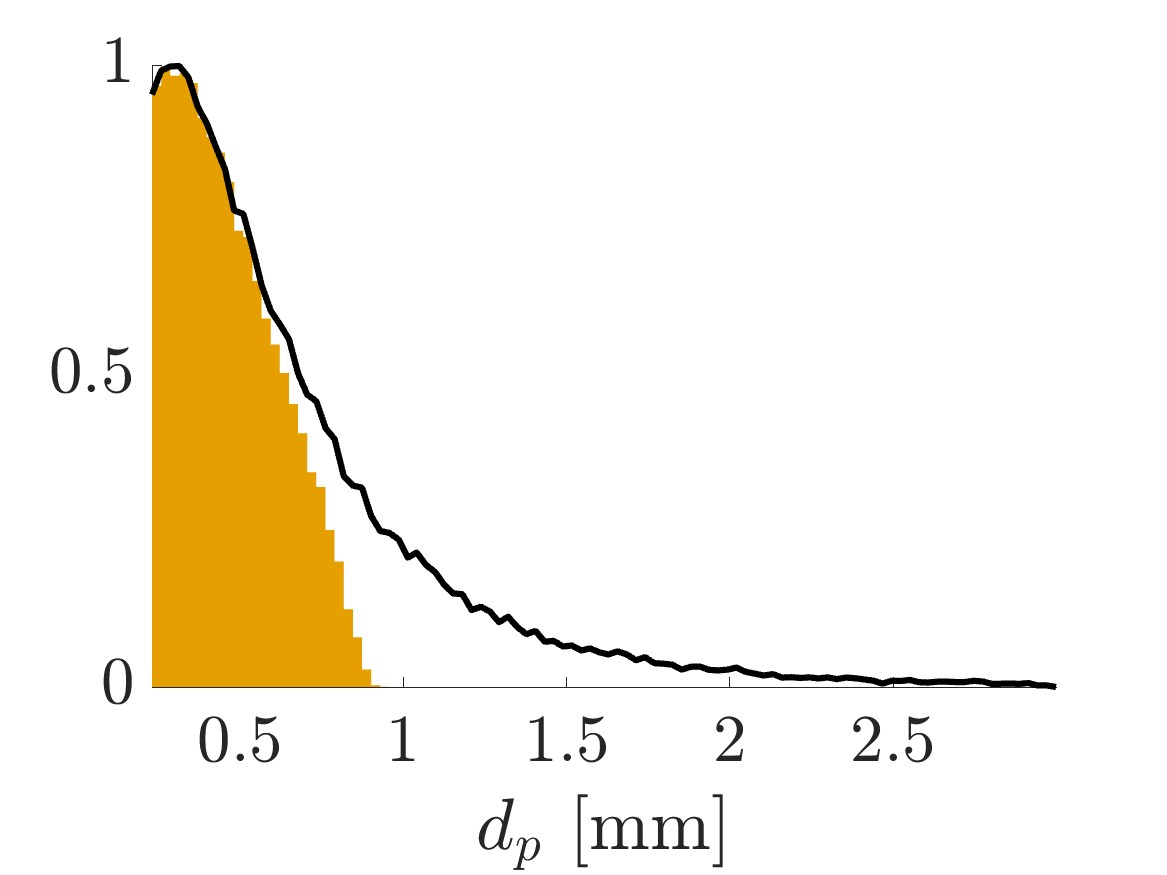} & 
\includegraphics[width=0.22\textwidth] {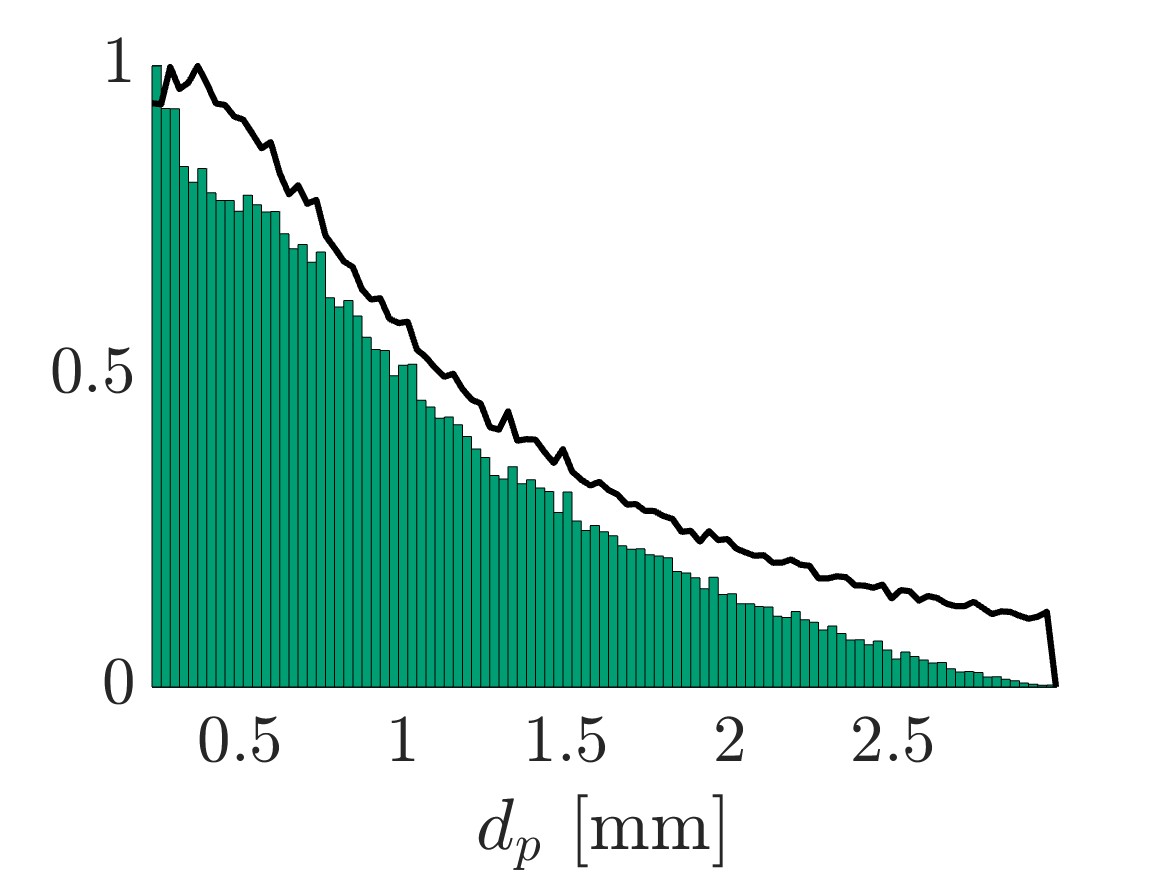}&
\includegraphics[width=0.22\textwidth]{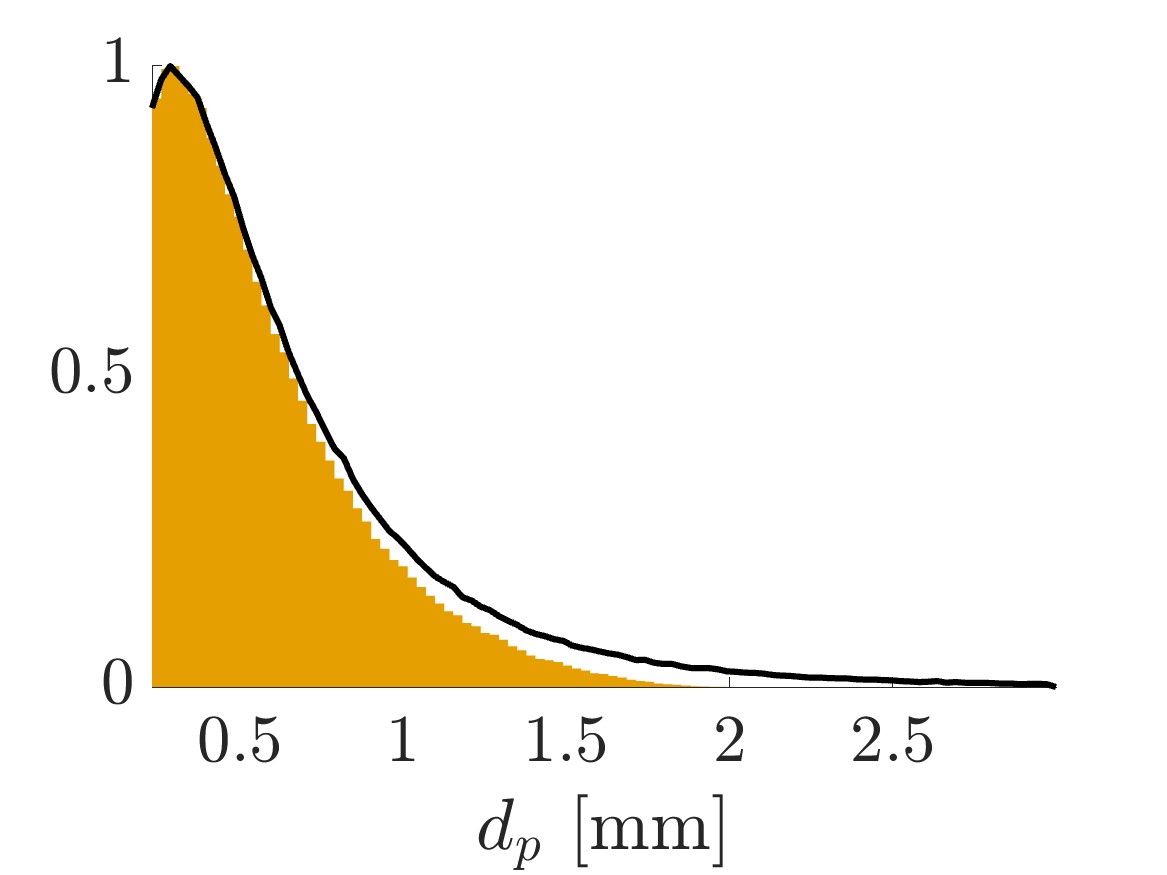}\\[2ex]
\rotatebox{90}{\hspace{1em} Region B} &\includegraphics[width=0.22\textwidth]{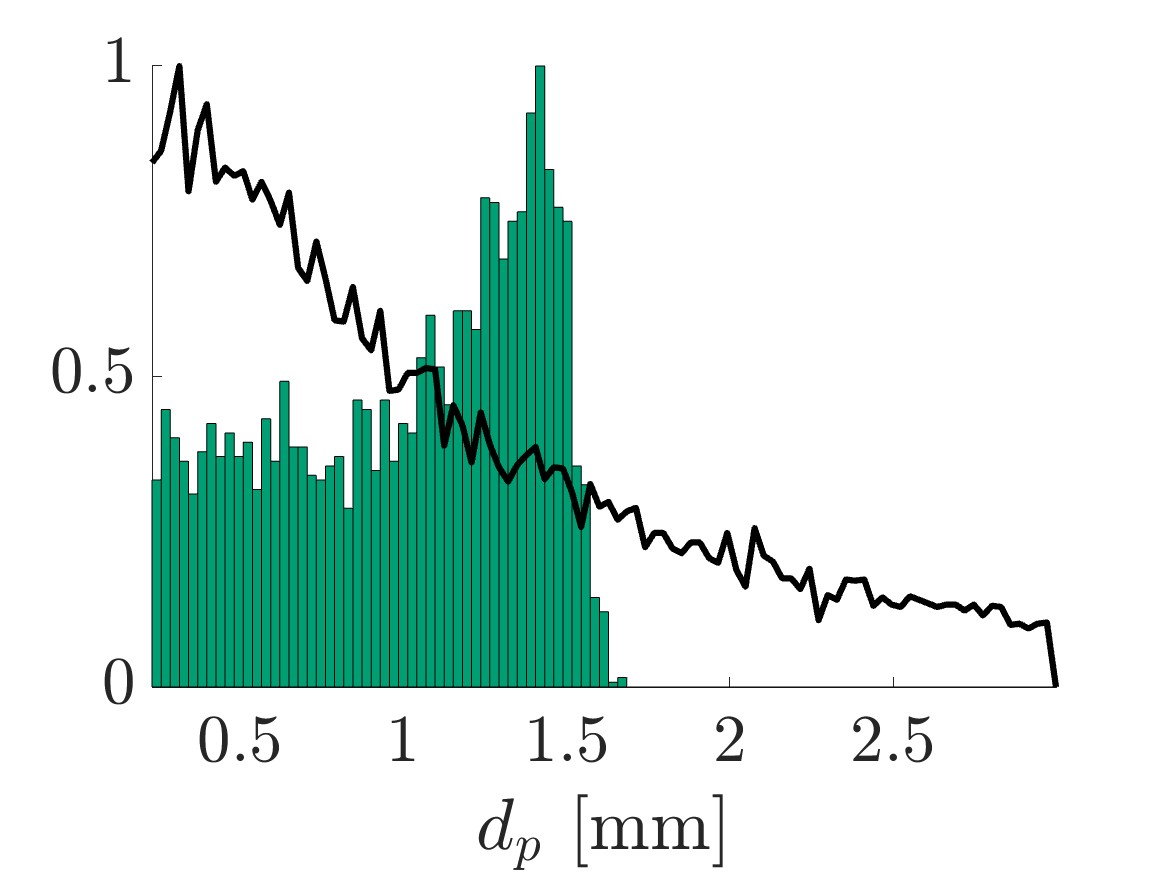} & 
\includegraphics[width=0.22\textwidth]{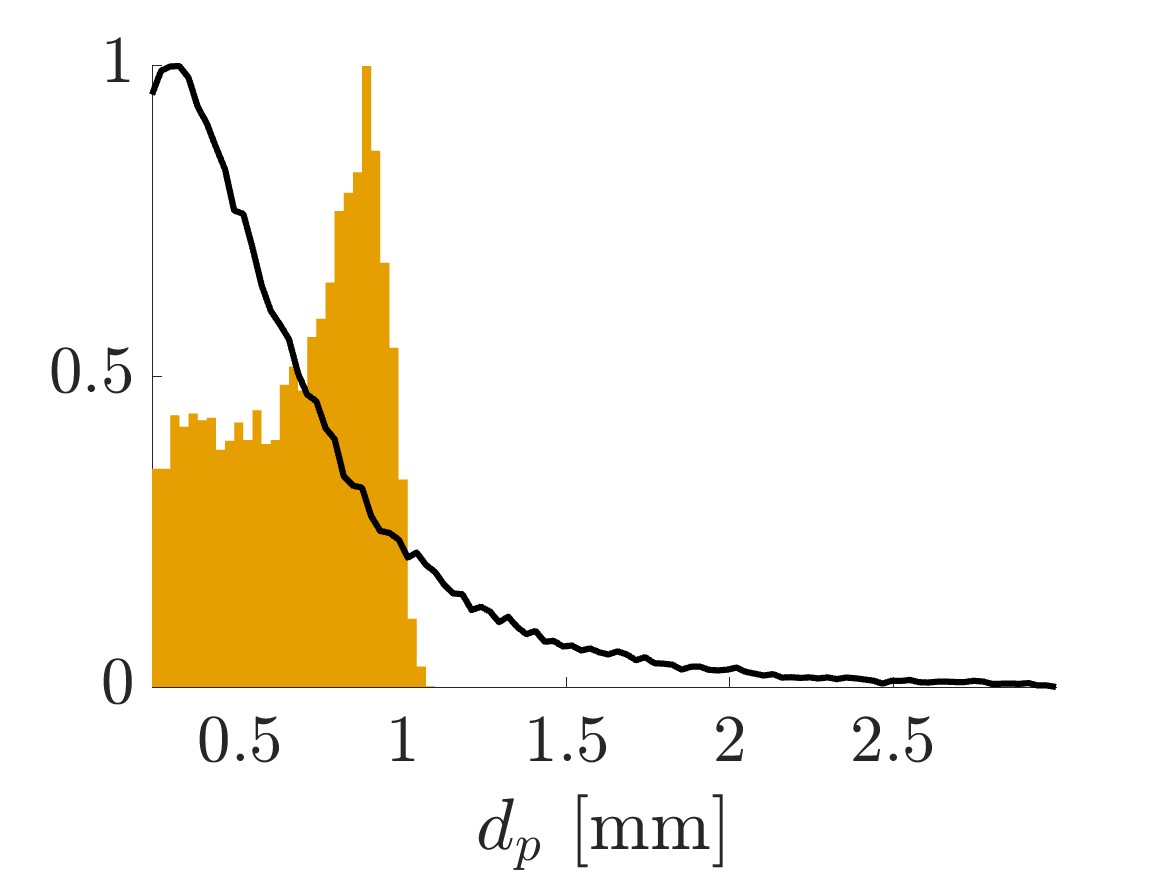} & 
\includegraphics[width=0.22\textwidth] {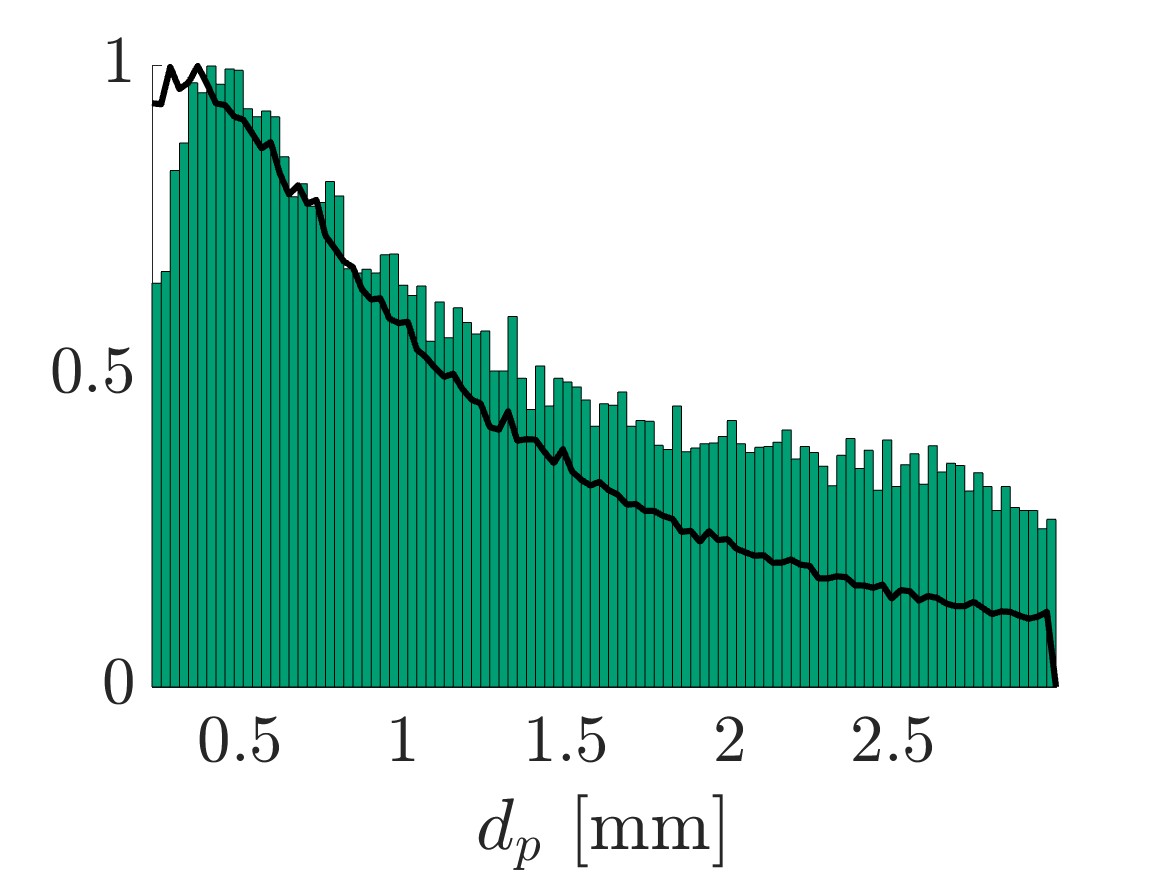}&
\includegraphics[width=0.22\textwidth]{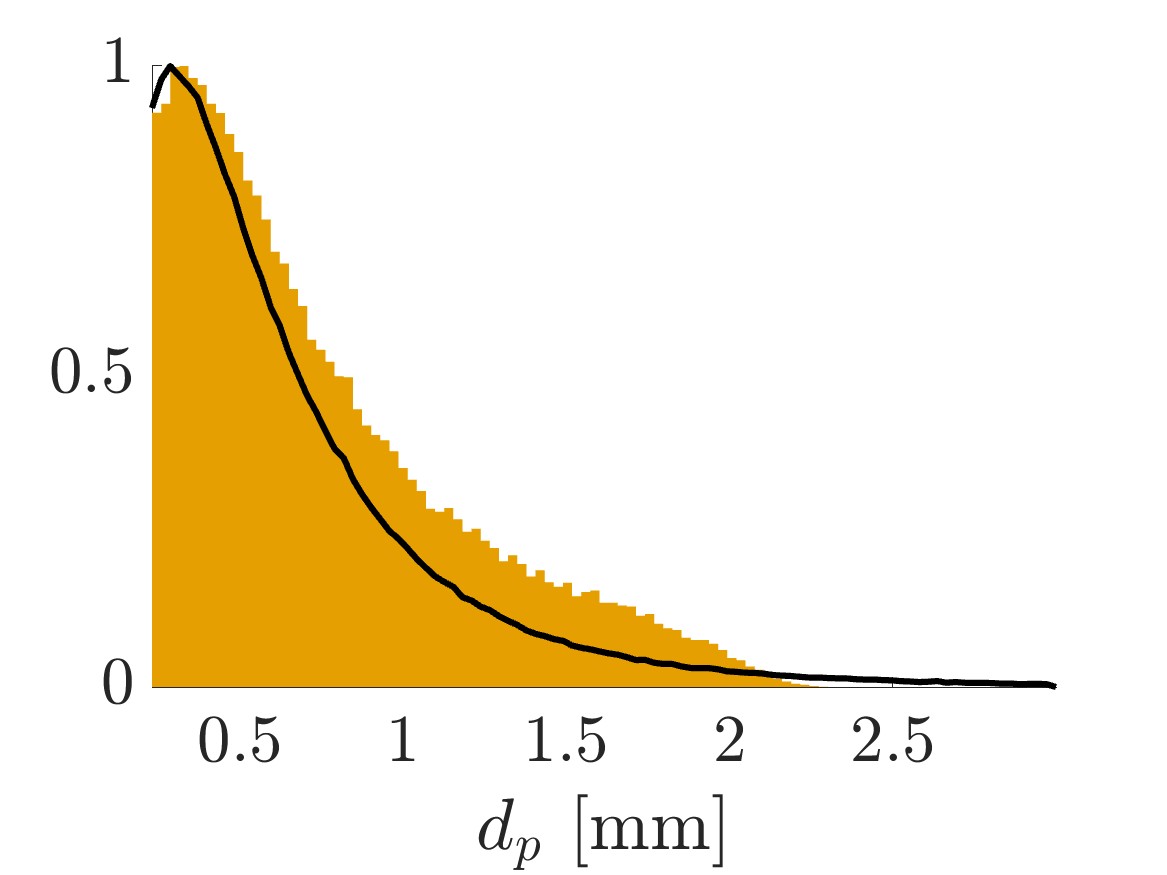}\\ [2ex]
\rotatebox{90}{\hspace{1em} Region C}& \includegraphics[width=0.22\textwidth]{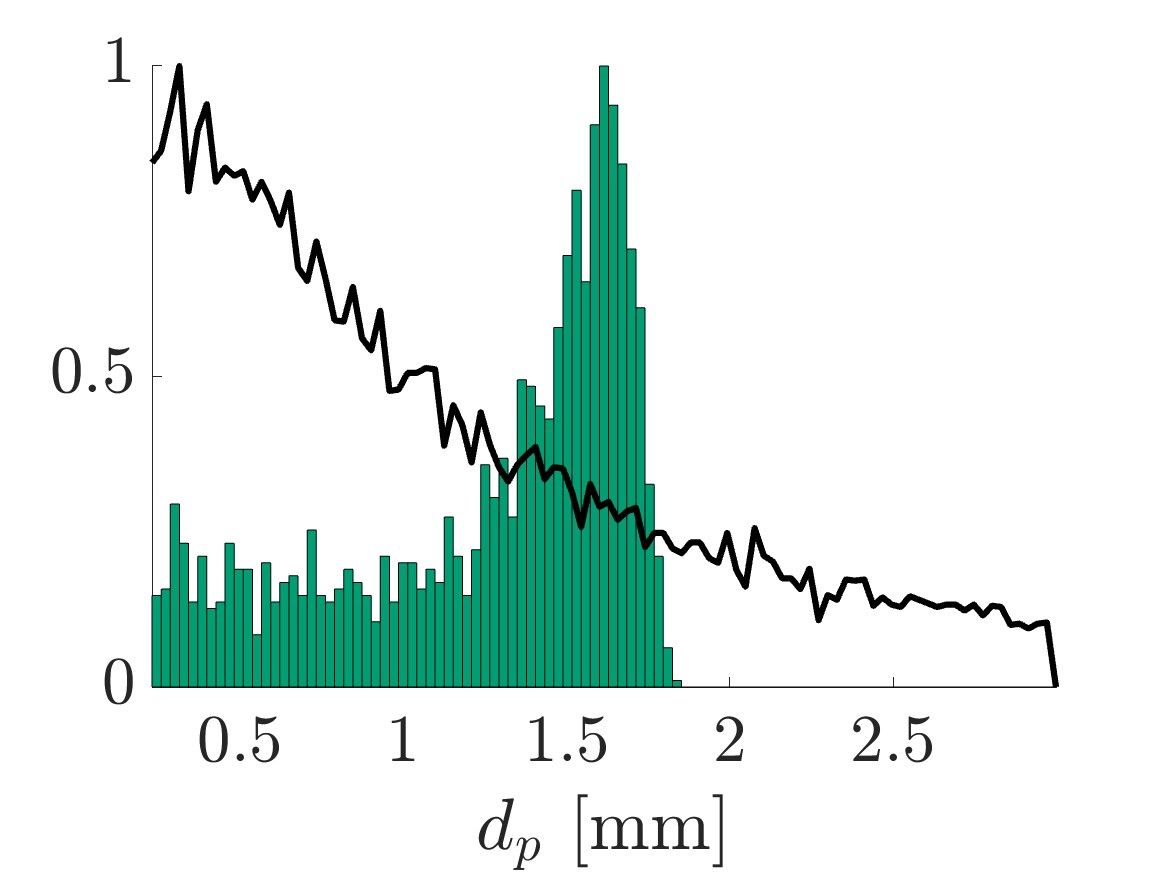} & 
\includegraphics[width=0.22\textwidth]{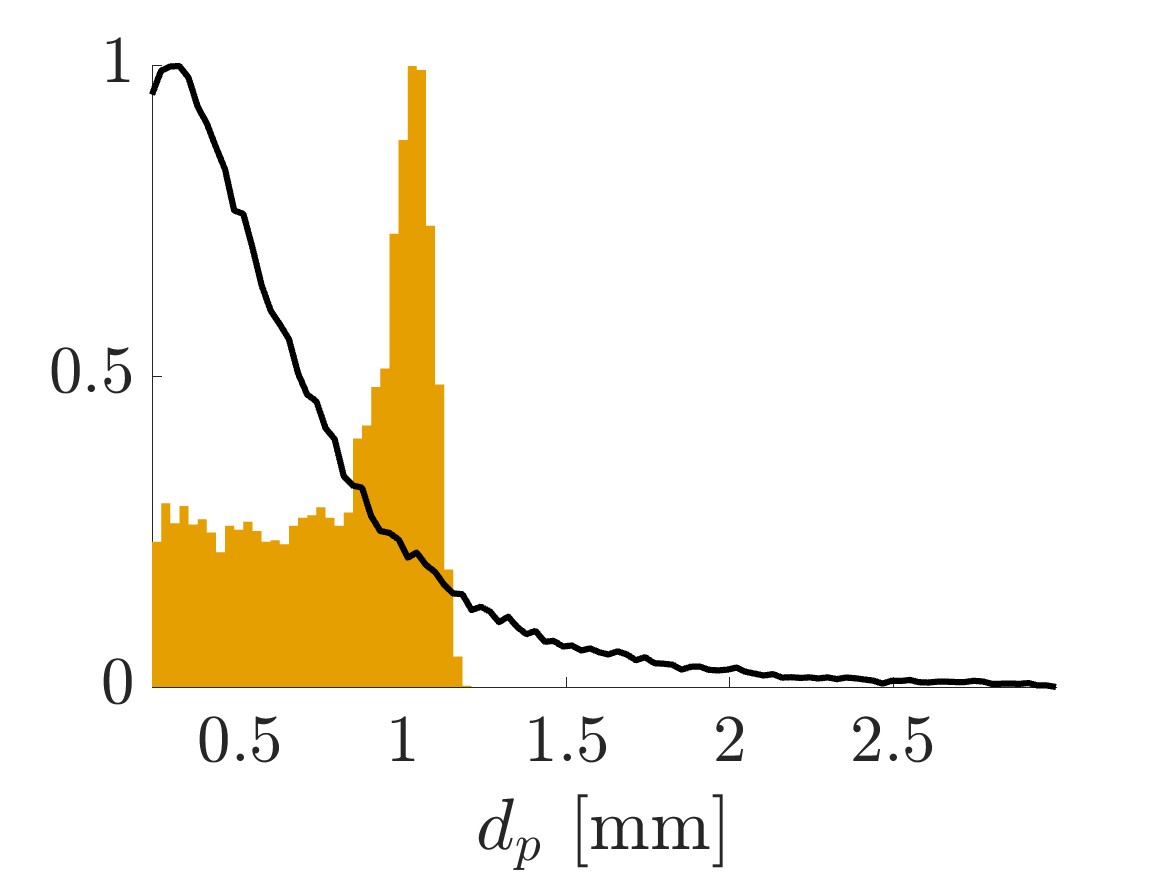} & 
\includegraphics[width=0.22\textwidth] {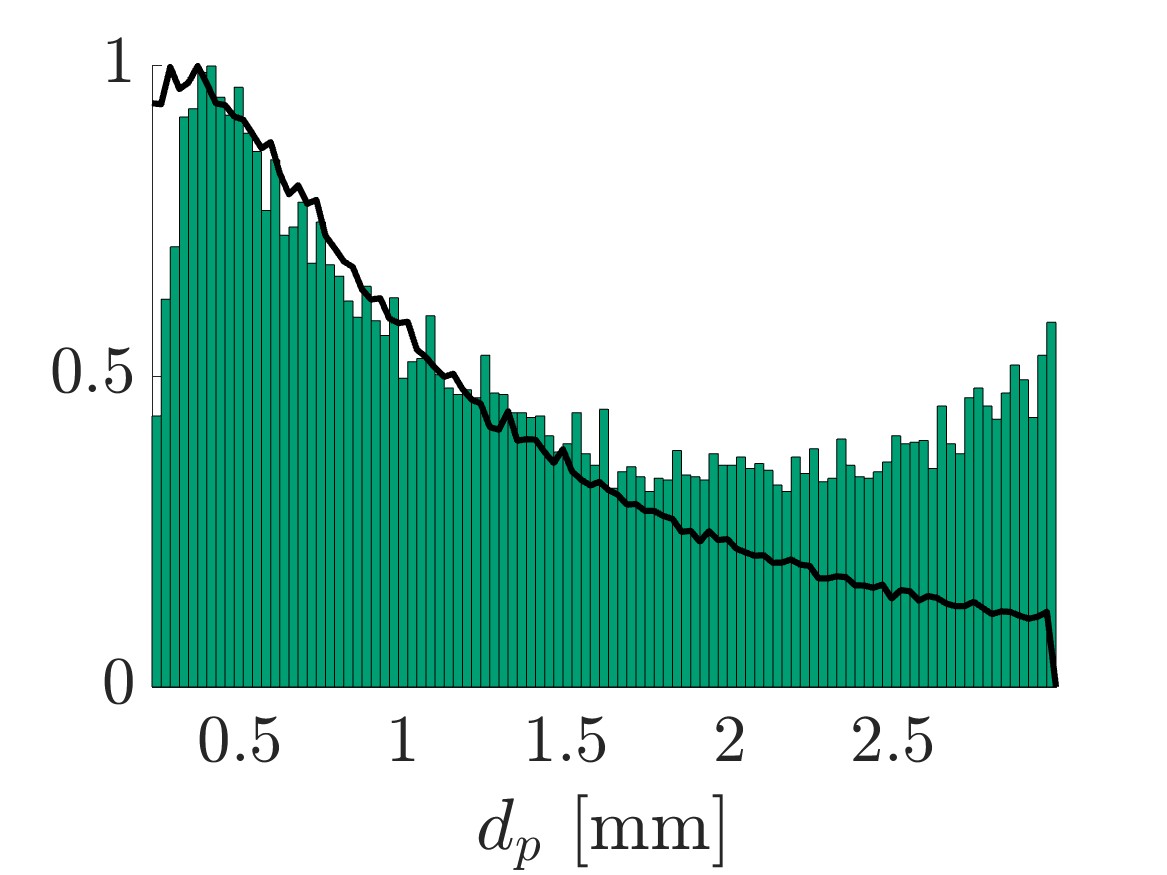}&
\includegraphics[width=0.22\textwidth]{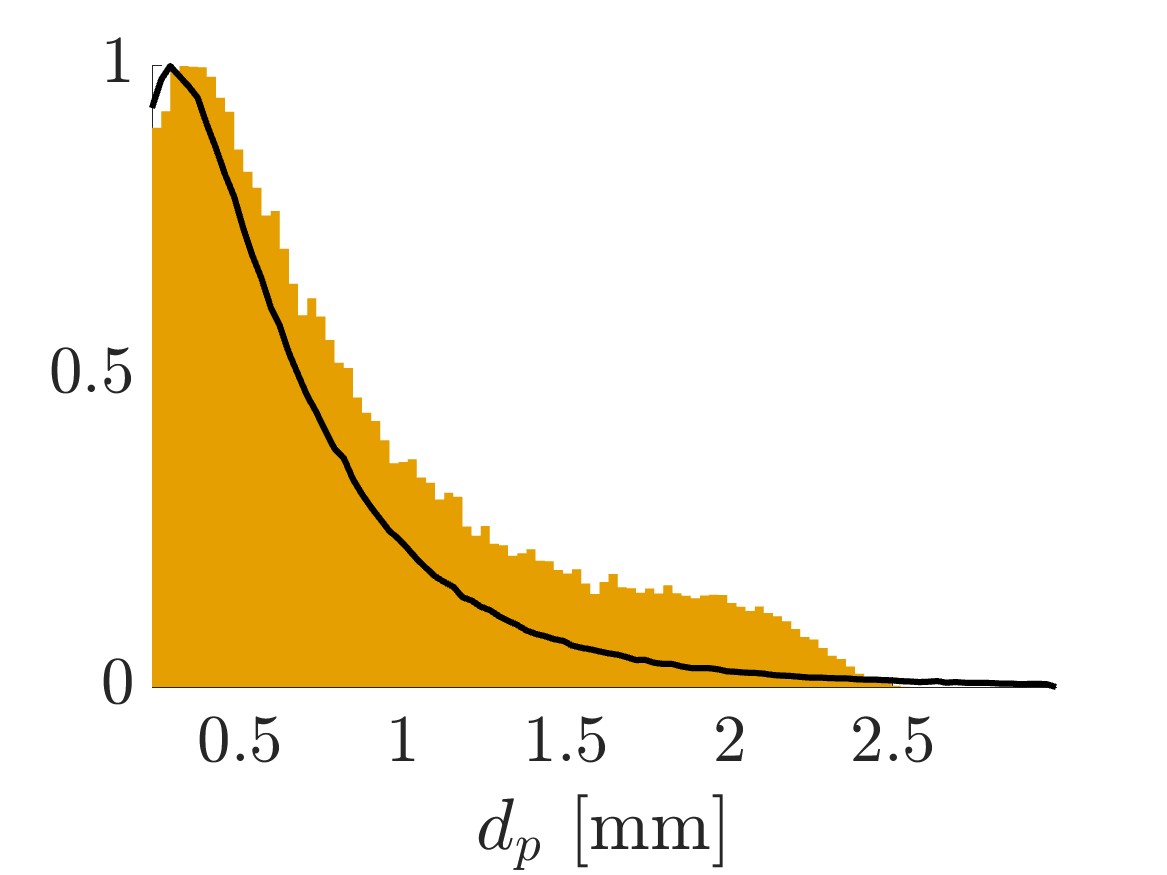}\\[2ex]
\rotatebox{90}{\hspace{1em} Region D}&\includegraphics[width=0.22\textwidth]{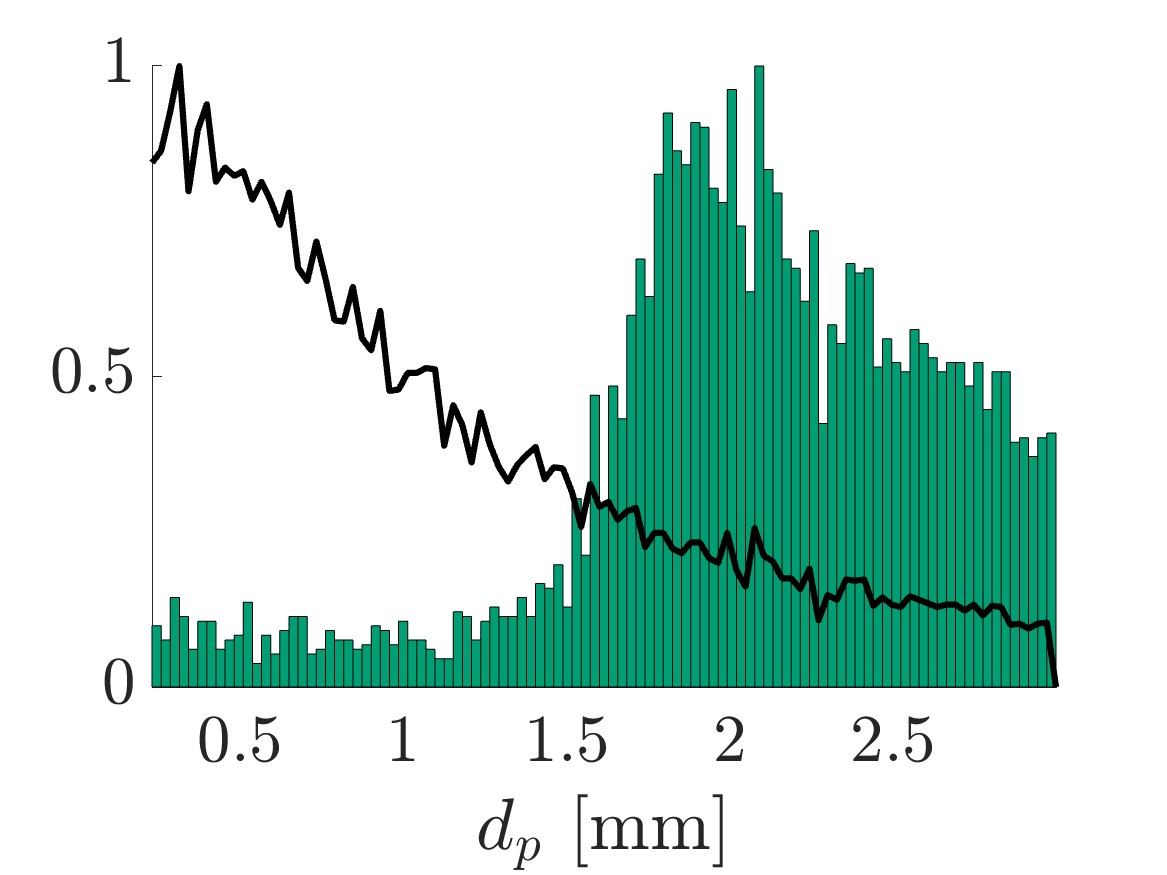} & 
\includegraphics[width=0.22\textwidth]{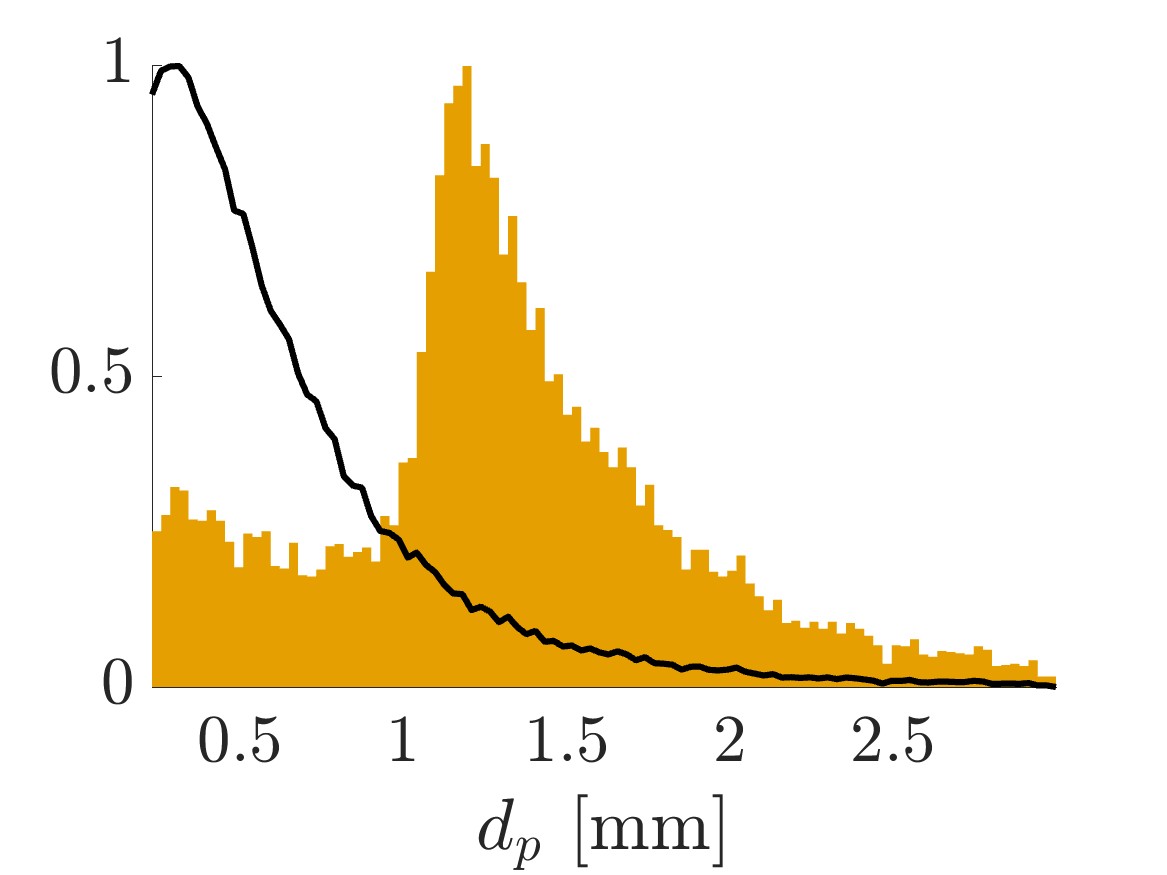} & 
\includegraphics[width=0.22\textwidth] {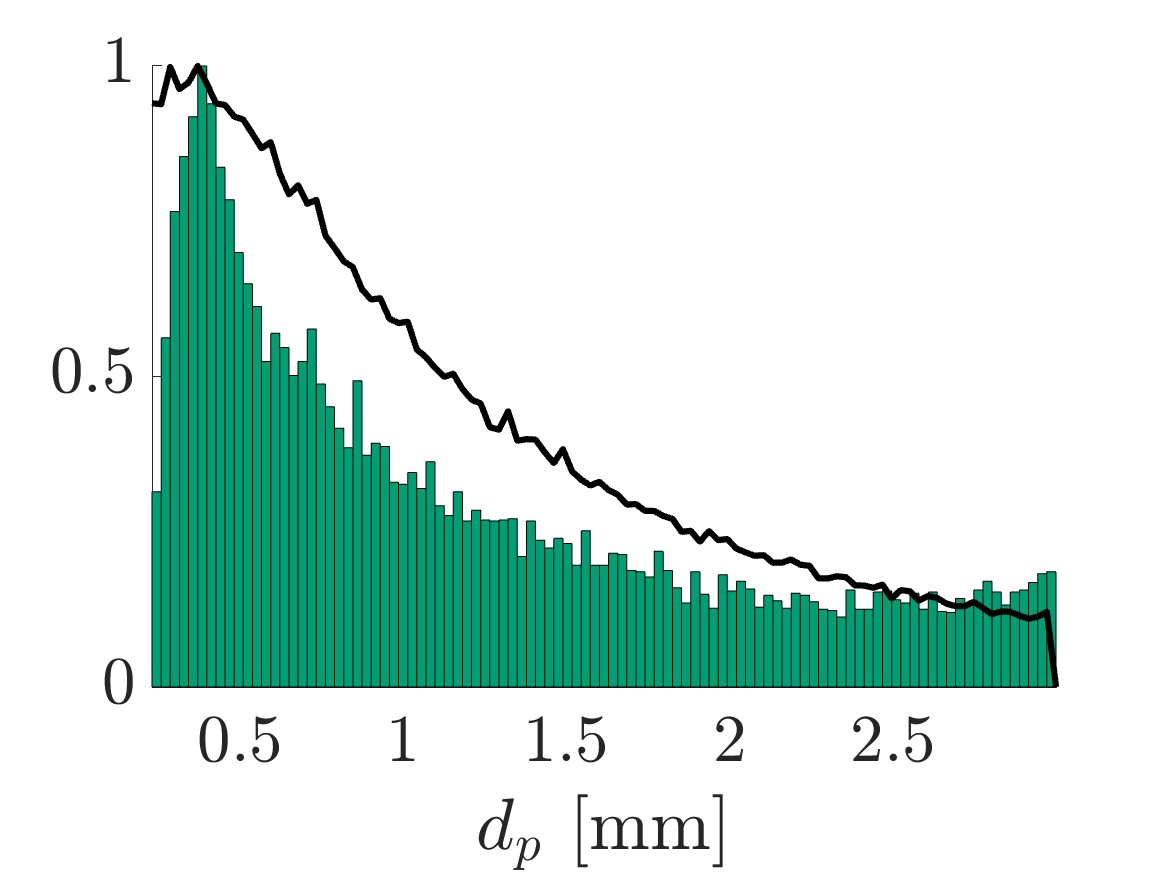}&
\includegraphics[width=0.22\textwidth]{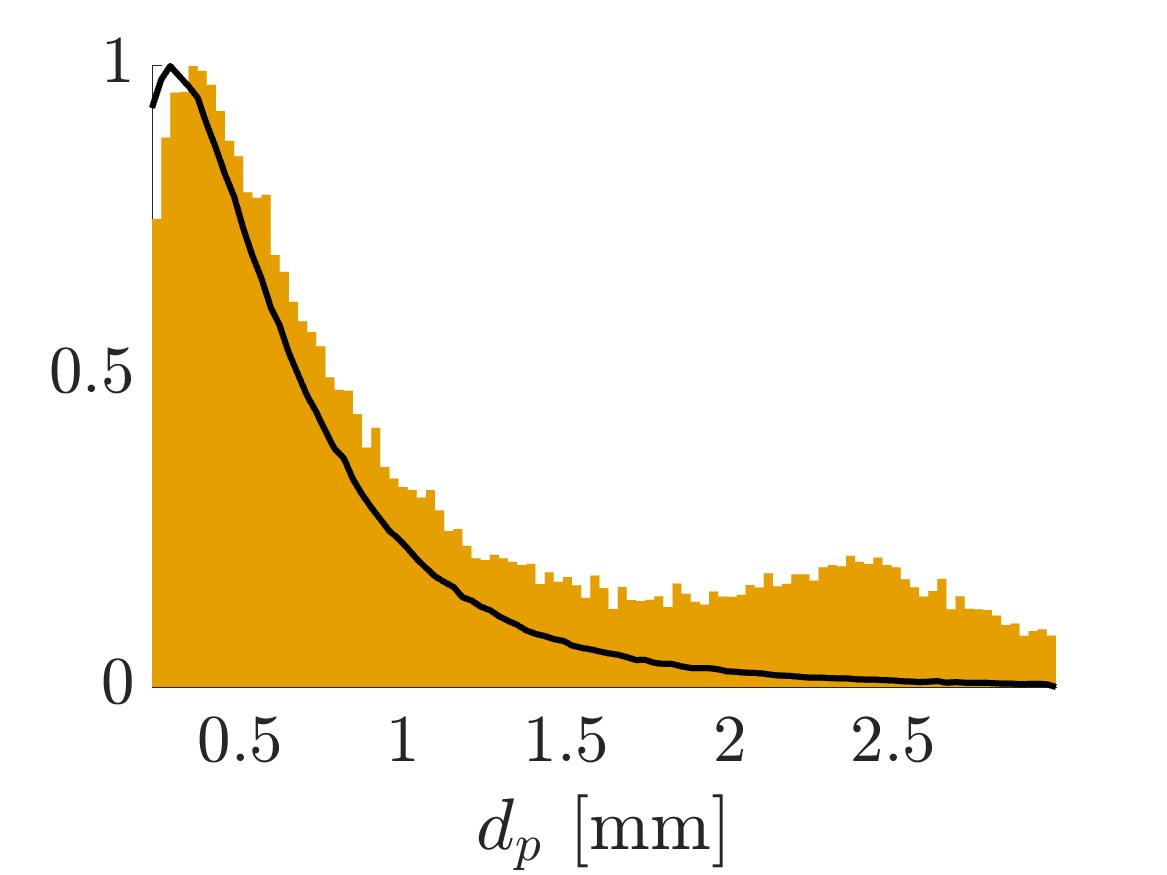}\\
\end{tabular} 
\caption{Summary of the particle size distributions based on local volume fraction for the polydisperse distributions $A$ (teal) and $B$ (mustard) at  $\langle \alpha_p \rangle = 0.01$ (left two columns) and $\langle \alpha_p \rangle = 0.10$ (right two columns). PDFs are computed based on local volume fraction cutoffs corresponding to Regions A, B, C and D, as noted. All plots show the normalized pdf (shaded bars) of the particle diameters in each region of the flow, with the normalized distribution of all the particles shown as a solid black line.}
\label{fig:pdfs_diameter_clustering}
\end{figure}

\begin{figure}
\centering
\begin{tabular}{c c c | c c}
& \multicolumn{2}{c}{ $\langle \alpha_p \rangle = 0.01$} & \multicolumn{2}{c}{ $\langle \alpha_p \rangle = 0.10$}\\
& Dist. $A$ & Dist. $B$ & Dist. $A$ & Dist. $B$\\ [1ex]
\hline
\rotatebox{90}{\hspace{1em} Region A} & \includegraphics[width=0.22\textwidth]{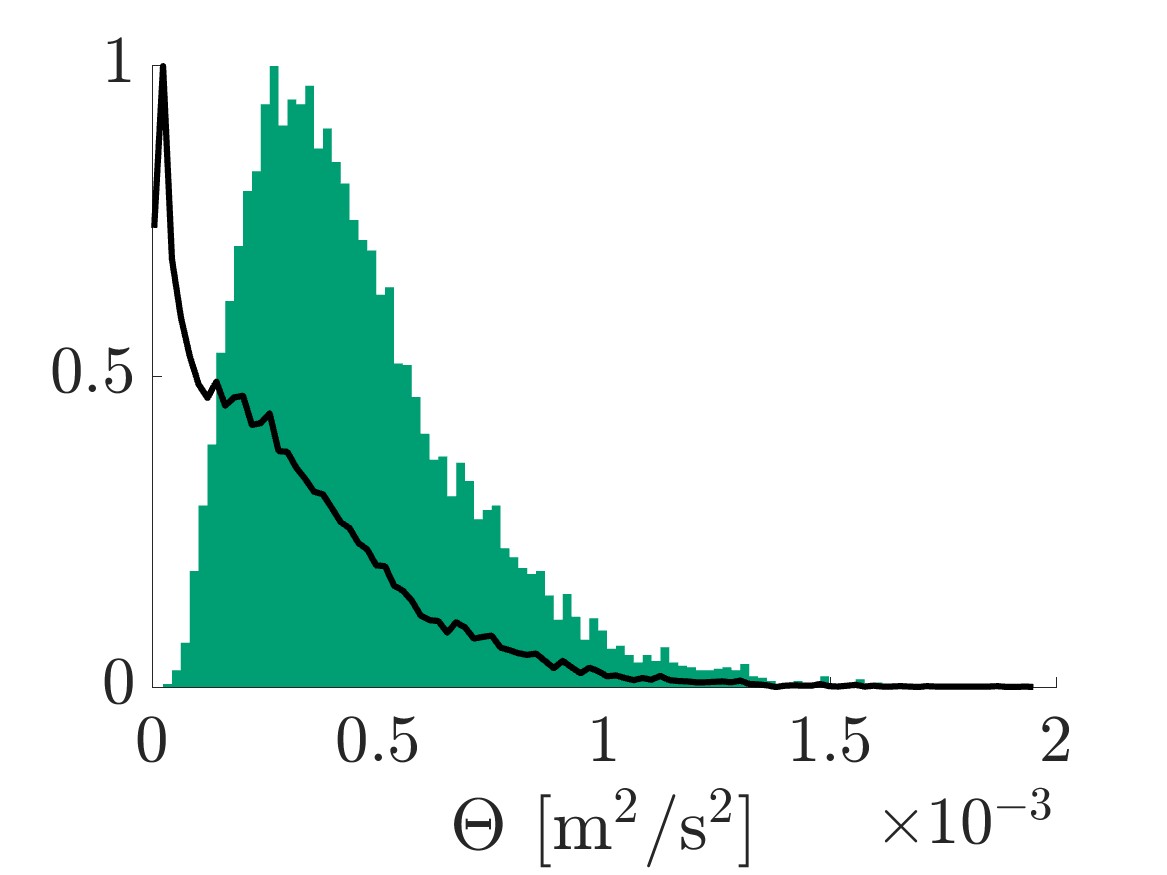} & 
\includegraphics[width=0.22\textwidth]{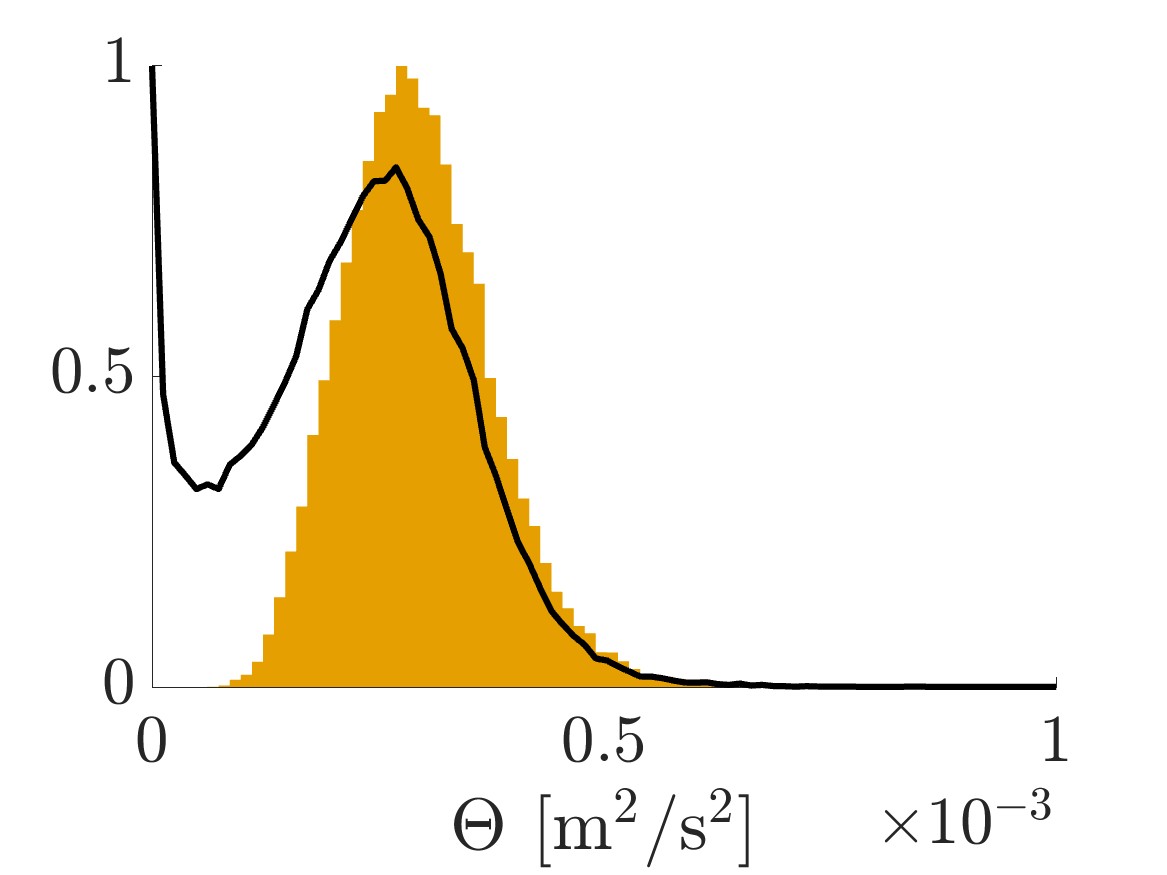} & 
\includegraphics[width=0.22\textwidth] {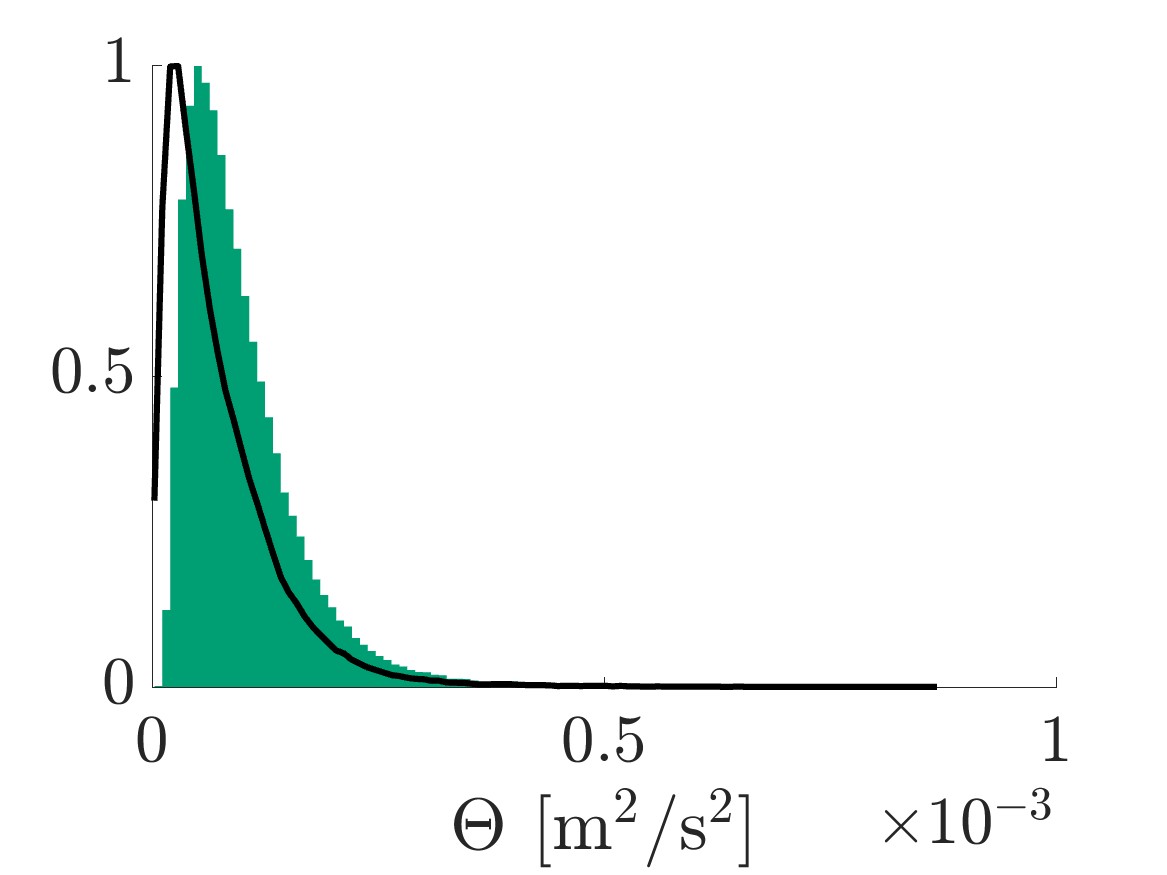}&
\includegraphics[width=0.22\textwidth]{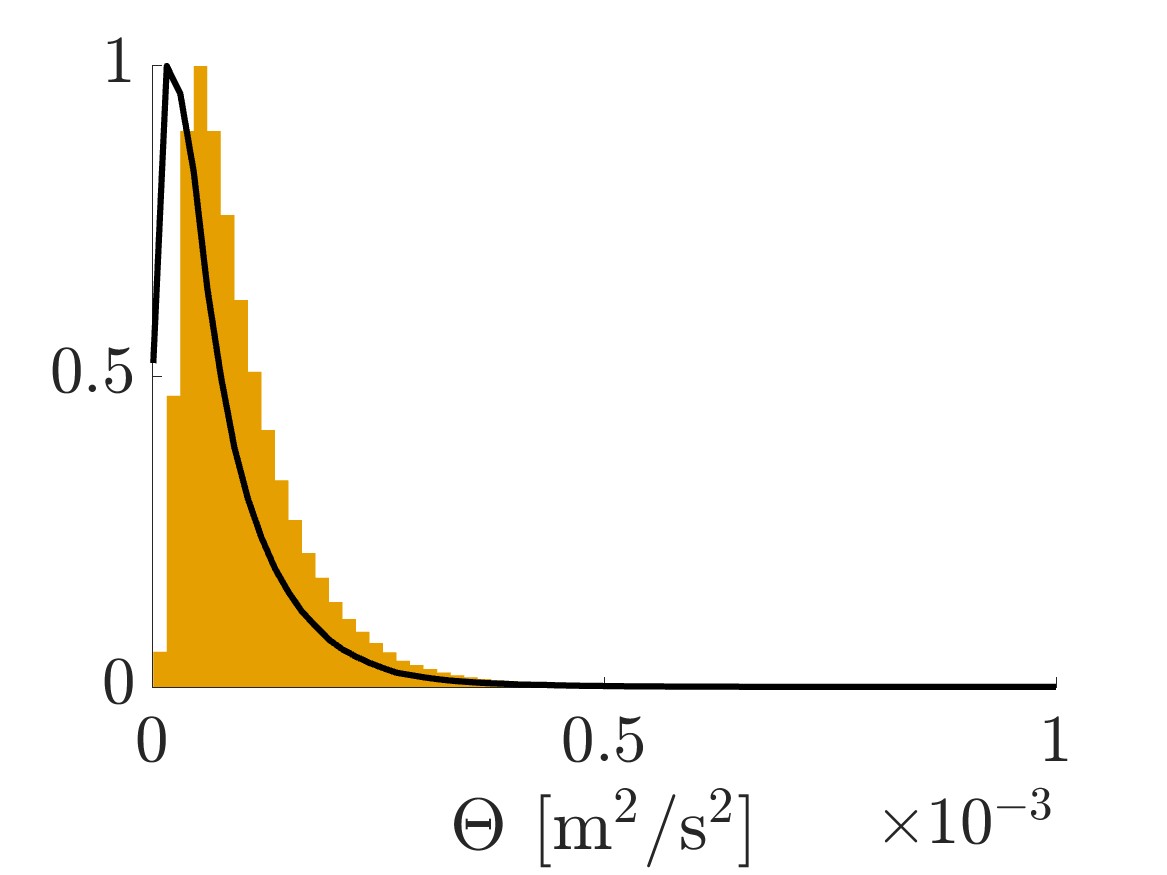}\\[2ex]
\rotatebox{90}{\hspace{1em} Region B} &\includegraphics[width=0.22\textwidth]{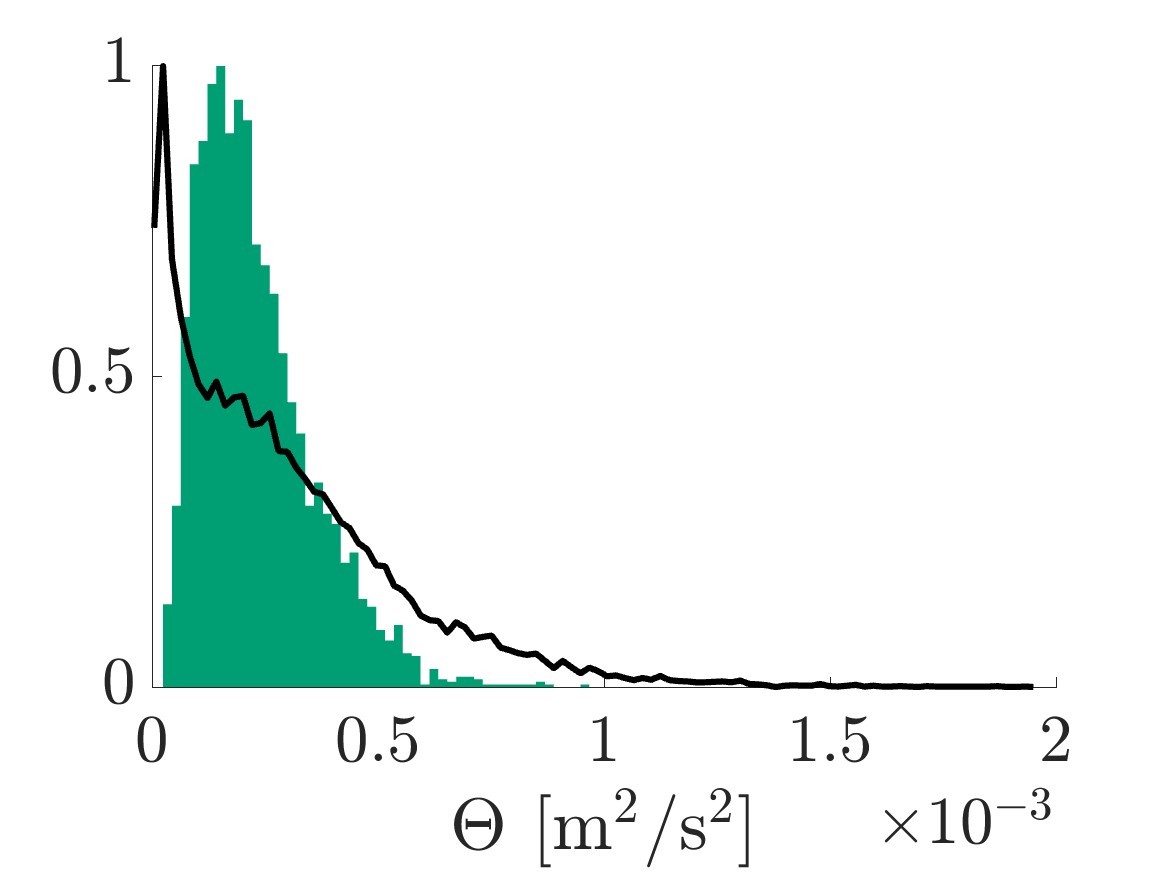} & 
\includegraphics[width=0.22\textwidth]{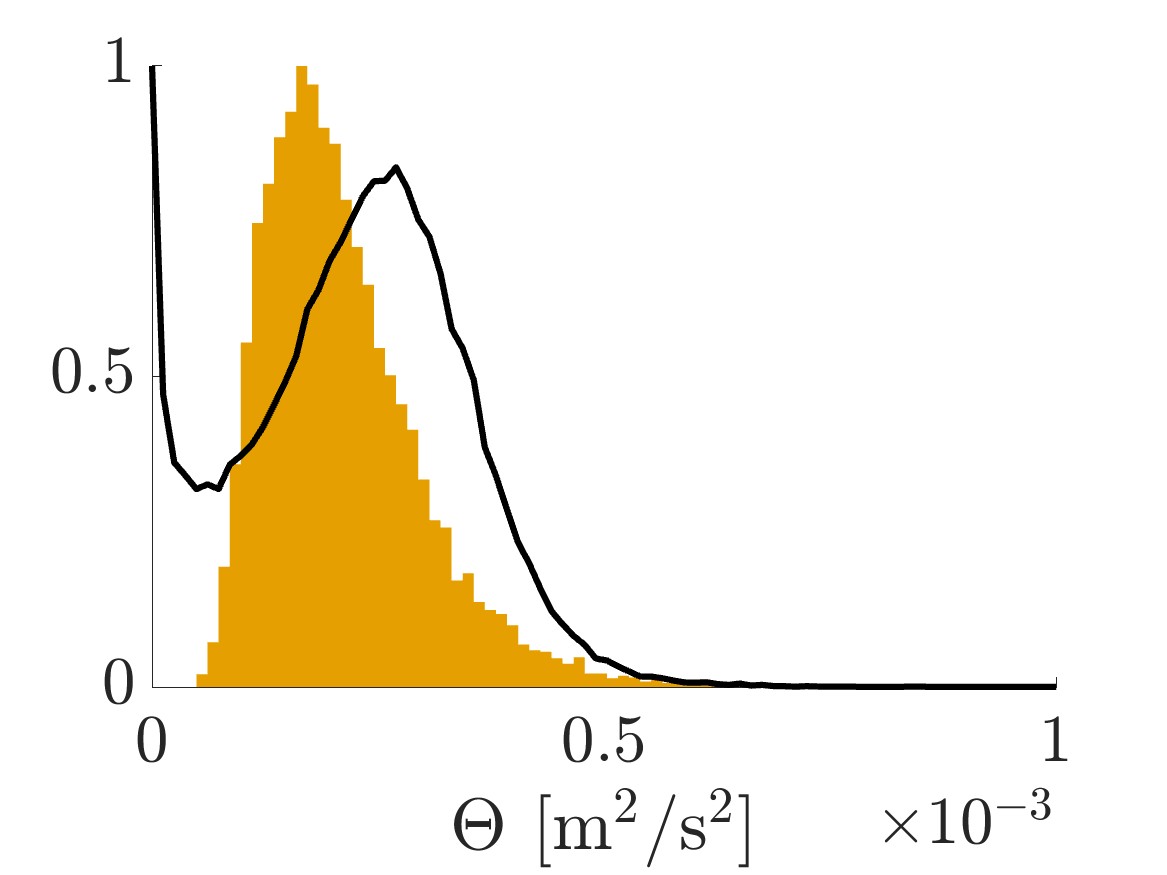} & 
\includegraphics[width=0.22\textwidth] {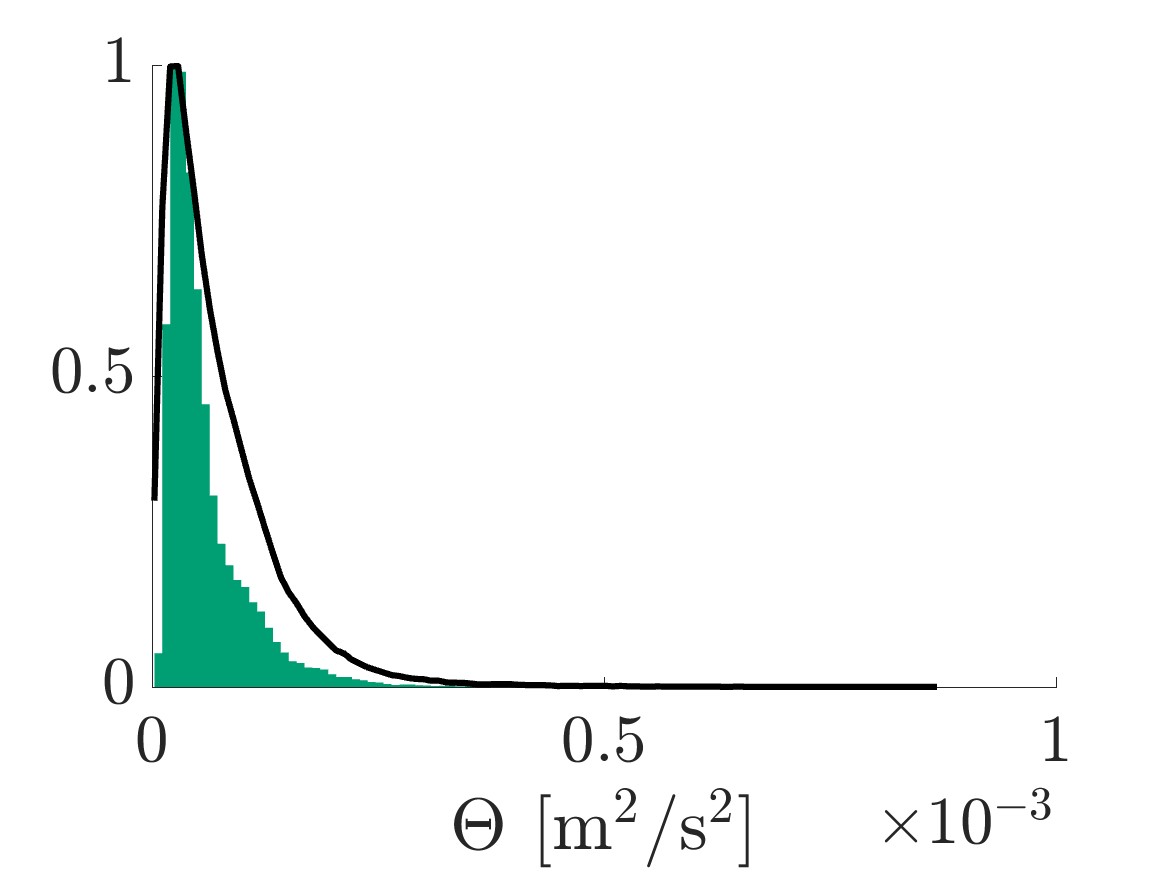}&
\includegraphics[width=0.22\textwidth]{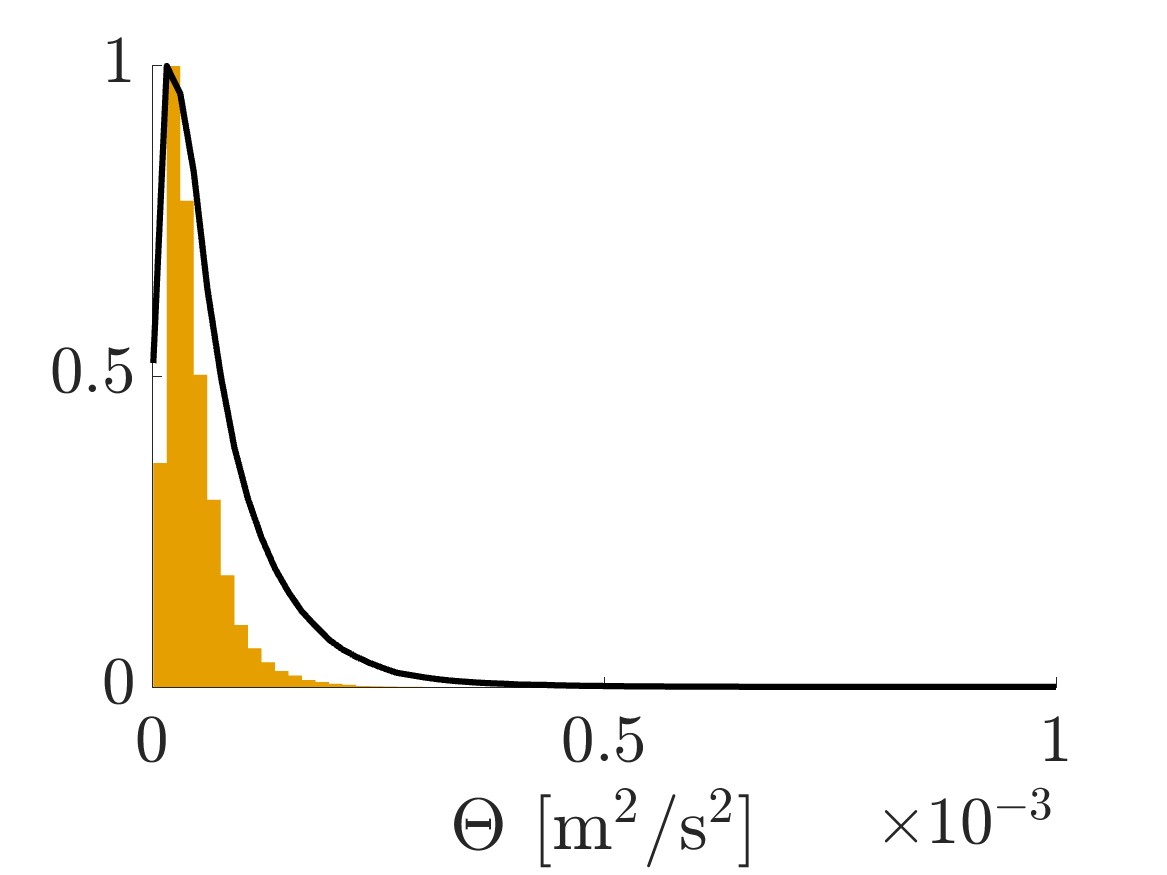}\\ [2ex]
\rotatebox{90}{\hspace{1em} Region C}& \includegraphics[width=0.22\textwidth]{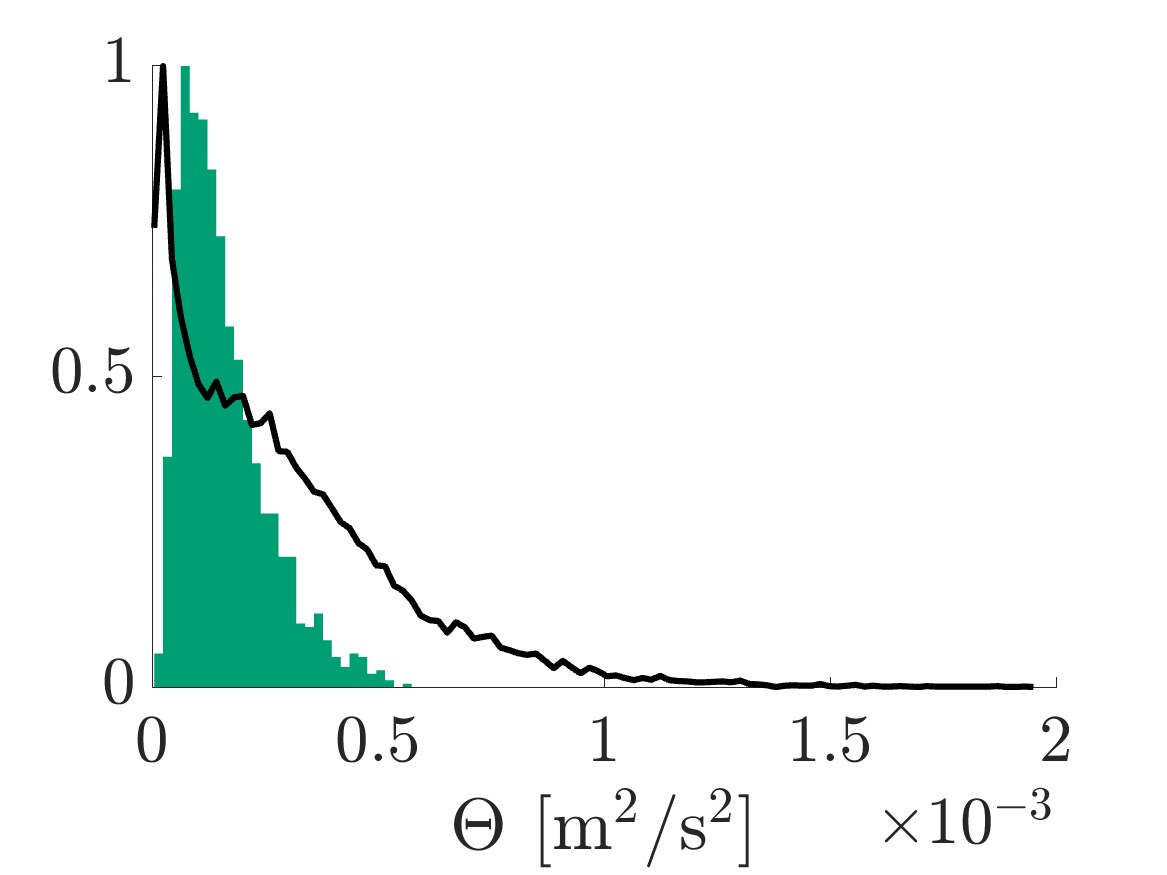} & 
\includegraphics[width=0.22\textwidth]{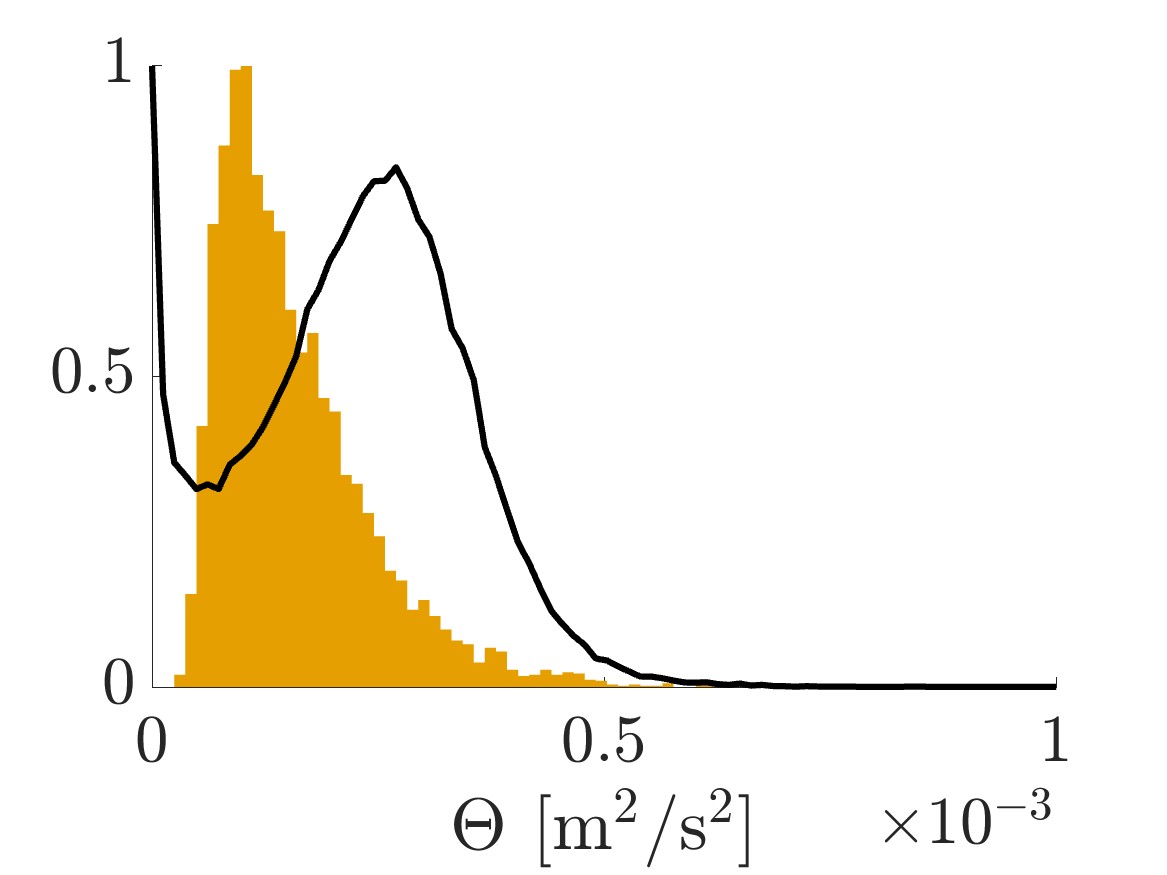} & 
\includegraphics[width=0.22\textwidth] {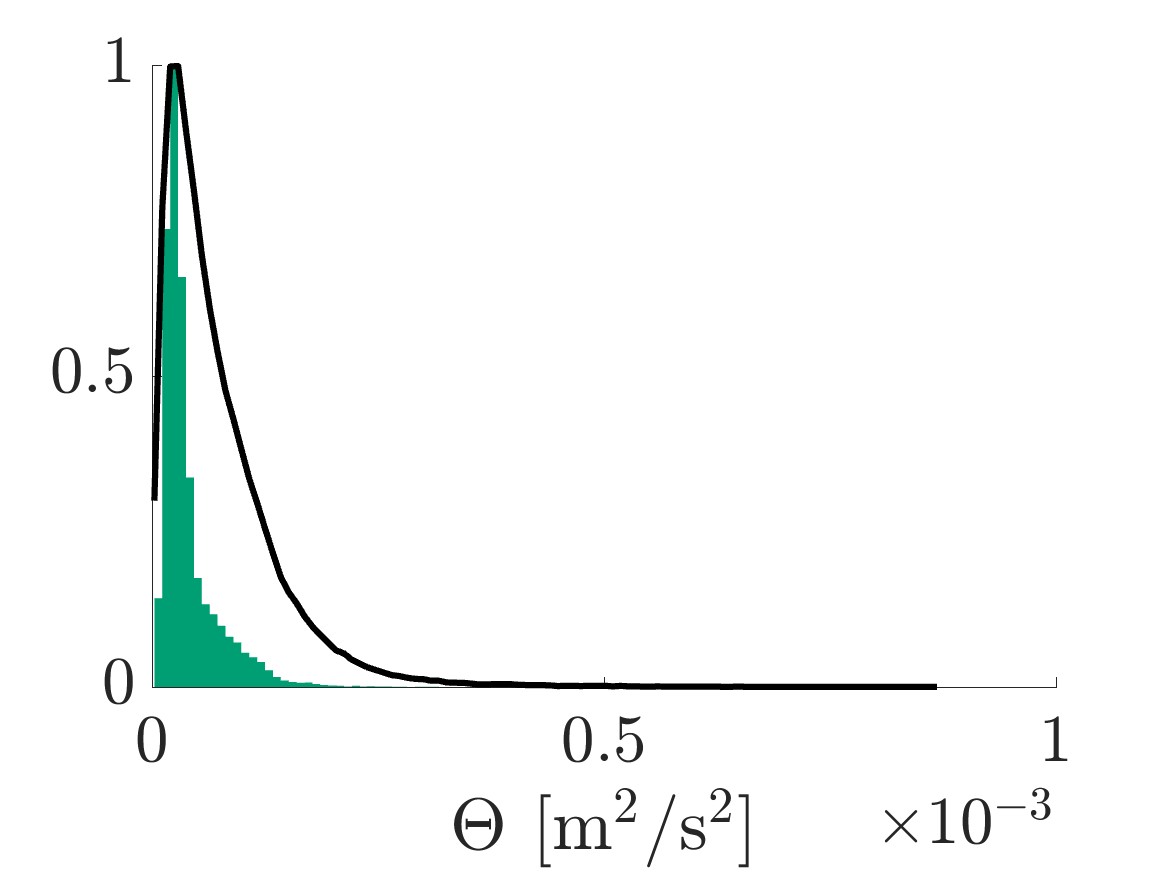}&
\includegraphics[width=0.22\textwidth]{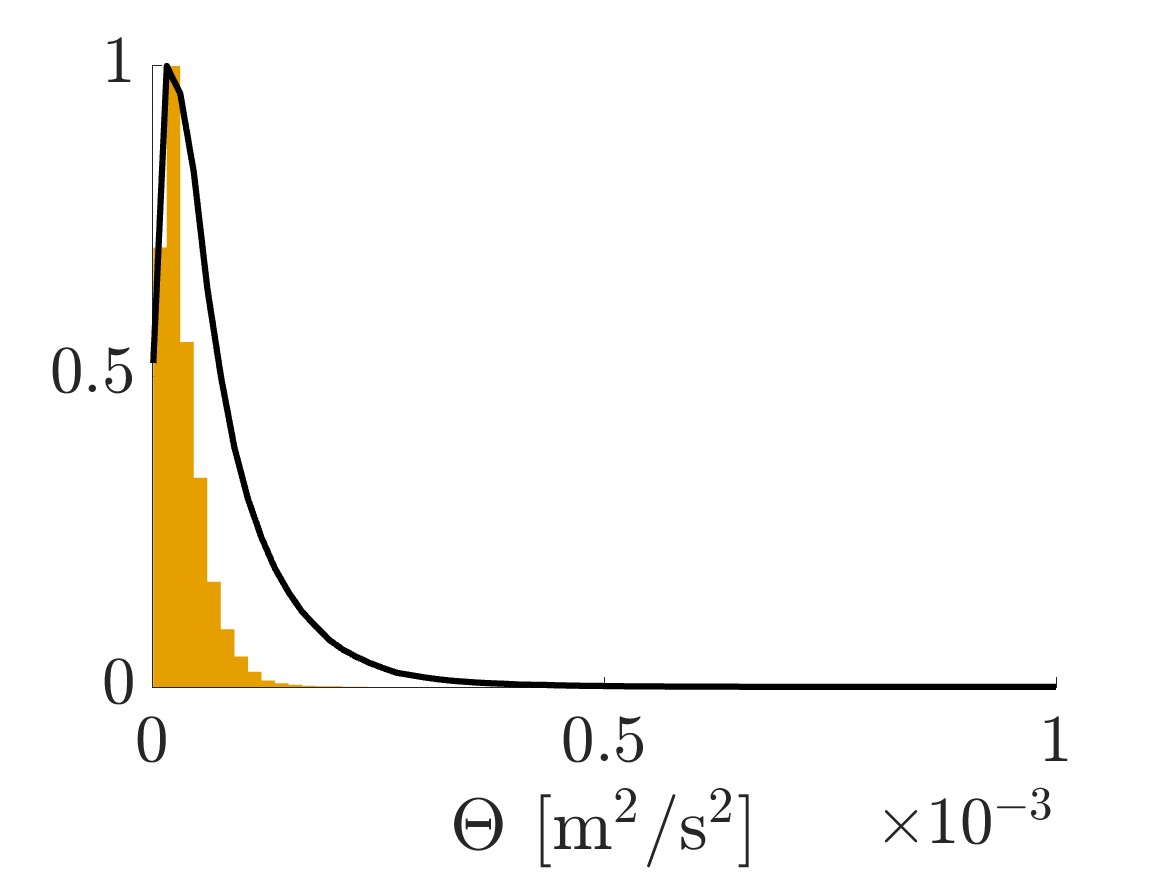}\\[2ex]
\rotatebox{90}{\hspace{1em} Region D}&\includegraphics[width=0.22\textwidth]{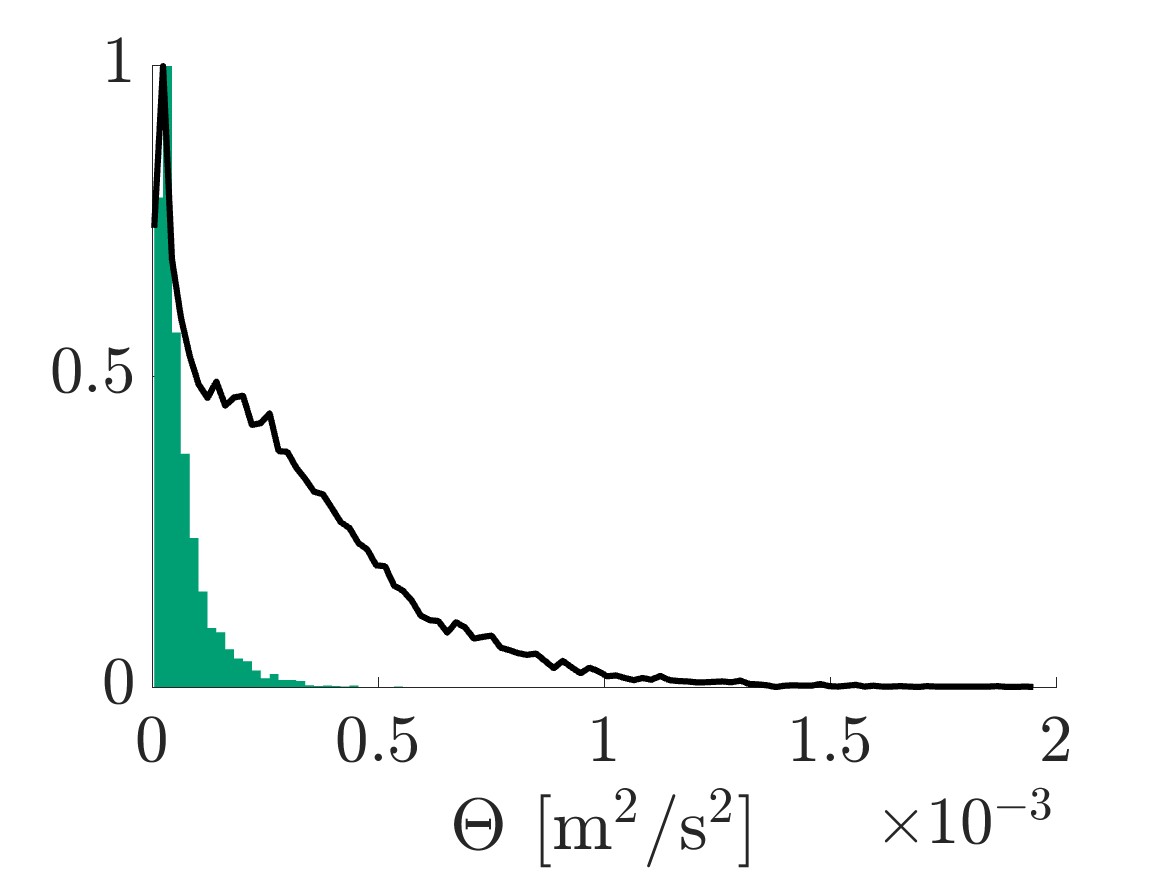} & 
\includegraphics[width=0.22\textwidth]{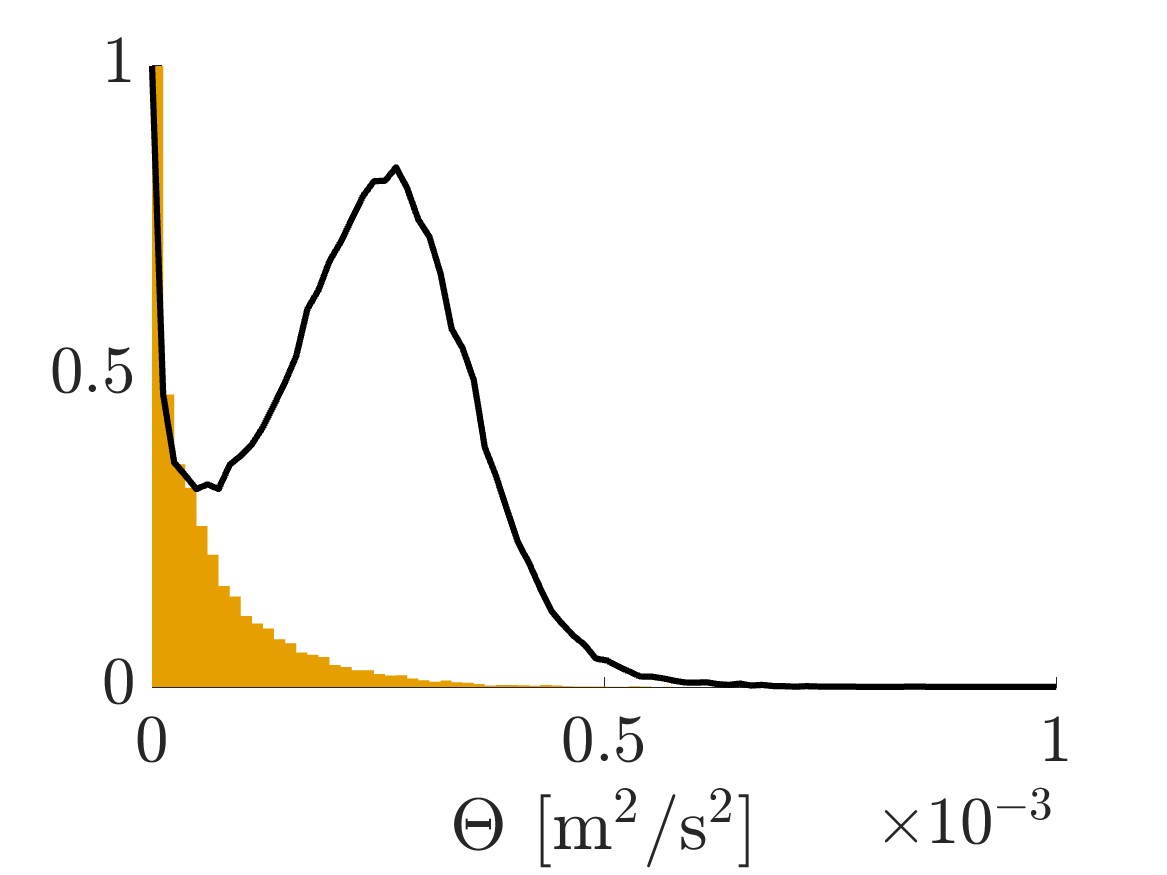} & 
\includegraphics[width=0.22\textwidth] {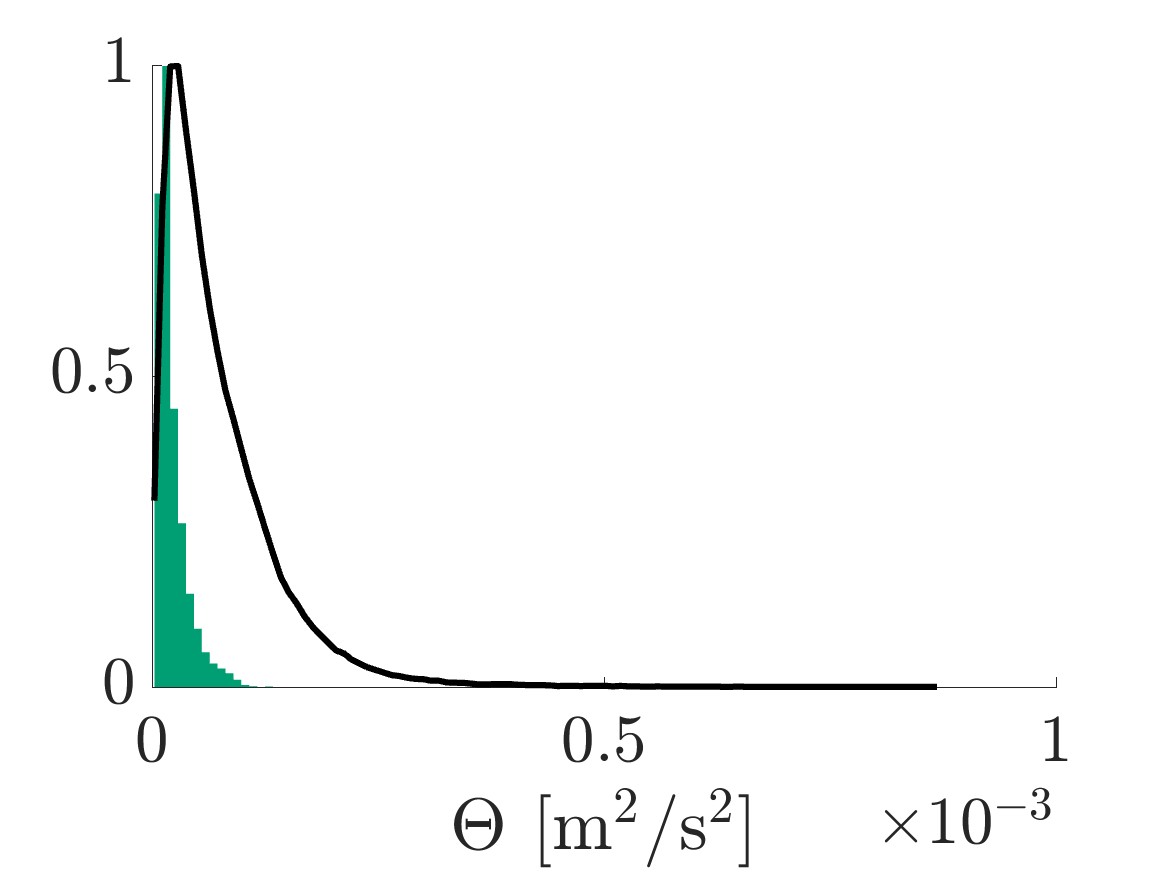}&
\includegraphics[width=0.22\textwidth]{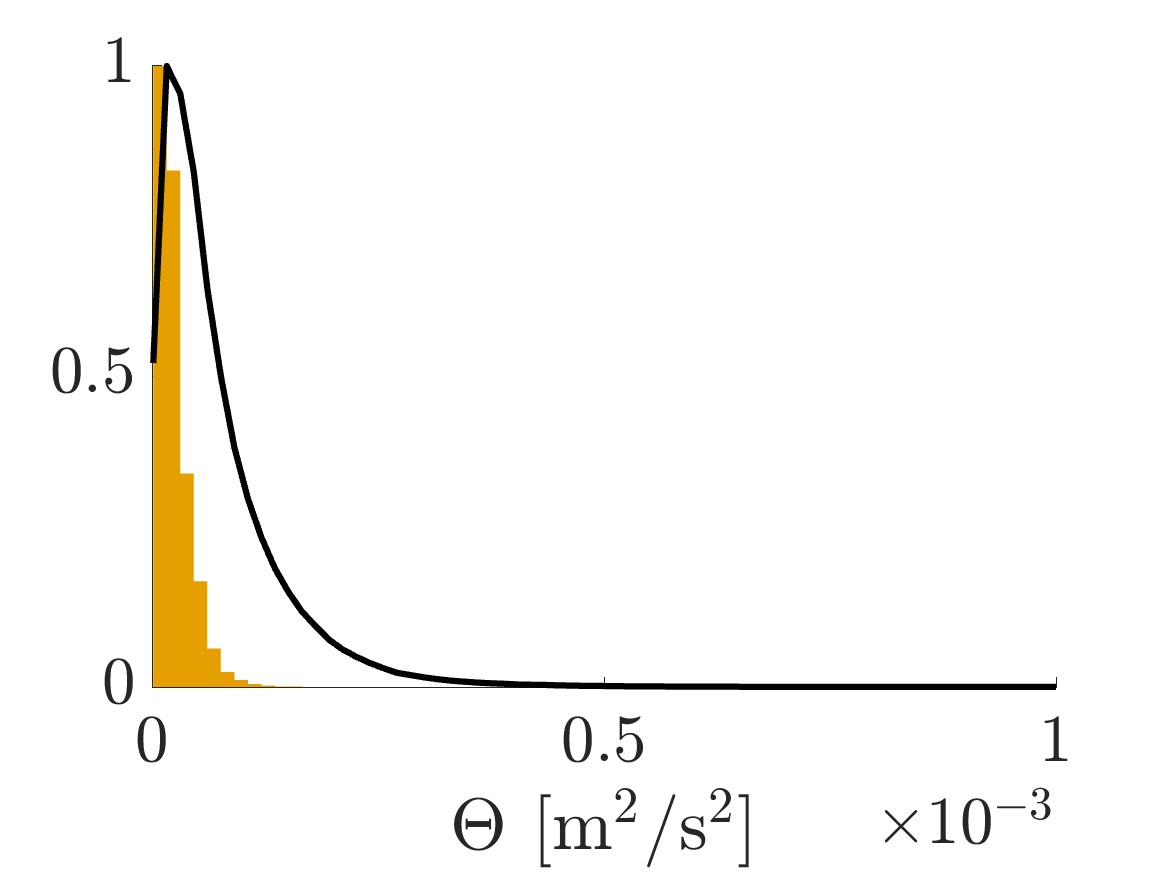}\\
\end{tabular} 
\caption{Summary of the granular temperature based on local volume fraction for the polydisperse distributions $A$ (teal) and $B$ (mustard) at  $\langle \alpha_p \rangle = 0.01$ (left two columns) and $\langle \alpha_p \rangle = 0.10$ (right two columns). Distributions are computed based on local volume fraction cutoffs corresponding to Regions A, B, C and D, as noted. All plots show the normalized distribution (shaded bars) of the particle diameters in each region of the flow, with the normalized distribution of all the particles shown as a solid black line.}
\label{fig:pdfs_theta_clustering}
\end{figure}

Finally, we consider the distribution of particle sizes (Fig.~\ref{fig:pdfs_diameter_clustering}) and granular temperature (Fig.~\ref{fig:pdfs_theta_clustering}) within each of the Regions A-D in the flow. Here, each particle is binned into regions according to the local Eulerian particle volume fraction interpolated to its center. Then, particles belonging to each of the four regions are binned by diameter resulting in a distribution for each group. In each of the subfigures of Fig.~\ref{fig:pdfs_diameter_clustering} and \ref{fig:pdfs_theta_clustering}, the distribution of all the particles in the domain are represented with a solid black line, which is consistent across the plots for each of Regions A-D. The normalized distribution of the particles belonging to each of the four regions is denoted with shaded bars. 

\begin{table}
\centering
    \begin{tabular}{ c c | c c c} 
  & &\multicolumn{3}{c}{$\alpha_p^{\text{thresh}}$}\\ [1ex]
  $\alpha_p$ & Distribution & $1.5\langle \alpha_p \rangle$ & $2.25 \langle \alpha_p \rangle$ & $3.0 \langle \alpha_p \rangle$ \\
   \hline
 \multirow{4}{*}{0.01} & $A_0$ & 249 & 79 & 6\\
 & $A_{\;}$ & 1,751 & 2,703 & 3,031 \\ 
 & $B_0$ & 18,287 & 6,545 & 1,066\\
 & $B_{\;}$ & 6,643 & 8,066 & 7,898 \\ [1ex]
  \multirow{4}{*}{0.10} & $A_0$ & 7 & 4 & 6\\
 & $A_{\;}$ & 3,066 & 1,855 & 517 \\ 
 & $B_0$ & 69 &91 & 72 \\
 & $B_{\;}$ & 7,663 & 9,214 & 8,334\\
    \end{tabular} 
    \caption{Summary of the number of connected regions containing volume fractions at thresholds of $(1.5, 2.25, 3.0)\langle \alpha_p \rangle$. Eulerian cells containing volume fractions above the prescribed threshold with two or connected cells are considered a `connected region.'}
    \label{tab:Connectivity}
\end{table}

Considering first the distribution of particle diameters by region, we make several observations. First, smaller particles are preferentially found in the most dilute regime (i.e. Region A) and this effect is enhanced for $\langle \alpha_p \rangle = 0.01$. Interestingly, for the dilute cases, all of the particles found in Region A are less than $D_{30}$, though this does not extend to the higher volume fraction cases. In Regions B and C, we note similar behavior for both Dist. $A$ and $B$ at $\langle \alpha_p \rangle = 0.01$. In these regions, there is a dramatic increase in the probability of mid-sized particles. As the density of the cluster increases, there is a much higher likelihood of encountering increasingly large particles and less likelihood of encountering very small particles. This is particularly true for Dist. $A$, where in Region D, almost all the particles are moderately large to very large. In Dist. $B$ we observe that in this region, there is a slight increase in the number of finer particles. This is likely attributable to the fact that Dist. $B$ has a great number of fine particles, and therefore more particles overall. These small particles can be more efficiently packed into a cluster and are also more likely to be entrained in clusters due to the greater number of small particles in the overall distribution.  

At higher volume fraction, Dist. $A$ exhibits a smoother increase in preference toward moderate to large-sized particles, particularly in Regions B and C. In fact, at high volume fraction, Region C is the most likely region to encounter very large particles. In Region D, there is a higher likelihood of finding very large particles, however, most notable is the reduction of mid-sized particles. At $\langle \alpha_p \rangle = 0.10$, Dist. $B$ also smoothly increases its preference toward larger particles as the local volume fraction increases, however, this increase persists through Region D.  

Importantly, we believe this preference for larger particles in Region D indicates that for polydisperse configurations of particles, it is the larger particles in the domain that give rise to cluster formation, very likely due to the relatively large regions of reduced drag that these particles provide for smaller particles. As previously mentioned, this strong preference for large particles at the center of clusters with smaller particles haloed around them which is more pronounced for lower volume fraction than at higher volume fraction, is likely observed because, at higher volume fraction, the regions of reduced drag stemming from large particles is obscured by the regions of reduced drag from the greater number of other nearby particles.   

As discussed in Sec.~\ref{sec:PhaseAveragedEqs}, granular temperature is an important measure of the evolution and consistency of clusters. For this reason, we also present the granular temperature distributions for each of Regions A-D. These distributions are shown in Fig.~\ref{fig:pdfs_theta_clustering}. Again, the solid lines represent the normalized domain-wide distribution and the shaded bars denote the normalized distributions for each of the regions. Here, we note that all four polydispersed configurations demonstrate similar general behavior in the sense that the granular temperature is preferentially high in the most dilute regions of the flow, and becomes increasingly preferentially small as the local volume fraction increases. A perhaps intuitive result, this indicates that particles in dilute regions of the flow attain a higher amount of uncorrelated energy compared with their counterparts in the cores of clusters. 

\subsection{A statistical description of particle settling}\label{sec:polySettling}
\begin{figure}
  \includegraphics[height = 0.8\textwidth]{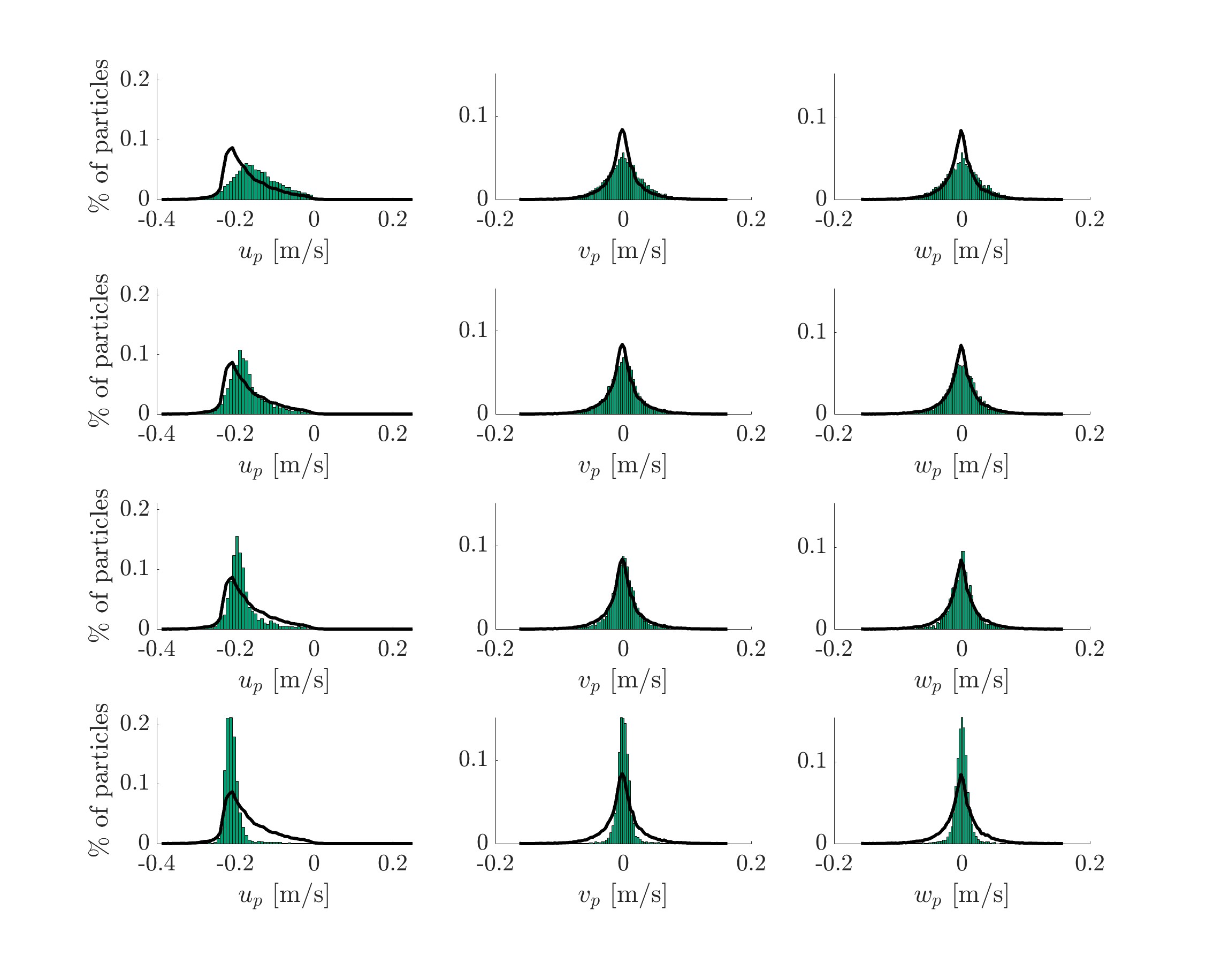}
    \caption{Distributions of particle velocities ($u_p, v_p, w_p$, from left to right) for Dist. A, $\langle \alpha_p \rangle = 0.01$. Distributions are shown for dilute to dense regions of the flow (top to bottom). The dark lines represent the full domain distribution of velocities.}
    \label{fig:DistA01_velocities}
\end{figure}

\begin{figure}
    \centering 
\includegraphics[width = 0.8\textwidth]{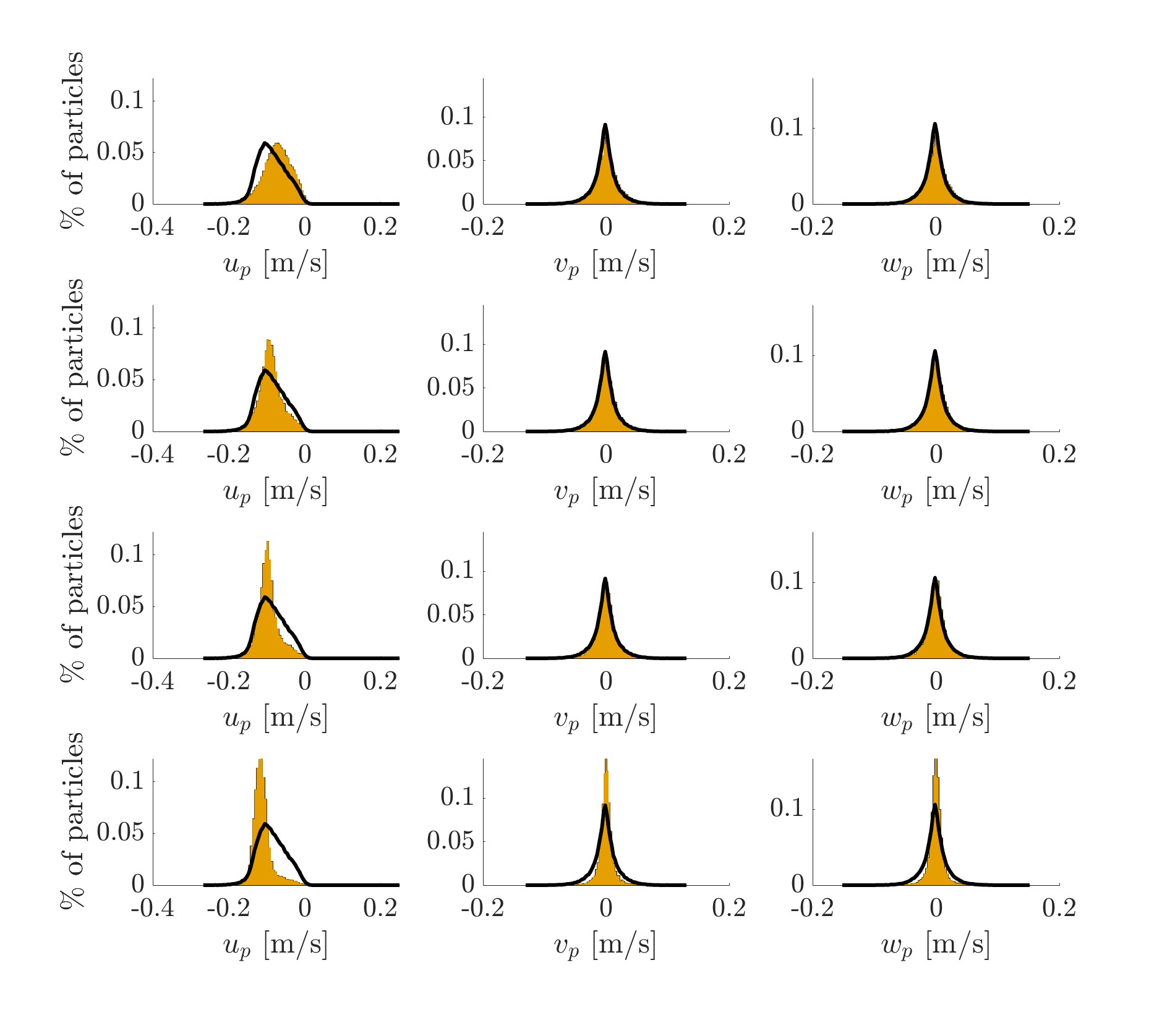}
    \caption{Distributions of particle velocities ($u_p, v_p, w_p$, from left to right) for Dist. B, $\langle \alpha_p \rangle = 0.01$. Distributions are shown for dilute to dense regions of the flow (top to bottom). The dark lines represent the full domain distribution of velocities.}
    \label{fig:DistB01_velocities}
\end{figure}

\begin{figure}
    \centering 
\includegraphics[width = 0.8\textwidth]{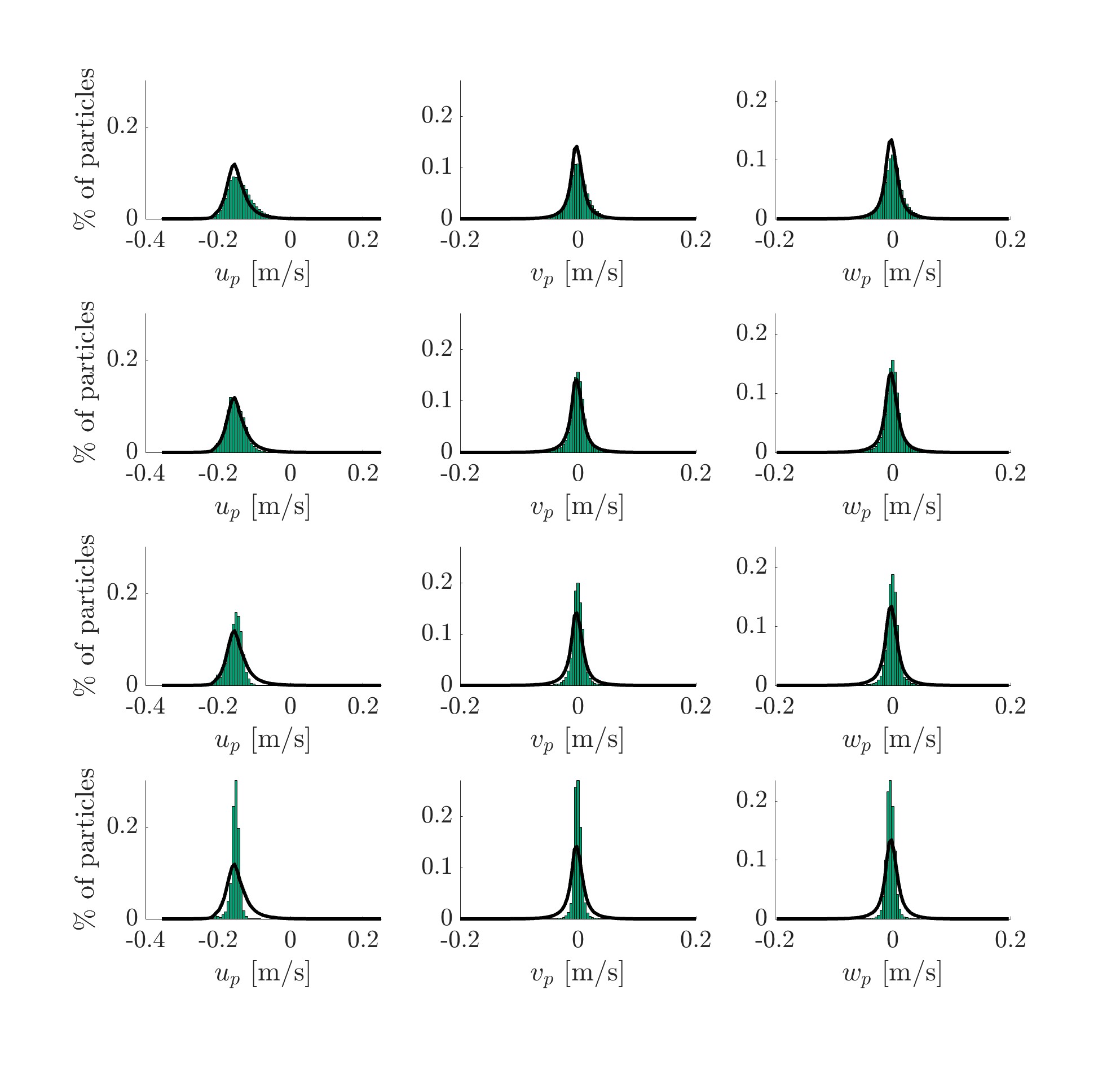}
    \caption{Distributions of particle velocities ($u_p, v_p, w_p$, from left to right) for Dist. A, $\langle \alpha_p \rangle = 0.10$. Distributions are shown for dilute to dense regions of the flow (top to bottom). The dark lines represent the full domain distribution of velocities.}
    \label{fig:DistA1_velocities}
\end{figure}

\begin{figure}
    \centering 
\includegraphics[width = 0.8\textwidth]{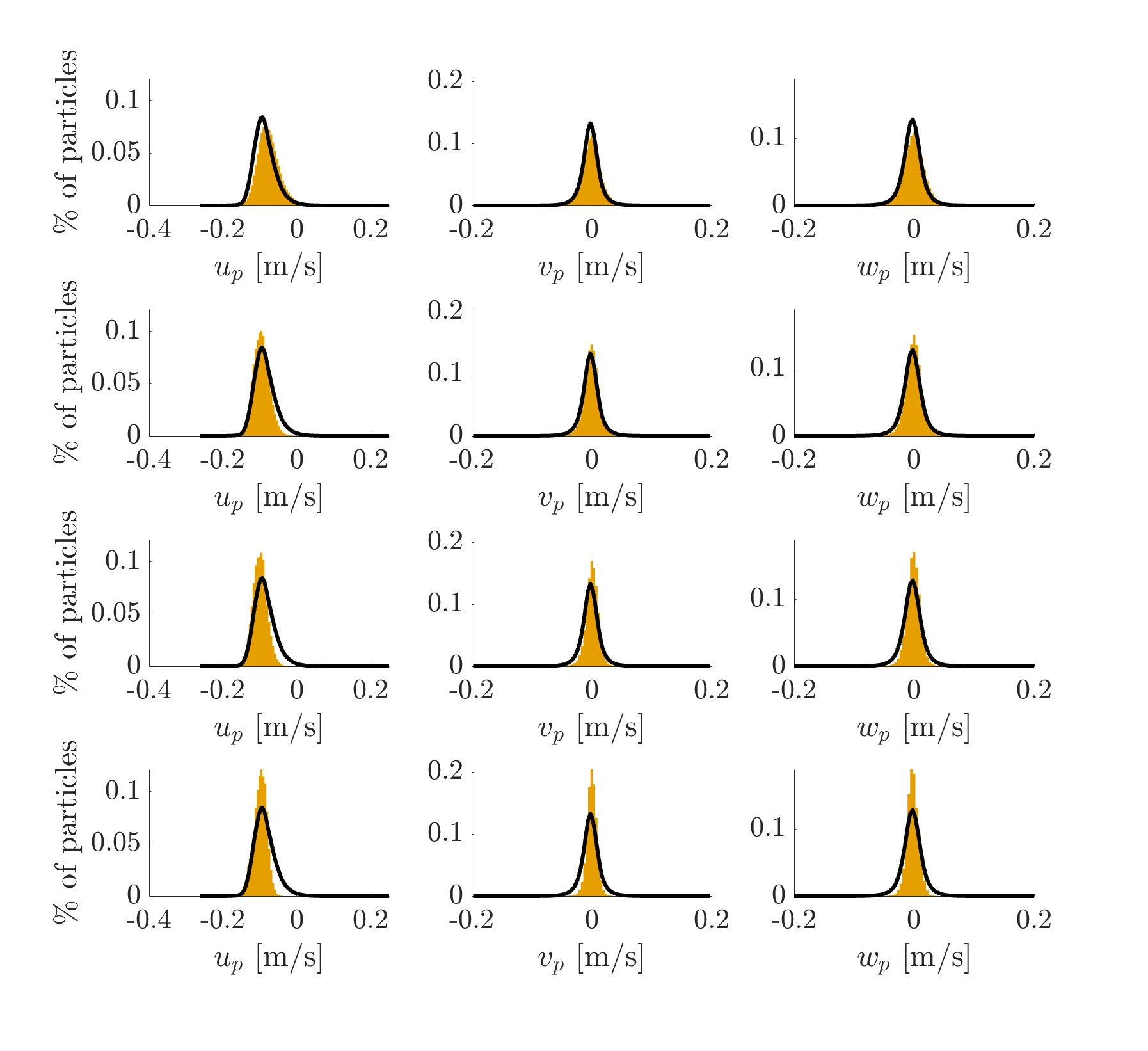}
    \caption{Distributions of particle velocities ($u_p, v_p, w_p$, from left to right) for Dist. B, $\langle \alpha_p \rangle = 0.10$. Distributions are shown for dilute to dense regions of the flow (top to bottom). The dark lines represent the full domain distribution of velocities.}
    \label{fig:DistB1_velocities}
\end{figure}

In this section, we now examine the settling behavior of each of the configurations under study, with a particular focus on how clustering behavior implicates settling. To begin this analysis, we consider the distributions of particle velocity in Regions A--D, in direct analogy with the discussion in the previous section. Figs.~\ref{fig:DistA01_velocities}-\ref{fig:DistB1_velocities} show these distributions, where once again, the solid lines denote the distribution for each component of velocity for the full domain and the shaded distributions denote the distribution for the given region of the flow. 

Beginning with a comparison of the stream-wise velocity, $u_p$, for the dilute cases (Figs.~\ref{fig:DistA01_velocities} and \ref{fig:DistB01_velocities}), we find that particles in the most dilute region of the flow (Region A) tend to have statistically smaller settling velocities. This is due to two primary factors: (1) the particles are generally lone and would be expected to have smaller velocities for this reason and (2) particles may be swept up in jet-bypassing, which causes much smaller velocities, and occasionally velocities that oppose gravity. As cluster density increases, particles in these regions are increasingly more likely to have larger velocities, due to their correlation with other particles and entrapment in coherent structures. Naturally, the particles that are found in the densest region (Region D), attain the largest settling velocities. 

In contrast with the dilute cases, the denser cases (Figs.~\ref{fig:DistA1_velocities} and \ref{fig:DistB1_velocities}) exhibit distributions similar to the global distribution for Regions A and B, and only moderately depart from this trend for Regions C and D. Again, in these denser regions, particles are statistically more likely to attain greater settling velocities, but the extent for this preference toward larger settling velocities is less prominent than for the dilute cases. 

In the cross-stream directions, particles in Region A are more likely to attain cross-stream velocities ($v_p$ and $w_p$) further from null, indicating that they are more susceptible to being entrained in the underlying turbulence of the flow. This observation is consistent across all four polydispersed configurations. As the particle density increases, particles increasingly have velocities very near zero. This is indicative of granular temperature that is higher on the outer regions of clusters, as compared with the interior, consistent with what was reported in the previous section. This overall behavior is mirrored across distributions $A$ and $B$ and at both mean volume fractions, however Dist. $A$ has a narrower band of cross-stream velocities in Region D compared with Dist. $B$, indicating that Dist. $A$ clusters are more `rigidly' packed and particles are less able to move translationally. 

While this discussion illuminates the behavior of particles in each of the regions of the flow based on local volume fraction, it is also instructive to connect settling behavior with particle size. In Sec.~\ref{sec:Clustering}, we have already demonstrated that particle size and clustering potential are correlated, thus connections between particle size and settling velocity are related through a particle's likelihood to be involved in clustering. 

When carrying out this analysis, we leverage the fact that at steady state, all the forces acting on the particles are in equilibrium. In particular, we draw attention to the drag force felt by the particles. As described in Sec.~\ref{sec:ParticleEqs}, the drag force on a particle is given as 
\begin{equation}
    \bm{F}^{(i)}_{\text{drag}} = \frac{m_p^{(i)} \alpha_f\lbrack \bm{x}_p^{(i)}\rbrack F_D}{\tau_p} \left(u_f[x_p^{(i)}] - u_p^{(i)}\right)
\end{equation}
where $F_D$ is the correction based on particle Reynolds number and local volume fraction~\citep{Tenneti2011}. Since settling velocity is related to the balance between drag (which opposes settling) and mass (see Eq.~\ref{eq:InterExch2}), when the drag force increases, settling velocity is hindered. Conversely, when drag is decreased, settling is enhanced. Since drag depends upon not only the particle size, but also the slip velocity and local volume fraction, we note that complex behavior occurs as particles become correlated. 

\begin{figure}
    \begin{tabular}{c c}
     (a) Dist. A, $\langle \alpha_p \rangle = 0.01$ & (b) Dist. A, $\langle \alpha_p \rangle = 0.10$ \\
    \includegraphics[width = 0.45\textwidth]{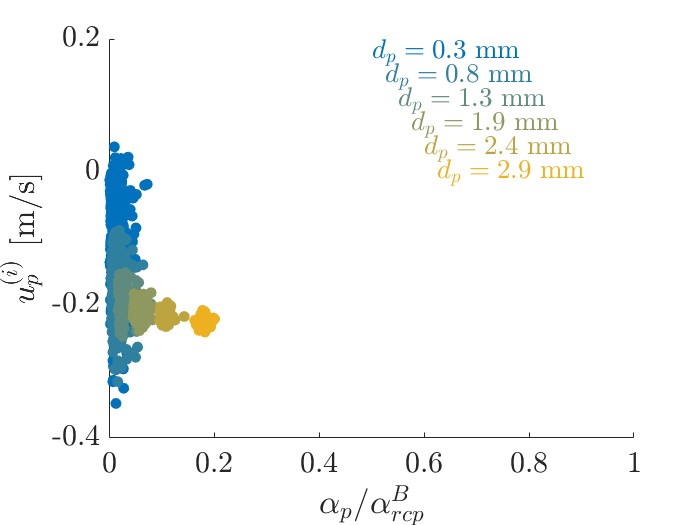} &\includegraphics[width = 0.45\textwidth]{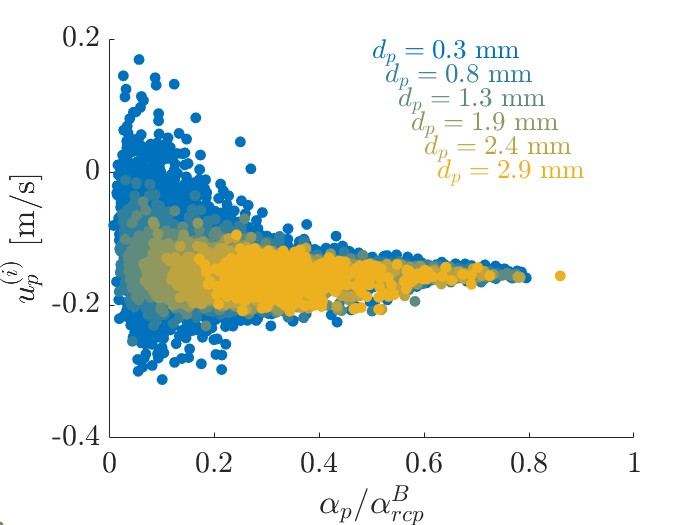} \\
    (c) Dist. B, $\langle \alpha_p \rangle = 0.01$ & (d) Dist. B, $\langle \alpha_p \rangle = 0.10$ \\
    \includegraphics[width = 0.45\textwidth]{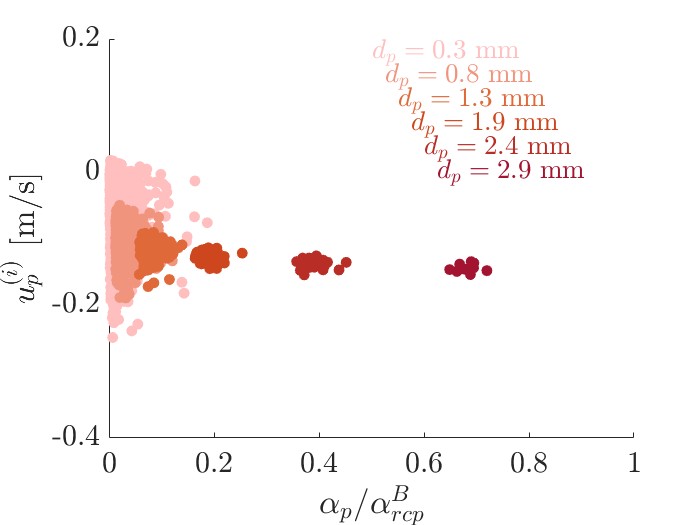} &\includegraphics[width = 0.45\textwidth]{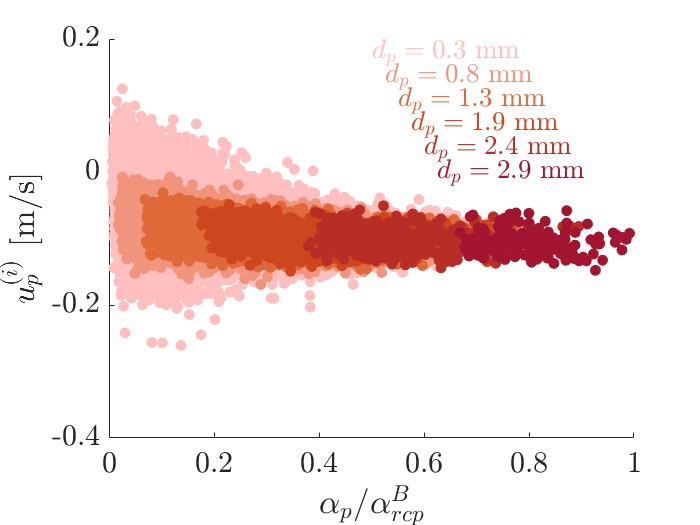} 
    \end{tabular} 
    \caption{Settling velocity of individual particles plotted against normalized local particle volume fraction for all four polydisperse cases under consideration. Colors correspond to six increasing diameters $d_p = (0.3, 0.8, 1.3, 0.9, 2.4, 2.9)$ [mm] and the color maps for distributions $A$ and $B$ used throughout. The data plotted represent the particles within the diameters listed $\pm 0.03$ mm.}
    \label{fig:settlingvsalpha}
\end{figure}

\begin{figure}
    \begin{tabular}{c c}
     (a) Dist. A, $\langle \alpha_p \rangle = 0.01$ & (b) Dist. A, $\langle \alpha_p \rangle = 0.10$ \\
    \includegraphics[width = 0.45\textwidth]{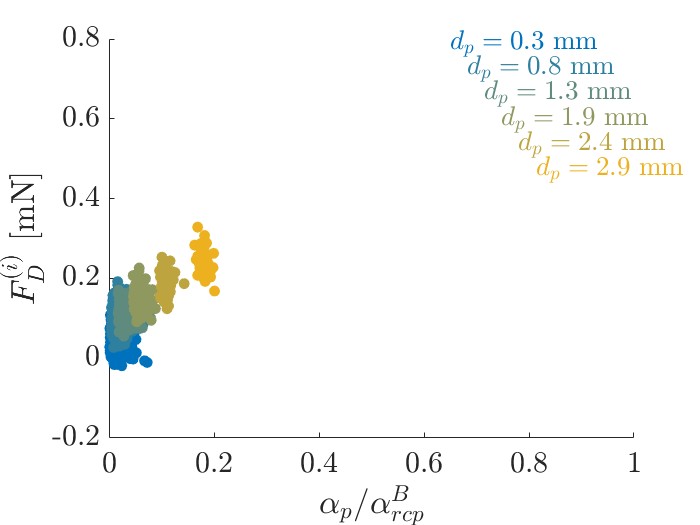} &\includegraphics[width = 0.45\textwidth]{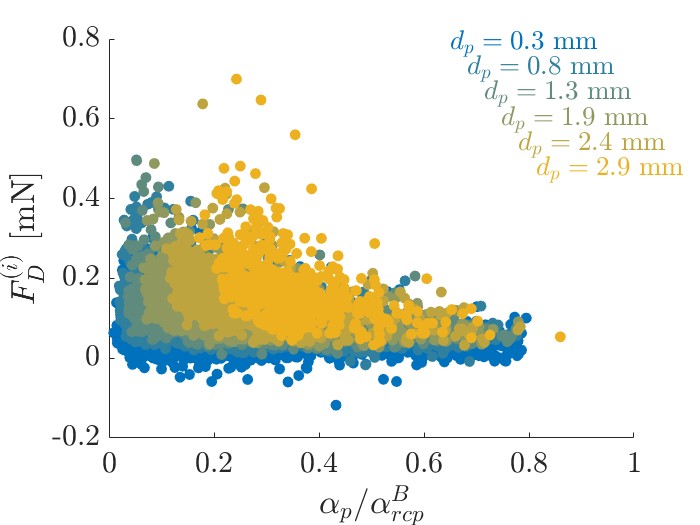} \\
    (c) Dist. B, $\langle \alpha_p \rangle = 0.01$ & (d) Dist. B, $\langle \alpha_p \rangle = 0.10$ \\
    \includegraphics[width = 0.45\textwidth]{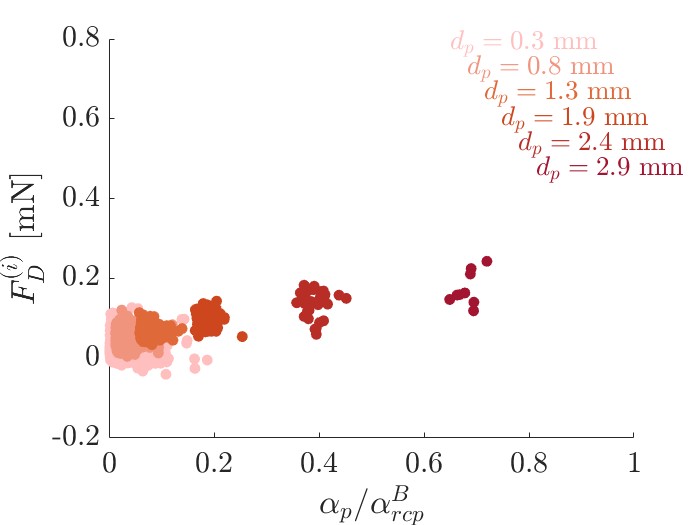} &\includegraphics[width = 0.45\textwidth]{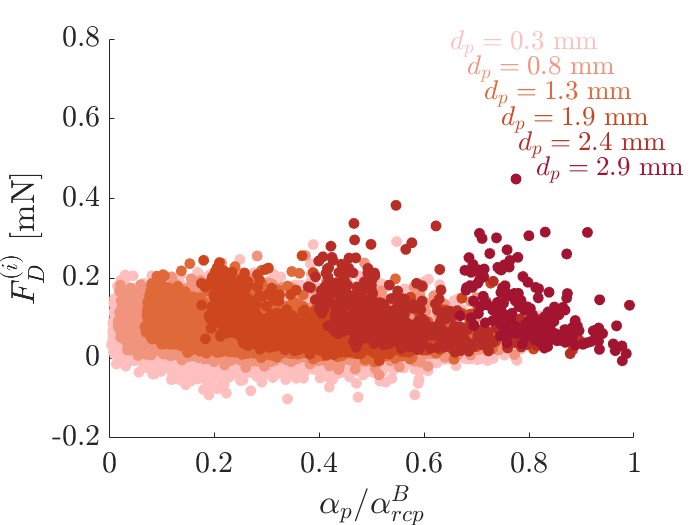} 
    \end{tabular} 
      \caption{Drag force on individual particles plotted against normalized local particle volume fraction for all four polydisperse cases under consideration. Colors correspond to six increasing diameters $d_p = (0.3, 0.8, 1.3, 0.9, 2.4, 2.9)$ and the color maps for distributions $A$ and $B$ are used throughout. The data plotted represent the particles within the diameters listed $\pm 0.03$ mm.}
    \label{fig:dragvsalpha}
\end{figure}

To illustrate this behavior, Figs.~\ref{fig:settlingvsalpha} and \ref{fig:dragvsalpha} show the relationship between individual particle settling velocities and drag forces against local volume fraction for six bins of particle size ranging from very fine to coarse. By examining particle settling velocity in this way, we note that the finest particles attain the widest range of settling velocities for all configurations and that these particles are located primarily in dilute regions of the flow for lower volume fraction configurations, and exist across all regions--from dilute to dense--in the higher volume fraction configurations. For the denser configurations, these small particles are also exclusively involved in entertainment in strong jet by-passing, as they are the only particles that achieve positive velocities. This effect is more pronounced in the higher mean volume fraction cases, due to the fact that the clusters formed are larger, thereby producing larger turbulent wakes. The fine particles that exist at increasingly large volume fraction have an increasingly narrow range of attained velocities, indicating that they have become increasingly correlated with surrounding particles in a cluster. 

At higher volume fraction, the largest particles exist over a substantially broader range of volume fractions in Dist. $A$ as compared with Dist. $B$. This suggests that particles are more `mixed' in Dist. $A$ than $B$, where large particles exist only in higher volume fraction areas. This indicates that the clusters formed in Dist. $B$ tend to contain large particles only in the cluster `cores', whereas in Dist. $A$, large particles are located throughout clusters. For the dilute configurations, we note that there exists a similar, but more pronounced effect of what we observe in their denser counterparts. In particular, particles of increasing size have substantially more overlap as volume fraction increases for distribution $A$, however, the larger particle sizes are extremely stratified in distribution $B$. This is suggestive of the existence of clusters formed by particles of like size in the case of larger particles, and clusters formed by the smaller to moderately sized particles.  

Fig.~\ref{fig:dragvsalpha} shows a similar analysis for the drag felt by each particle. Here, we find that the force of drag increases with both volume fraction and particle diameter for the dilute cases. This is due to the fact that clusters are smaller and composed of fewer particles, making these particles more susceptible to experiencing increased drag due to a higher slip velocity. In contrast, at higher mean volume fraction, when the clusters are larger, only the outer particles in the cluster--those particles that compose the outer `shell' of the cluster--will observe this effect, while the inner core will experience \emph{reduced} drag due to the number of particles surrounding them and the resulting reduced slip velocity.  

\begin{figure}
    \begin{tabular}{c c}
     (a) Dist. A, $\langle \alpha_p \rangle = 0.01, d_p^{\text{crit}} = 1.64$ mm & (b) Dist. A, $\langle \alpha_p \rangle = 0.10, d_p^{\text{crit}} = 1.38$ mm\\
    \includegraphics[width = 0.45\textwidth]{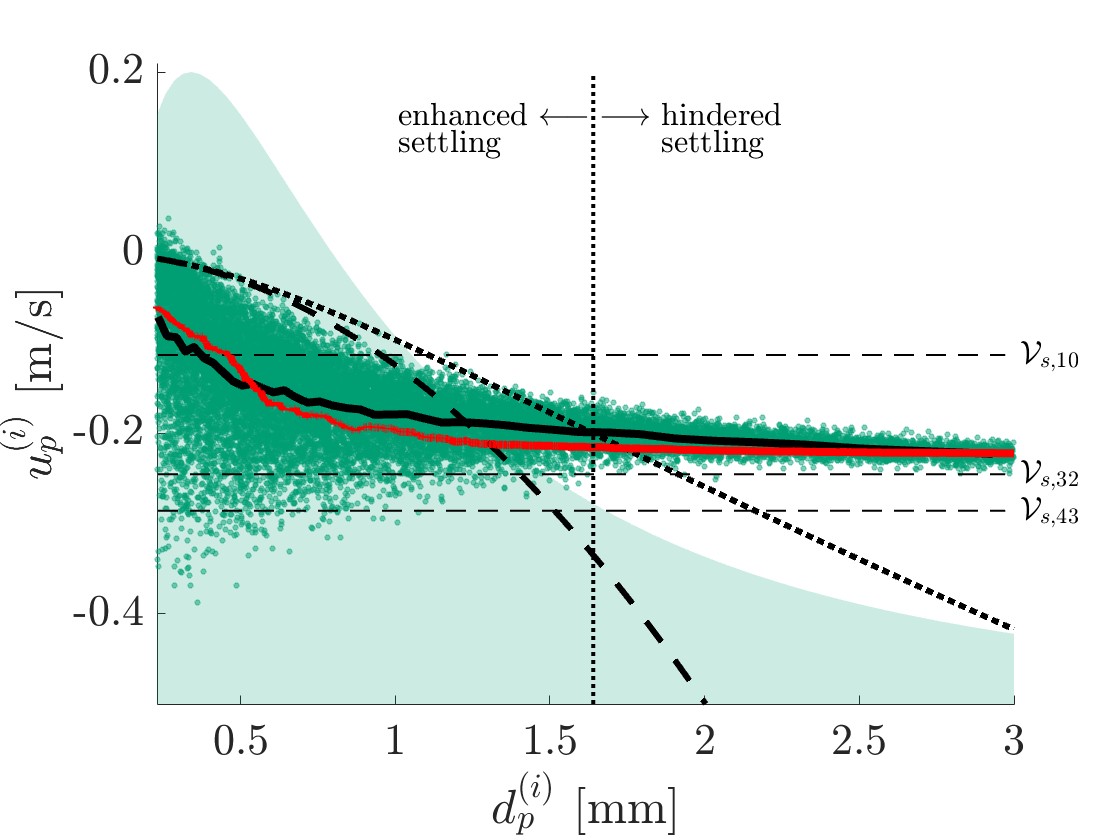} &\includegraphics[width = 0.45\textwidth]{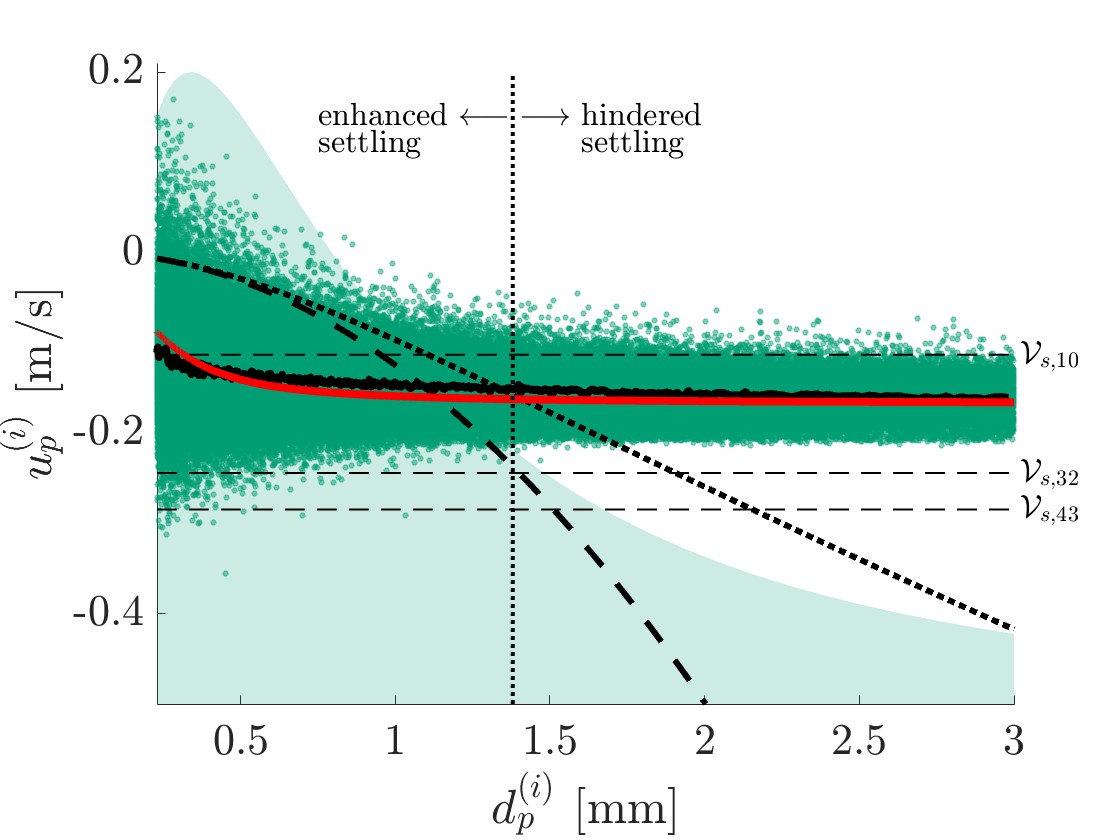} \\
    (c) Dist. B, $\langle \alpha_p \rangle = 0.01, d_p^{\text{crit}} = 1.10$ mm & (d) Dist. B, $\langle \alpha_p \rangle = 0.10, d_p^{\text{crit}} = 1.01$ mm \\
    \includegraphics[width = 0.45\textwidth]{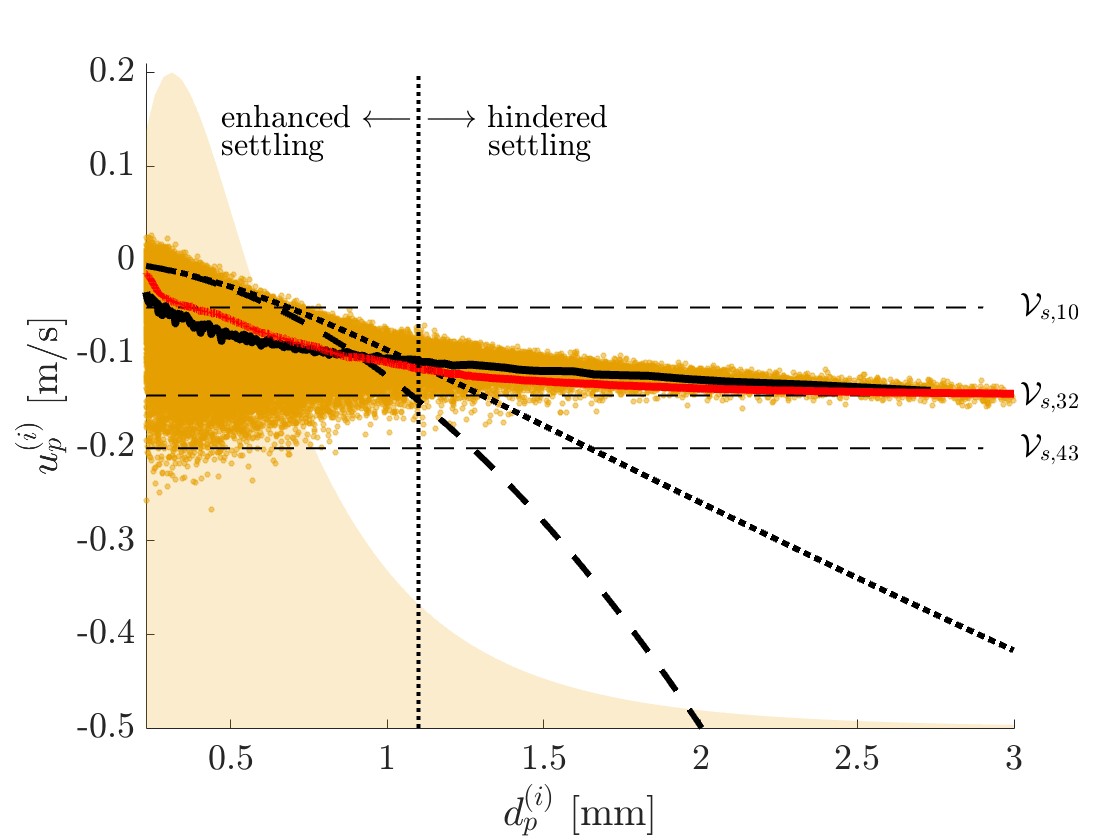} &\includegraphics[width = 0.45\textwidth]{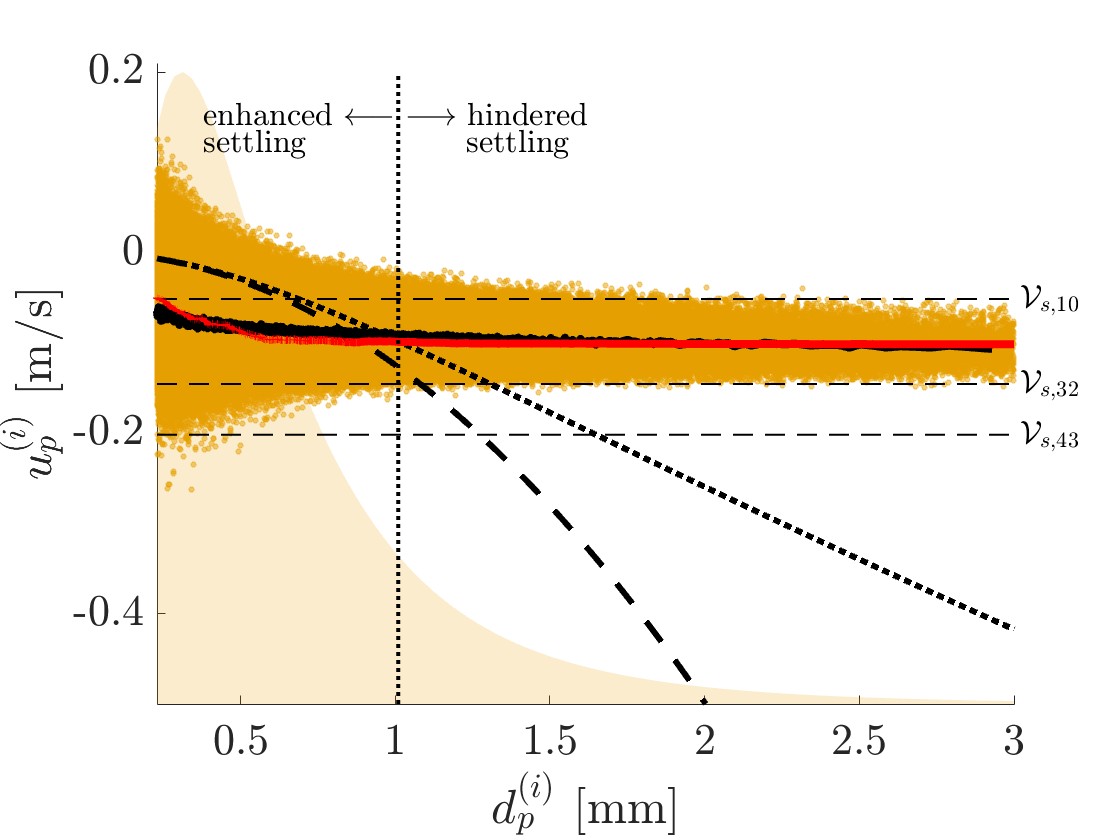} \\
    \end{tabular} 
    \caption{Settling velocity of particles with respect to particle diameter (dots) for all four polydisperse configurations (top figures). The pdf of particle diameters are shown in light shading in the background of each subfigure. The horizontal dashed lines represent the predicted settling velocity for particles corresponding to  $D_{10}$, $D_{32}$ and $D_{43}$ using Eq.~\ref{eq:Vs}. The heavy solid line is the mean velocity as a function of particle diameter. The vertical line delineates the diameter at which settling is enhanced for smaller particles and hindered for larger particles when using Eq.~\ref{eq:Vs} as a predictor (shown as a densely dotted line). The loosely dashed line represents the prediction of Stokes velocity as a function of particle size and the red solid is the prediction of settling velocity according to the proposed model shown in Eqs.~\ref{eq:VsNew} and \ref{eq:CDNew}.  The bottom of each figure shows the distribution of settling velocity corresponding to the coarse particle bins shown beneath.}
    \label{fig:settlingvsdp}
\end{figure}

Finally, we consider the settling velocity as a function of particle size, directly. To this end, Fig.~\ref{fig:settlingvsdp} plots the settling velocity against diameter for each particle in the system. A solid dark line represents the moving average of settling velocity as a function of diameter. Here, we note that, on average, the settling velocity of particles increases with increasing particle size before eventually approaching an asymptotic limit. This is observed for all the configurations considered in this work. While this is the trend on average, we also note that the smallest particles experience the widest range of velocities, with some particles having positive velocities (indicative of being swept up in `jet bypassing' events--the upward moving gas that results from a large cluster moving downward) and others having high downward velocities, due to entrainment in clusters. This phenomenon is observed across all four polydisperse configurations, however, we note that the settling velocities are more widespread for Dist. $A$ compared with $B$. We also note that the spread of particle velocities for very fine particles is more widespread at $\langle \alpha_p \rangle = 0.10$. 

An analysis of the settling behavior of gas-solid flows would be incomplete without a discussion on and comparison with existing predictions. The Stokes settling velocity, $\mathcal{V}_0 = \tau_p g$ with $\tau_p = \rho_p d_p^2/(18\mu_f)$ as previously defined, has been traditionally used in many works to predict settling behavior in multiphase flows~\citep{Shankar2021,Dietrich1982,Basson2009}, often in the absence of other \emph{a priori} models. As an improvement to this na\"{i}ve prediction, which makes gross oversimplifications, contemporary work has focused on extending depth-averaged equations that were originally used in the context of shallow-water systems to gas-solid flows~\citep{de'Michieli2023,Esposti2020,Breard2023}. To this end, \citet{de'Michieli2023} generalized the depth-averaged equations to include a dispersed phase and employed an Eulerian-Eulerian approach to simulate a wide range of volcanic phenomena. As a result of this work, a depth-averaged settling velocity, denoted as $\mathcal{V}_s$, with a model for the drag coefficient originally developed using kinetic theory arguments~\citep{Gidaspow1994} was proposed. This model is given as
\begin{align} 
\mathcal{V}_s &= \sqrt{ \frac{4}{3C_D}g d_p  \left(\frac{\rho_p-\rho_f}{\rho_f}\right)} \label{eq:Vs}\\
C_D &= \begin{cases}
\frac{24}{\text{Re}}(1 + 0.15\text{Re}^{0.687}) &  \text{Re}\leq 1000 \\
        0.44 &  \text{Re} > 1000,
    \end{cases}
\end{align}
where we note that Re$= \mathcal{V}_s d_p/\nu_f$, which requires Eq.~\ref{eq:Vs} to be solved numerically. 

When using either of these expressions to predict mean settling behavior, it is necessary to choose a mean diameter as argument. Several statistical mean diameters can be chosen for polydisperse assemblies of particles (see \ref{appendix:Diameters}), and the predictions of both of these models with three different mean diameters as argument ($D_{10}$, $D_{20}$ and $D_{32}$) are summarized in Tab.~\ref{tab:stokessetllingcomp}. In examining these predictions, we note that the $\mathcal{V}_{s}$ prediction is a more reliable predictor for the monodispersed configurations, however still not consistently accurate. For the polydispersed configurations, using the surface mean diameter, $D_{20}$, and the model for $\mathcal{V}_s$ provides the best approximation for mean settling behavior. The use of the surface mean diameter for more accurate results is perhaps intuitive, as clustering and drag are intimately connected and drag is related to the degree of surface area contact between fluid and particles.   

\begin{table}
\centering
    \begin{tabular}{c c c | c c c | c c c}
$\langle \alpha_p \rangle$ & Dist. & $\langle u_p \rangle$ & $\mathcal{V}_{0,10}$ &  $\mathcal{V}_{0,20}$ &$\mathcal{V}_{0,32}$ & $\mathcal{V}_{s,10}$ & $\mathcal{V}_{s,20}$ & $\mathcal{V}_{s,32}$\\ 
\hline 
\multirow{4}{*}{0.01} & $A_0$ & -0.20 & -0.37 & --- & --- & -0.21 & --- & --- \\
& $A_{\;}$ &  -0.17 & -0.15 & -0.21 & -0.45 & -0.11 & -0.15 & -0.24\\
& $B_0$ & -0.05 & -0.05 & --- & --- & -0.04 & --- & --- \\
 & $B_{\;}$ &  -0.08 & -0.06 & -0.08 & -0.21 & -0.05 & -0.07 & -0.14\\ [2ex]
 \multirow{4}{*}{0.10} & $A_{0}$ & -0.16 & -0.37 & --- & --- & -0.21 & --- & --- \\
  &$A_{\;}$ & -0.14  & -0.15  & -0.21 &-0.45 &-0.11  &-0.15  & -0.21\\
  & $B_0$ & -0.06 & -0.05 & --- & --- & -0.04 & --- & --- \\
 & $B_{\;}$ & -0.08  & -0.06 & -0.08 & -0.21 & -0.05 & -0.07 & -0.15\\
    \end{tabular} 
    \caption{Summary of the mean settling velocity for each configuration compared with Stokes law and the settling law of \citet{de'Michieli2019} with several mean diameters used as input. Note that for monodisperse assemblies, the single value for each settling law is reported under the columns corresponding to the result using the $d_{10}$ diameter as argument. All velocities are shown in meters per second. }
    \label{tab:stokessetllingcomp}
\end{table}

While $\mathcal{V}_{s,20}$ is reasonably predictive for \emph{global mean} settling behavior, it is important to note that both $\mathcal{V}_0$ and $\mathcal{V}_s$ fall short when predicting the settling behavior as a function of particle diameter. To illustrate this, Fig.~\ref{fig:settlingvsdp} shows the predictions of both Stokes (loosely dashed line) and the model of \citet{de'Michieli2019} (densely dashed line) as continuous functions of particle diameter.

While the model described by Eq.~\ref{eq:Vs} demonstrates a clear improvement over a Stokes assumption, neither model captures the appropriate mean settling behavior of particles over the range of diameters considered. Additionally, the concavity of these predictions is in direct opposition to the Euler--Lagrange data. The existing models suggest that the settling velocity should continue to increase with increasing diameter, whereas our data suggests that settling velocity approaches an asymptotic limit, due to cluster formation. Interestingly, we notice that there exists a critical particle diameter such that particles smaller than $d_p^{\text{crit}}$ exhibit \emph{enhanced} settling, and particles larger than $d_p^{\text{crit}}$ exhibit \emph{hindered} settling. This behavior is directly connected to clustering behavior. Smaller particles on average experience reduced drag due to their proximity to or presence within clusters, thus resulting in enhanced settling. Conversely, larger particles experience hindered settling because of their entrainment in coherent structures much larger in size than the particles themselves. This then results in particles that feel increased drag compared with the drag they would feel as a lone particle thus yielding hindered settling velocities. 

For both distributions $A$ and $B$, the critical diameter that delineates between enhanced and hindered settling is larger at $\langle \alpha_p \rangle = 0.01$ than for $\langle \alpha_p \rangle = 0.10$. Interestingly, these values correspond approximately to $d_{20}$ and $d_{30}$ for Dist. $A$ and $B$ at $\langle \alpha_p \rangle = 0.10$, respectively, perhaps indicating that for a distribution that favors fine particles, the surface area contact between the phases is more important than for distributions that contain more moderately sized particles, as in distribution $A$. At $\langle \alpha_p \rangle = 0.01$, $d_p^{\text{crit}}$ corresponds to values between $d_{30}$ and $d_{32}$ for both distributions. This suggests that for more dilute suspensions, higher-order statistical means are more relevant for predicting settling behavior. We also note that the maximum settling velocity is slightly higher for the cases at lower volume fraction for both distributions $A$ and $B$. We postulate that this is due to the nature of the cluster structures. In the denser suspensions, the clusters have a larger cross-sectional area which in turn results in a larger drag on the cluster itself. 

In light of this data, we propose an improved model in which the expression for the settling velocity, $\mathbb{V}_s$, retains the same functional expression as for $\mathcal{V}_s$, 
\begin{align} 
\mathbb{V}^{(i)}_s &= \sqrt{ \frac{4}{3C_D}g d_p  \left(\frac{\rho_p-\rho_f}{\rho_f}\right)} \label{eq:VsNew} 
\end{align}

but the expression for $C_D$ is modified as 
\begin{equation}\label{eq:CDNew}
    \mathbb{C}_D = C\left\lbrack \frac{24}{D\text{Re}_p}\left(0.2 + 0.01 \left( D \text{Re}_p\right)^{0.9}\right) +0.35 \left(D\text{Re}_p\right) \right\rbrack - \frac{2}{\left(D\text{Re}_p\right)^2 + 0.09} +  \frac{E\;\mathcal{W}}{1+\text{Re}_p^2}
\end{equation}
where $C$, $D$ and $E$ are constant coefficients specified using the highly resolved data in this study and summarized in Tab.~\ref{tab:ModelCoeffs}.

\begin{figure}
    \begin{tabular}{c c} 
    Dist. $A$, $\langle \alpha_p \rangle = 0.01$ & Dist. $A$, $\langle \alpha_p \rangle = 0.10$\\
    \includegraphics[width = 0.45\textwidth]{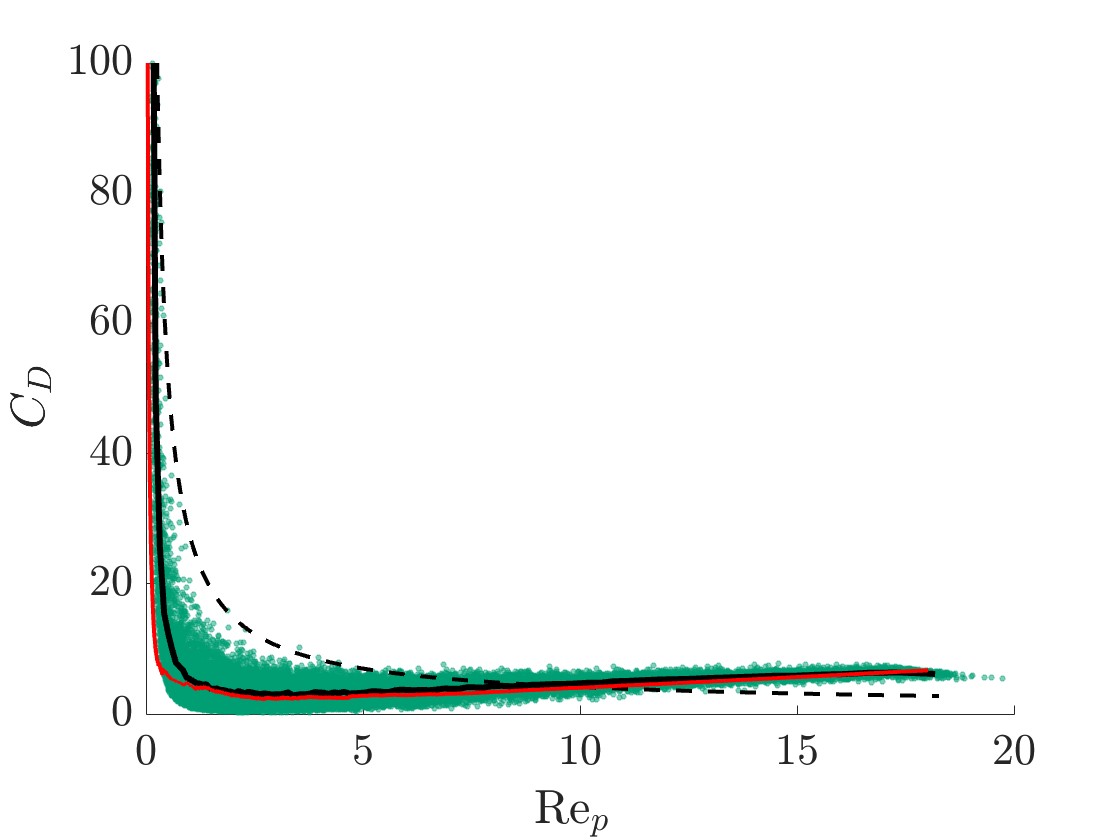} & \includegraphics[width = 0.45\textwidth]{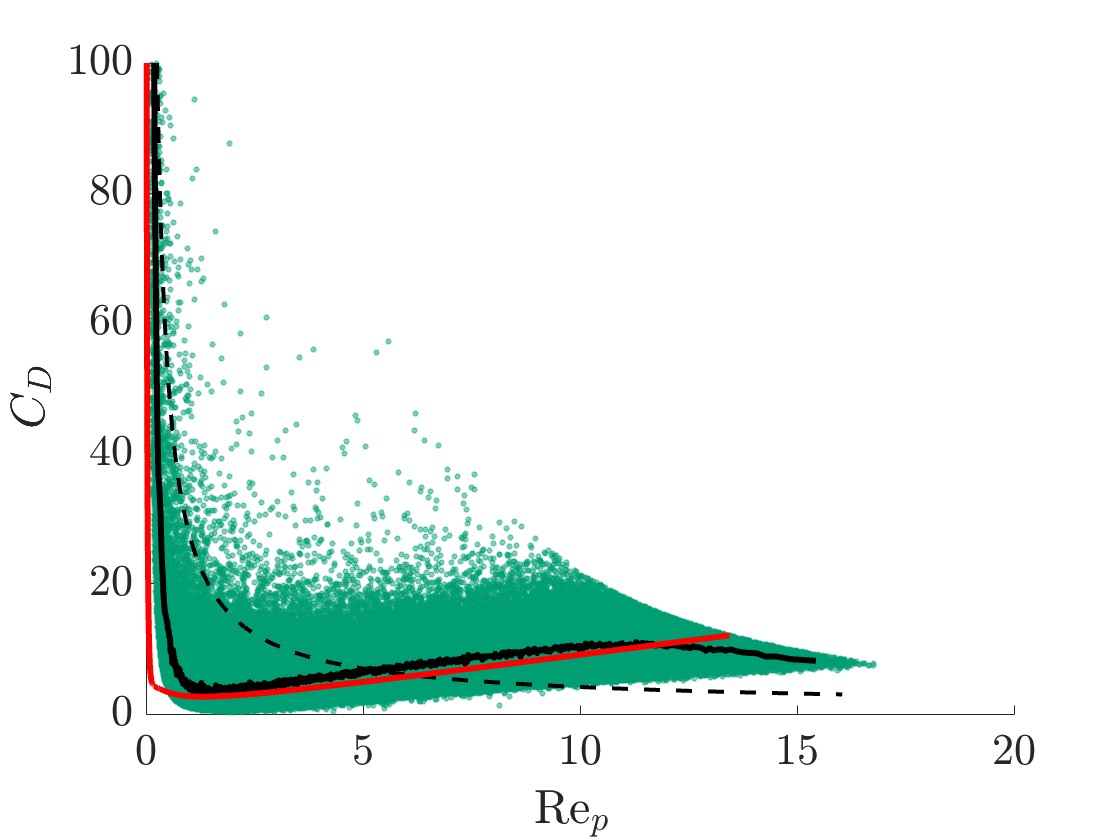}\\
        Dist. $B$, $\langle \alpha_p \rangle = 0.01$ & Dist. $B$, $\langle \alpha_p \rangle = 0.10$\\
    \includegraphics[width = 0.45\textwidth]{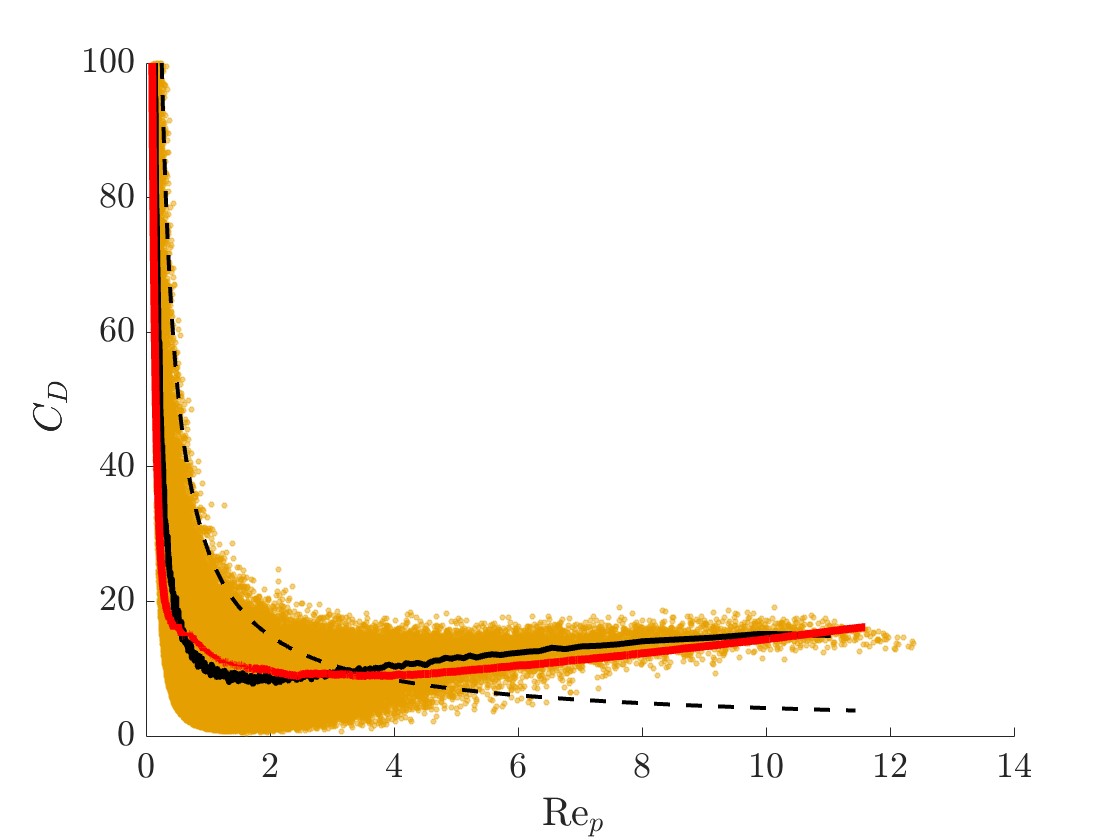} & \includegraphics[width = 0.45\textwidth]{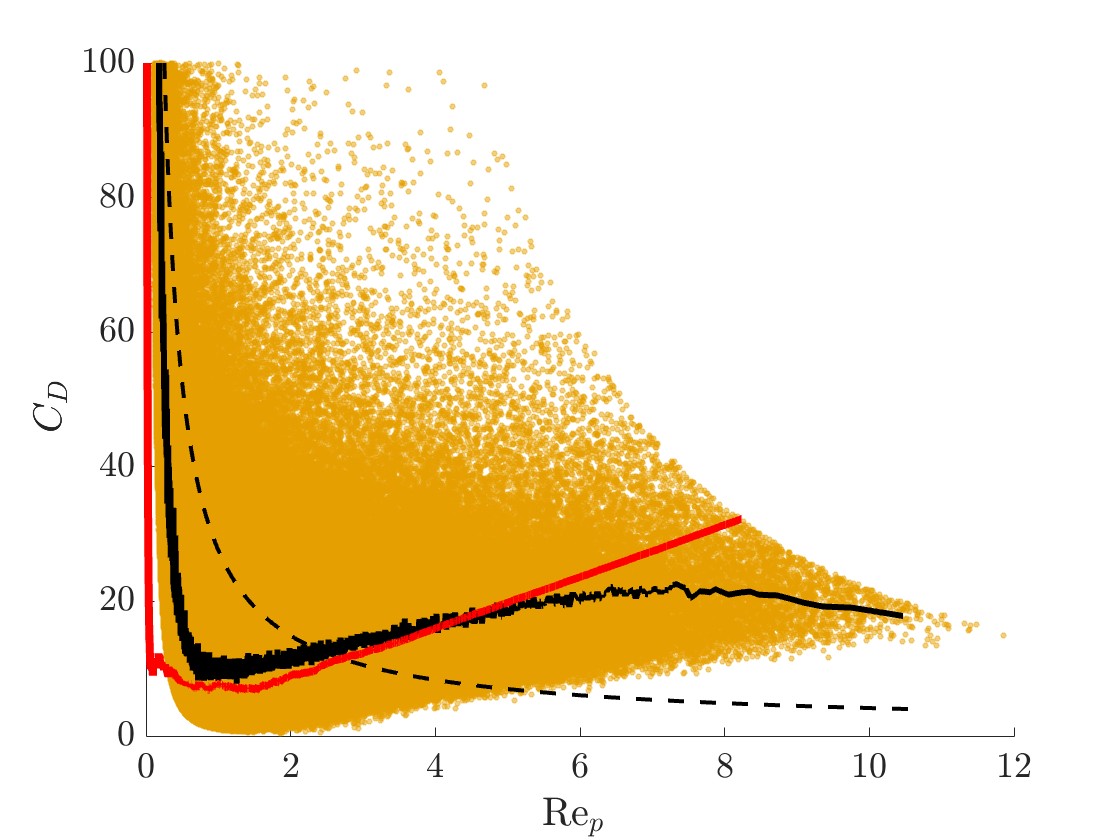}\\
    \end{tabular}
    \caption{Coefficient of drag required to ensure $\mathbb{V}_s^{(i)} = u_p^{(i)}$, using the expression in Eq.~\ref{eq:VsNew}. Values for individual particles are shown as shaded dots, the mean of this data as a function of particle diameter is shown as a dark black line and the models for $C_D$ are shown as dashed lines (black dashed is the baseline model of \citep{de'Michieli2019} and the red dashed is the proposed new model shown in Eq.~\ref{eq:VsNew}. }
    \label{fig:CdvsDp}
\end{figure}

To develop this improved model, we begin by computing the coefficient of drag, $C_D$, required to result in the settling velocity observed in the highly resolved simulations, as prescribed by Eq.\ref{eq:VsNew}. These results are shown in Fig.~\ref{fig:CdvsDp} for all the particles in the configuration. The dark shaded line represents the mean of the data as a function of particle size. In this study, we did not observe Reynolds numbers greater than $1000$, and as such have left the upper limit equal to the historical model of $C_D = 0.44$. 

\begin{table}
\centering
\begin{tabular}{c c c c c}
$\langle \alpha_p \rangle$ & Distribution & $\langle u_p \rangle$  &$\mathcal{V}_0$ & $\mathbb{V}_s$ \\ 
\hline 
 \multirow{2}{*}{0.01} & $A_0$ & -0.201 & -0.367 & -0.200 \\
 & $B_0$ & -0.047 & -0.050 &  -0.047 \\ [1.75ex]
  \multirow{2}{*}{0.10} & $A_0$ & -0.146 & -0.367 &-0.144  \\
 & $B_0$ & -0.056 & -0.050 & -0.056\\
\end{tabular} 
\caption{Predictions of the proposed model summarized alongside the mean settling velocity from the Euler--Lagrange studies and the Stokes prediction for the monodisperse assemblies. All velocities are shown in meters per second.}
\label{tab:MonoPredictions}
\end{table}

Here, we note that the model for $C_D$ closely follows the mean behavior of the highly resolved data, with the exception of deviations at very small Reynolds numbers for all configurations studied and at high Reynolds numbers in the denser suspensions. Deviations at low Reynolds numbers were required to ensure accurate predictions of the settling velocity. This stems from the fact that the way in which $C_D$ was computed does not take into account differences in the mean slip between the phases. This effect is particularly important for the small particles. Additionally, we observe that the behavior for $C_D$ at higher Reynolds number is not well captured by the proposed model for dense suspensions, however the settling velocity for larger particles is not sensitive to these deviations in the drag coefficient.

\begin{figure}
\centering
    \includegraphics[width=0.7\textwidth]{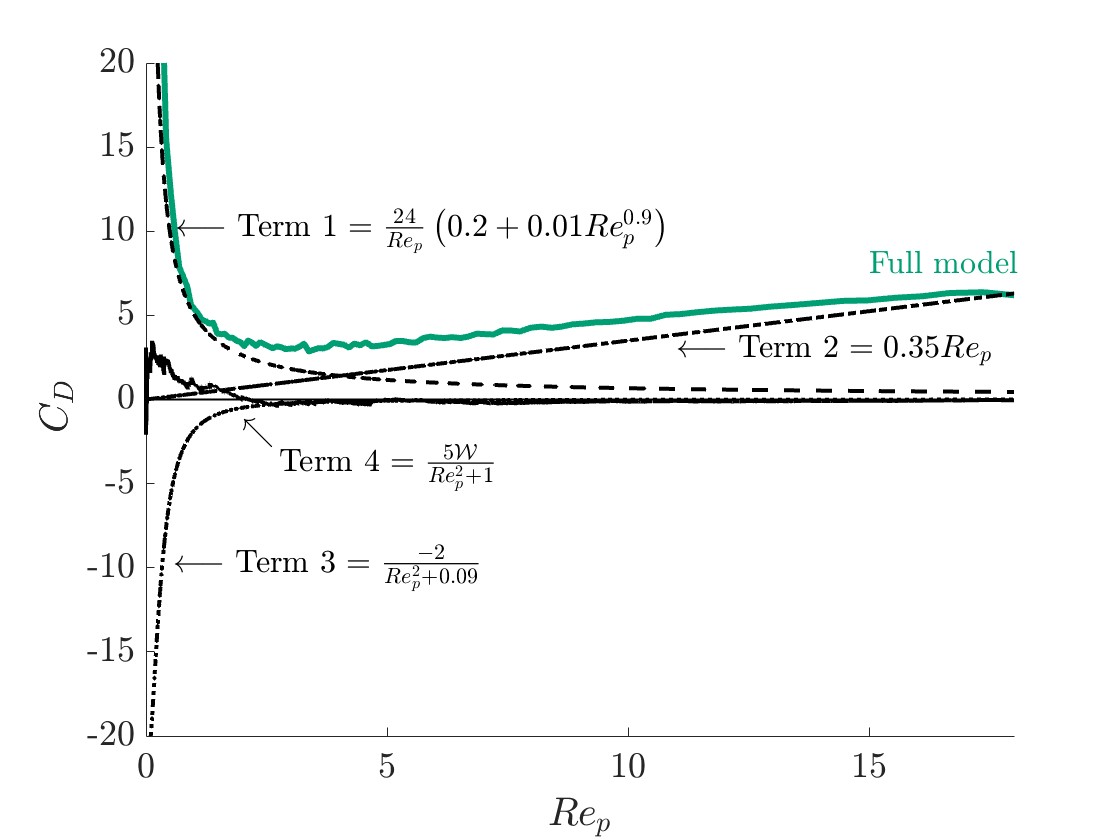}
    \caption{Exemplary case (Dist. $A$, $\langle \alpha_p \rangle = 0.01$) demonstrating the relative contribution of each of the terms in the proposed coefficient of drag model in Eq.~\ref{eq:VsNew}.}
    \label{fig:CDModelDemo}
\end{figure}

In the proposed model for $C_D$ (Eq.~\ref{eq:CDNew}), the first term is a modification of the original definition of $C_D$ from \citet{Gidaspow1994}. The second, linear term captures hindered settling for larger particles and the third term captures enhanced settling for smaller particles. Finally, the fourth term introduces stochasticity into the model through $\mathcal{W}$, a Wiener process that depends upon the particle Reynolds number and is implemented numerically as shown in Fig.~\ref{fig:Wiener}. The relative contribution to the drag coefficient for each of these terms is demonstrated for the model corresponding to Dist.$A$ in Fig.~\ref{fig:CDModelDemo}. 

\begin{figure}
\centering
\begin{verbatim}
    T = max(Rep); dt = T/Npart; 
    dW = zeros(1,Npart); 
    W = zeros(1,Npart); 
    
    dW(1) = sqrt(dt)*randn 
    W(1) = dW(1); 
    for j = 2:Npart 
        dW(j) = sqrt(dt)*randn; 
        W(j) = W(j-1) + dW(j); 
    end
\end{verbatim}
\caption{Summary of the numerical implementation of the Wiener process. This code snippet is written in the style of Matlab, where `randn' represents a normally distributed random number bounded by [0, 1].  } 
\label{fig:Wiener}
\end{figure}

The settling velocity predictions resulting from this model are shown in Fig.~\ref{fig:settlingvsdp} as solid red lines for the polydisperse configurations and summarized in Tab.~\ref{tab:stokessetllingcomp} for the monodispersed assemblies of particles. 

Since this work considered only two volume fractions and two distributions of particle diameter, we leave the values for the model coefficients summarized in Tab.~\ref{tab:ModelCoeffs} for each configuration, rather than proposing closures for each based on flow parameters. A study of additional points in the parameter space is required before a robust functional dependence of the coefficients on polydisperse and volume fraction can be developed. In particular, we postulate that $C = f(\sigma, \mu, d_{10}, d_{20}, d_{30}, d_{32}, d_{43}, ...)$ and $D = f(\langle \alpha_p \rangle)$, however this is left for future work.
\begin{table}
    \centering
\begin{tabular}{c c c c c}
$\langle \alpha_p \rangle$ & Distribution & $C$ & $D$ & $E$\\
\hline 
 \multirow{4}{*}{0.01} & $A_0$ & 1.25 & 1.00 &  \\
 & $A_{\;}$ & 1.00 & 1.00 & 5.00 \\
 & $B_0$ & 14.00 & 1.00 &\\
 & $B_{\;}$ & 3.50 & 1.00 & 5.00\\[1.75ex]
  \multirow{4}{*}{0.10} & $A_0$ & 1.25 & 3.00 & \\
  & $A_{\;}$ & 1.00 & 2.50 & 0.10\\
  & $B_0$ & 10.00 & 3.00 & \\
 & $B_{\;}$ & 2.20 & 5.00 & 5.00 \\
\end{tabular}
\caption{Summary of the model coefficients for the eight polydisperse configurations studied.}
\label{tab:ModelCoeffs}
\end{table}

\section{Conclusions}
\label{sec:conclusions} 
The work discussed here represents, to the authors' knowledge, the most extensive study of clustering and settling behavior of strongly coupled gas-solid flows at the mesoscale. In particular, we investigate the effects of polydispersity on particle clustering and settling behavior using highly-resolved Euler--Lagrangian simulations of two polydispersed distributions of particles and two analogous monodispersed distributions of particles. These four particle distributions were studied at two volume fractions ($\langle \alpha_p \rangle = 0.01$ and $\langle \alpha_p \rangle= 0.1$) and all simulation parameters were sampled to align with values typical of PDCs. 

Due to the strong coupling between the phases and the presence of a gravitational body force, coherent structures in the form of clusters spontaneously emerge, thereby altering settling behavior as compared with the uncorrelated initial dispersion of particles. To date, the extent to which polydispersity is implicated in clustering structure and consequently on settling behavior has been largely unquantified.     
To this end, this work demonstrated that mass loading--which has historically been used as a \emph{a priori} estimate for predicting clustering--is insufficient for predicting the degree of clustering, particularly for polydispersed particles. This owes to the fact that mass loading is by definition agnostic to how mass is partitioned throughout a volume. Based on the data collected in this study, we propose an alternate \emph{a priori} predictor for the degree of clustering expected in gas-solid flows, which we refer to as surface loading. This quantity leverages the mean surface diameter of particles and is shown to predict both the degree of clustering and mean settling velocity through two new models proposed in this work. 

Additionally, this work identified that the surface mean diameter, $D_{20}$, is the best mean diameter to choose for use with existing models, such as the settling velocity model of \citet{de'Michieli2019} for the prediction of \emph{mean} flow behavior. This is perhaps intuitive due to the intimate connection between granular surface area and drag, and the relationship between drag and clustering. While using this diameter produces improved predictions for global mean settling behavior, however, we demonstrate that existing models fail to capture the settling behavior across a distribution of particle diameters. Importantly, we observe that fine-grained particles experience \emph{enhanced} settling and coarse particles experience \emph{hindered} compared with existing models. Further, existing models predict that settling velocity continues to increase with particle size, which is in contrast with our observation that settling velocity increases initially with increasing particle diameter, but quickly slows and approaches an asymptotic limit. This is attributed entirely to the existence of heterogeneity in the flow and how large particles initiate and become correlated through clustering. To this end, we propose a new model for the coefficient of drag that produces accurate settling velocity predictions as a function of particle diameter. Further, our proposed model can be immediately implemented in existing solvers, such as in \citet{de'Michieli2019}.  

Although this work represents the most highly-resolved study of polydisperse clustering and settling behavior to date and an initial step toward improved reduced-order models, future efforts are needed to build robustness into the models proposed herein. In particular, additional volume fractions and polydispersed assemblies of particles are required to formulate more comprehensive closures for the model coefficients proposed in this work.

%\backsection[Supplementary data]{\label{SupMat}Supplementary material and movies are available at \\https://doi.org/10.1017/jfm.2019...}

\section{Acknowledgements} 
This work was supported by funding provided by the National Science Foundation (award number 2346972) and the National Aeronautics and Space Administration (NASA), under award number 80NSSC20M0124, Michigan Space Grant Consortium (MSGC). E.C.P.B. was supported by a NERC Independent Research Fellowship (NE/V014242/1).  The computing resources and assistance provided by Oakland University are greatly appreciated. 

%\backsection[Data availability statement]{The data that support the findings of this study are openly available in [repository name] at http://doi.org/[doi], reference number [reference number]. See JFM's \href{https://www.cambridge.org/core/journals/journal-of-fluid-mechanics/information/journal-policies/research-transparency}{research transparency policy} for more information}

%\backsection[Author ORCIDs]{Authors may include the ORCID identifers as follows.  F. Smith, https://orcid.org/0000-0001-2345-6789; B. Jones, https://orcid.org/0000-0009-8765-4321}

%\backsection[Author contributions]{Authors may include details of the contributions made by each author to the manuscript'}

\appendix

\section{A brief note on distribution terminology} \label{Appendix:sorting}

While the applications of polydisperse gas-solid flows are far-reaching, ranging from natural to industrial flows, the parameters under study in this work are drawn from PDCs. In the geoscience community, it is common to quantify the distribution of sedimentary particles using a convention referred to as the `$\phi$ scale.' Since this convention is less common in other areas of science and the applications of this study are far-reaching outside the geoscience community, we present here a very brief discussion connecting the $\phi$ scale with a standard lognormal distribution defined in terms of millimeters. 

As such, the $\phi$ scale was developed on the notion that sediment behavior is a function of particle diameter squared. Thus, $\phi \equiv -\log_2(d_p)$ is used as the basis for this scale~\citep{krumbein1936application}. It is important to note that in the definition of $\phi$ the particle diameter, $d_p$, has units of millimeters. Since the basis of the definition of $\phi$ is log$_2$, then the parameter $\phi_{\text{sorting}}$ describes the \emph{normal} distribution of the particles in terms of $\phi$. This also implies that the distribution of particle diameter in units of mm is lognormal. 

To traverse these two definitions for particle distribution, one can make use of the following relationships: 
\begin{align}
\langle d_p \rangle &= 2^{-\phi} \\
\mu &= \ln(\langle d_p\rangle) = \ln(2^{-\phi})\\
\sigma &= \phi_{\text{sorting}}^2 \ln(2)^2 
\end{align}
where $\langle d_p \rangle$ represents the mean particle diameter in millimeters and $\mu$ and $\sigma$ represent the standard lognormal parameters, such that the probability distribution function for particle diameter is defined as
\begin{equation}
f_{d_p} = \frac{1}{d_p \sigma \sqrt{2\pi}}\exp{\left( -\frac{(\ln d_p - \mu)^2}{2\sigma^2}\right)}
\end{equation}

\section{Statistical diameters} \label{appendix:Diameters}

When considering assemblies of particles of varying size, as is done in this work, it is often informative to quantify how particles are distributed by using the following statistical diameters: 

\begin{enumerate}
\item $D_{10}$ represents the mean diameter in the usual, arithmetic sense as 
\begin{equation}
\label{D10}
D_{10} = \frac{\sum\limits_{i=1}^{N_p} d_p^{(i)}}{N_p}
\end{equation}
This expression can be equivalently expressed in terms of mass fraction, where $x_j$ and $n_j$, are the mass fraction and number of particles of size $d_j$ and $J$ represents the total number of bins the particles are divided into, 
\begin{equation}
D_{10} = \frac{\sum\limits_{j=1}^{J} d_j}{\sum\limits_{j=1}^{J} {3 x_j}/{(4 \rho_p d_j^3})}
\end{equation}
\item $D_{20}$ represents the surface mean diameter. In other words, this is diameter of a monodisperse assembly of $N_p$ particles with the same total surface area of the polydisperse assembly. It is defined as  
\begin{equation}
\label{D20}
D_{20} = \sqrt{\frac{\sum\limits_{i=1}^{N_p} \left(d_p^{(i)}\right)^2}{N_p}}
\end{equation}
In terms of mass fraction, this can also be defined as, 
\begin{equation}
D_{20} = \sqrt{\frac{\sum\limits_{j=1}^{J} x_j/d_j}{\sum\limits_{j=1}^{J} x_j/(d_j^3)}}
\end{equation}
\item $D_{30}$ represents the volume mean diameter. Similar to the surface mean diameter, this value represents the diameter of a monodisperse assembly of $N_p$ particles with the same total volume as the polydisperse assembly. It is given as 
\begin{equation}
\label{D30}
D_{30} = \left(\frac{\sum\limits_{i=1}^{N_p} \left(d_p^{(i)}\right)^3}{N_p}\right)^{1/3}
\end{equation}
Equivalently in terms of mass fraction, 
\begin{equation}
D_{30} = \left(\frac{1}{\sum\limits_{j=1}^{J}\left( x_j/d_j^3\right)}\right)^{1/3}
\end{equation}
\item $D_{32}$ denotes the surface moment mean diameter, commonly referred to as the Sauter mean diameter. This is the diameter required for a monodisperse assembly of $N_p$ particles to have the same ratio of volume to surface area as the polydisperse assembly. 
\begin{equation}
\label{D32}
D_{32} = \frac{\sum\limits_{i=1}^{N_p} \left(d_p^{(i)}\right)^3}{\sum\limits_{i=1}^{N_p} \left(d_p^{(i)}\right)^2} = \frac{D_{30}^3}{D_{20}^2}
\end{equation}
In terms of mass fraction, this is defined as 
\begin{equation}
    D_{32} = \left(\sum\limits_{j=1}^{J} (x_j/d_j)\right)^{-1}
\end{equation}
\item $D_{43}$ denotes the volume moment mean diameter. This quantity is an indicator of which particle sizes contain a majority of the particle volume. 
\begin{equation}
\label{D43}
D_{43} = \frac{\sum\limits_{i=1}^{N_p} \left(d_p^{(i)}\right)^4}{\sum\limits_{i=1}^{N_p} \left(d_p^{(i)}\right)^3}
\end{equation}
Similarly, the definition based on mass fraction is given as
\begin{equation}
D_{43} = \sum\limits_{j=1}^{J}\left( x_j d_j\right) 
\end{equation}
\end{enumerate} 

The values corresponding to each of these statistical mean diameters for distributions $A$ and $B$ are summarized in Tab.~\ref{tab:rcpValues}. 

\bibliographystyle{jfm}
\bibliography{JFM_Foster_Breard_Beetham}

\end{document}